\documentclass[11pt]{article}

\usepackage{fullpage}

\usepackage{epsf}
\usepackage{fancyhdr}
\usepackage{graphics}
\usepackage{graphicx}
\usepackage{psfrag}

\usepackage[linesnumbered,ruled]{algorithm2e}% http://ctan.org/pkg/algorithm2e
\DontPrintSemicolon

\usepackage{color}

\usepackage{setspace}

\usepackage{amsthm}
\usepackage{amsfonts}
\usepackage{amsmath}
\usepackage{amssymb,bbm}
\usepackage[colorlinks, allcolors=blue]{hyperref}
\usepackage{tabularx}%

\usepackage{enumitem}
\usepackage{booktabs}
\usepackage{caption,subcaption}

\usepackage{mathtools}

\usepackage{tikz}
\usetikzlibrary{calc, shapes, backgrounds}
\usepackage{subcaption} %for side by side figures.

\numberwithin{equation}{section}
\theoremstyle{plain}

\newtheorem{theorem}{Theorem}[section]
\newtheorem{lemma}[theorem]{Lemma}

\newtheorem{conj}[theorem]{Conjecture}

\newtheorem{proposition}{Proposition}
\usepackage{color}
\usepackage{amssymb}
\usepackage{comment}
\usepackage{pdfpages}
\usepackage{graphicx}% Include figure files
\usepackage{dcolumn}% Align table columns on decimal point

\numberwithin{equation}{section}

\def \be{\begin{equs}}
\def \ee{\end{equs}}

\def \E{\mathbb{E}}

\newcommand{\abss}[1]{\left| #1 \right |}
\newcommand{\parenth}[1]{\left( #1 \right)}
\newcommand{\samples}{\ensuremath{n}}
\newcommand{\obs}{\samples}
\newcommand{\brackets}[1]{\left[ #1 \right]}

\newcommand\mle{\text{mle}}
\usepackage{afterpage}

\usepackage{amsmath}

\usepackage[authoryear, round]{natbib}
\usepackage{fullpage}

\usepackage{xr}
\begin{document}

\title{Weak separation in mixture models\\and implications for principal stratification\thanks{email: \texttt{afeller@berkeley.edu}. AF and LM gratefully acknowledge financial support from the Spencer Foundation through a grant entitled ``Using Emerging Methods with Existing Data from Multi-site Trials to Learn About and From Variation in Educational Program Effects,'' and from the Institute for Education Science (IES Grant \#R305D150040). NSP is partially supported by an ONR  grant. We would like to thank Isaiah Andrews, Peter Aronow, Peter Bickel, Alex D'Amour, Peng Ding, Fabrizia Mealli, Christian Robert, Don Rubin, Dylan Small, Aaron Smith, Weixin Yao, and members of the Spencer group for helpful comments and discussion, as well as seminar participants at the Atlantic Causal Inference Conference and Joint Statistical Meetings. All opinions expressed in the paper and any errors that it might contain are solely the responsibility of the authors.}}
\author{Avi Feller, Evan Greif, Nhat Ho, Luke Miratrix, and Natesh Pillai \\[0.5em] UC Berkeley and Harvard University}
\date{August 2019}

\maketitle
\thispagestyle{empty}
\pagenumbering{gobble}

\begin{abstract}
Principal stratification is a widely used framework for addressing post-randomization complications. After using principal stratification to define causal effects of interest, researchers are increasingly turning to finite mixture models to estimate these quantities. Unfortunately, standard estimators of mixture parameters, like the MLE, are known to exhibit pathological behavior. We study this behavior in a simple but fundamental example, a two-component Gaussian mixture model in which only the component means and variances are unknown, and focus on the setting in which the components are weakly separated. 
In this case, we show that the asymptotic convergence rate of the MLE is quite poor, such as $O(n^{-1/6})$ or even $O(n^{-1/8})$.
We then demonstrate via theoretical arguments as well as extensive simulations that, in finite samples, the MLE behaves like a threshold estimator, in the sense that the MLE can give strong evidence that the means are equal when the truth is otherwise. We also explore the behavior of the MLE when the MLE is non-zero, showing that it is difficult to estimate both the sign and magnitude of the means in this case. 
We provide diagnostics for all of these pathologies and apply these ideas to re-analyzing two randomized evaluations of job training programs, JOBS II and Job Corps. Our results suggest that the corresponding maximum likelihood estimates should be interpreted with caution in these cases. 
\end{abstract}

\clearpage
\pagenumbering{arabic}
\onehalfspacing
%\doublespacing

%%%
%%% INTRODUCTION
%%%
\section{Introduction}
\label{sec: intro}

Finite mixture models are notorious for giving pathological results~\citep{Redner:1984}; indeed, Larry Wasserman has called finite mixtures the ``Twilight Zone of Statistics''~\citep{Wasserman_blog}. Our motivation for this paper is to understand how the pathological features of \emph{weakly separated} finite mixture models affect inference for component means, especially with respect to estimating causal effects in the principal stratification framework, an important example of such inference.
 
Principal stratification is a widely used approach for addressing post-randomization complications, including noncompliance with treatment assignment~\citep{Frangakis:2002wp}. Typically, the goal is to estimate causal effects within partially latent subgroups known as principal strata. While there are many possible ways to estimate these principal causal effects, the most common approach is via finite mixture models, treating the unknown principal strata as mixture components~\citep{Imbens}. To date, scores of applied and methodological papers have relied on finite mixtures to estimate causal effects, both explicitly and implicitly.

To present our main results, we construct a simple two-parameter model that captures the essential features of the problem: maximum likelihood estimation for the component means and variances in a two-component location-scale mixture of Gaussian distributions,
\begin{equation}
\label{eq: mix}
Y_i \stackrel{\text{iid}}{\sim} \pi N(\mu_0, \sigma_{0}) + (1 - \pi) N(\mu_1, \sigma_{1}),
\end{equation}
where the mixing proportion, $\pi \in (0,1)$, is assumed to be known.
%we assume $\pi \leq 1/2$ for simplicity, but analogous results always follow for $\pi \in (1/2,1)$).

While the two-component finite mixture model in \eqref{eq: mix} is a toy example in some settings, it is a fundamental structure in many causal inference problems. For instance, in the canonical example of noncompliance in a randomized trial \citep{angrist1996identification}, individuals randomly assigned to the treatment group who actually receive the treatment are a mixture of Compliers and Always Takers. Assuming that individual outcomes follow a Normal distribution yields the mixture model in \eqref{eq: mix}. 
Thus understanding the difficulties of component-specific inference are vital to estimating parametric principal stratification models.

The asymptotic properties of the MLE for the component means in Equation~\eqref{eq: mix} are well established in two settings. First, when the difference in means, $\Delta \equiv \mu_1 - \mu_0$, is fixed, the MLE has strong asymptotic guarantees, including consistency and parametric convergence~\citep{everitthand, chen2016consistency}. Second, when the mixture is degenerate, \textit{i.e.}, $\Delta = 0$, the MLE has at most $O(n^{-1/4})$ convergence~\citep{Chen, Jonas-2017}. This is closely related to the problem of testing the number of components in a finite mixture~\citep{mclachlan2004finite}.

\begin{figure}[bt]
\centering
		 \begin{subfigure}[b]{0.5\textwidth}
				\includegraphics[width = \textwidth]{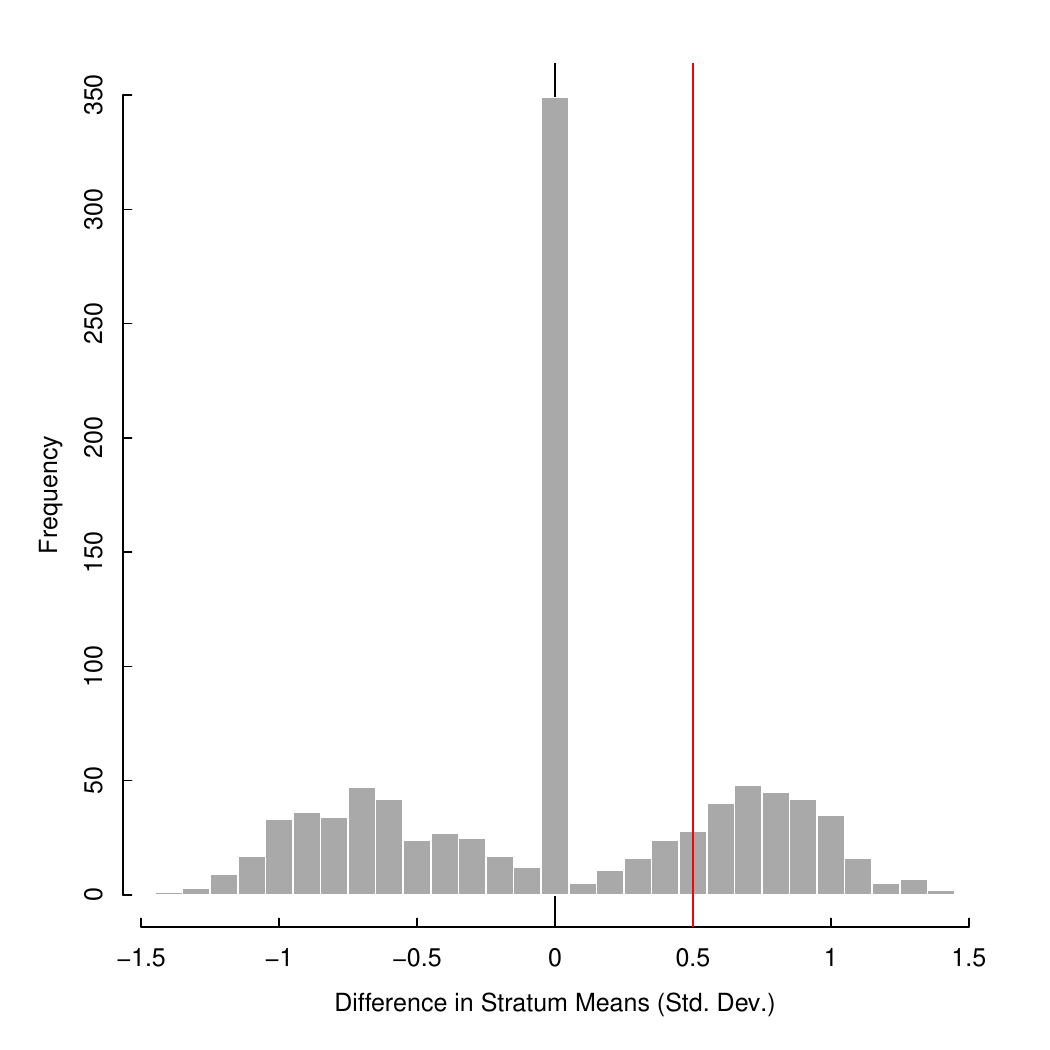}
				\caption{\textit{JOBS II:} $N = 132$, $\pi = 0.45$}
				\label{fig:jobs_mle_example_50SD}
		\end{subfigure}%
		\;\begin{subfigure}[b]{0.5\textwidth}
				\includegraphics[width = \textwidth]{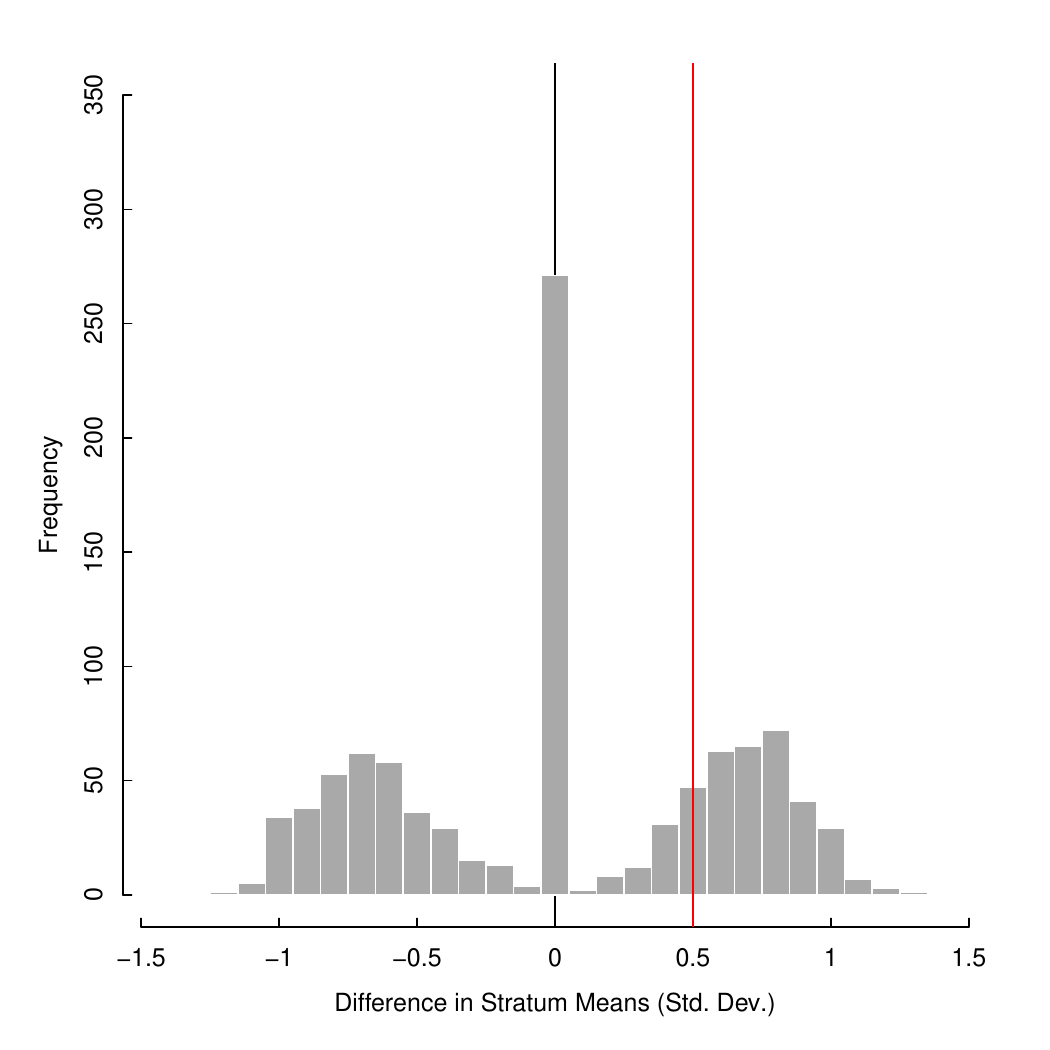} 
				\caption{\textit{JobCorps:} $N = 3,371$, $\pi = 0.06$}
				\label{fig:jobcorps_mle_example_50SD}
		\end{subfigure}
\caption{Distribution of $\widehat{\Delta}^\mle$ for 1000 fake data sets designed to reflect the JOBS II and JobCorps studies. Data sets were generated from the two-component homoskedastic Normal mixture model in Equation~\eqref{eq: mix} with $\Delta = 0.5$ and, respectively, (a) $N=132$ and $\pi=0.45$ and (b) $N = 3,371$ and $\pi = 0.06$.\label{fig:mle_examples}
}
\end{figure}

In this paper, we focus on the behavior of the MLE when $\Delta$ is \emph{small but not zero}. This ``intermediate sample size regime" is an  important case in practice and is especially relevant for principal stratification models. To set the stage, Figure~\ref{fig:mle_examples} shows the distribution of the MLE of $\Delta$ for 1000 synthetic data sets generated from Equation~\eqref{eq: mix} for two settings.
The sample sizes and mixing proportions match those in our two key principal stratification examples, JOBS II and Job Corps. The assumed difference in component means is $\Delta = 0.5$ standard deviations, which is quite large for many social science applications but smaller than in textbook examples of well-separated components. In both cases, the distribution of the simulated MLEs is markedly non-Normal. 
Both distributions have three notable features. First, there is a large point mass at zero. Second, a considerable portion of simulated MLEs have the opposite sign from the truth. Finally, simulated MLEs that are non-zero and have correct sign are not centered at the true value. To emphasize, these features are not due to model mis-specification: we estimate the MLE using the true model.

\subsection{Main contributions of our paper}
In this paper, we give theoretical explanations for some of the practical difficulties encountered in estimation in two component finite mixture models, as shown in Figure \eqref{fig:mle_examples}, and, based on our findings, suggest guidance for practice.

We first, in Section \ref{sec:asymptotics}, study the asymptotic properties of the MLE of the two component model  \eqref{eq: mix} in the ``intermediate sample size regime'' when $\Delta \rightarrow 0$
as $n \rightarrow \infty$. 
This framework adequately captures weakly separated mixture components in relation to the sample size.
Even for the basic model \eqref{eq: mix}, not much seems to be known about the convergence rate of the MLE in this regime, especially when $\sigma_{0}$ and $\sigma_{1}$ are unknown. 
We first establish the convergence rate of the MLE, resulting in several interesting findings for the model in \eqref{eq: mix} when $\sigma_0 = \sigma_1$.
When $\sigma \equiv \sigma_0 = \sigma_1$ is known and $\pi \neq \frac{1}{2}$, the convergence rate can and does reach $O(n^{-1/6})$ up to logarithmic factors. 
%(For $\pi = \frac{1}{2}$, the difference can only be estimated up to a sign due to identifiability issues.)
%\footnote{For $\pi = \frac{1}{2}$, the difference can only be estimated up to a sign due to identifiability issues.}  
This is \emph{worse} than the rate set for the degenerate case where $\Delta=0$, suggesting that small but non-zero separations are particularly difficult to estimate well.
In such scenarios, our theoretical results explain the empirically observed difficulties in estimating $\Delta$ shown in Figure \ref{fig:mle_examples}. 
For $\pi = \frac{1}{2}$, we can only estimate the difference up to a sign due to identifiability issues.
In this case, the convergence rate for estimating the magnitude of the parameter is a more rapid --- yet still slow --- $O(n^{-1/4})$.

When $\sigma$ is not known the worse-case convergence rate of the MLE remains $O(n^{- 1/ 6})$ for the $\pi \neq \frac{1}{2}$ setting but falls to $O(n^{-1/ 8})$ for the $\pi = \frac{1}{2}$ setting --- an order of magnitude worse than when $\sigma$ was assumed known.
These results are quite novel and delicate to derive, as we have to carefully account for the interaction between the location and scale parameters.
Interestingly, the results together show that while the convergence is faster for the symmetric case than the asymmetric case in the known variance regime, it is slower in the unknown variance regime.

After presenting our convergence results, we turn to the practical difficulties in estimating $\Delta$ and formalize the phenomenon of the large point mass at zero shown in Figure \ref{fig:mle_examples}. We call this phenomenon \textit{pile up}. Specifically, we show via a mix of simulations and theoretical arguments that, in certain intermediate sample size regimes, $\widehat{\Delta}^{\text{mle}} = 0$ with very high probability even though $\Delta \neq 0$. Thus, the MLE behaves like a threshold estimator analogous to the classic Hodges estimator~\citep[see][]{van2000asymptotic}. 
We then show that pile up occurs when the overall mixture variance is less than the within-component variance. 
To the best of our knowledge, we are the first to document this pile-up phenomenon in finite mixtures.

Next, we turn to using higher-order mixture moments for diagnosing pathologies with the MLE. First, we use these moments to bound the probability of pile up given either the realized data set or population parameters. We then discuss the classic problem of choosing the correct mode in a bimodal likelihood and argue that it is particularly difficult here. We show that this problem corresponds to estimating the sign of $\Delta$ (\textit{i.e.}, the relative ordering of $\mu_0$ and $\mu_1$) and demonstrate how to use the third moment of the mixture distribution to assess the probability that this occurs. We combine these results with extensive simulations to show that, across a range of reasonable settings, the sign of the MLE for $\Delta$ is no better at predicting the true sign than a coin flip. 

We finally apply these mixture results to estimating principal stratification models in two randomized evaluations of job training programs, JOBS II~\citep{vinokur1995impact} and JobCorps~\citep{Schochet:2008kj}. These two examples have been the focus of several prominent papers using finite mixtures for principal stratification~\citep[\textit{e.g.},][]{Zhang:2009ku, Mealli:2013be, Frumento:2012gz} and highlight two main use cases for this framework. 
For both data sets, we slightly simplify the problem to isolate the pathologies of the finite mixtures. 
We then assess the observed mixture distributions using the diagnostics we propose and find that pathologies are quite likely. 
Consequently, we do not have high confidence in the quality of the maximum likelihood estimates of $\widehat{\Delta}^{\text{mle}} = 0$ for JOBS II and an implausibly large $\widehat{\Delta}^{\text{mle}}$ for JobCorps. Our overall findings suggest that finite mixture models should be used with caution in settings such as these.

Overall, the implications for parameter estimation in finite mixtures are both novel and important. In particular, there is a longstanding consensus in finite mixture modeling that the MLE can behave poorly when components are not well separated~\citep{Redner:1984}. Indeed, several experienced researchers have told us that estimating component-specific parameters is ``hopeless'' in the settings we consider. While we agree with this assessment, we argue that there are no clear guidelines for researchers in practice. In particular, how do researchers know when components are separated ``enough'' and what happens if they are not? This is especially important because, in settings with insufficient information in the data, the MLE gives a very plausible value of zero rather than `NA.' We believe that the framework we lay out here is an important next step towards deeper understanding of these issues.

\paragraph{Paper plan.}
Section~\ref{sec:asymptotics} describes the asymptotic behavior of the MLE under weak separation. 
Section~\ref{sec:weak_separation} explores the non-asymptotic behavior of the MLE and characterizes pile up. 
Section~\ref{sec:mom_main} uses the mixture moments for constructing diagnostics for the MLE.
Section~\ref{sec:causal} gives a brief overview of the principal stratification framework and the connection to finite mixture models as well as an analysis of JOBS II. 
Section~\ref{sec:discussion} provides additional discussion on implications for practice and possible research directions. Finally, the supplementary materials address several points that go beyond the main text, including proofs.

\paragraph{Notation.} For any two densities $p$ and $q$ (with respect to Lebesgue measure $\mu$), the variational distance between $p$ and $q$ is given by $V(p,q) = (1/2)\int \abss{p-q}d\mu$. Additionally, the squared Hellinger distance between $p$ and $q$ is given by $h^{2}(p,q)= (1/2)\int \parenth{p^{1/2}-q^{1/2}}^2d\mu$. Furthermore, the expression $a_{n} \gtrsim b_{n}$ is used to denote $a_{n} \geq C b_{n}$ for some $C$ that is independent of $n$.

\subsection{Related literature and previous work \label{sec:related}}
There is a vast literature on inference in finite mixture models, dating back to the seminal work of~\citet{Pearson:1894dt}. For thorough reviews, see \citet{everitthand},~\citet{Redner:1984},~\citet{Titterington},~\citet{mclachlan2004finite}, and~\citet{mclachlan2019finite}.~\citet{fruhwirth2006finite} focuses on the Bayesian paradigm;~\citet{Lindsay:1995bm} gives an overview of moment estimators; and~\citet{moitra_book} discusses relevant results from machine learning. We briefly highlight several relevant aspects of this literature. 

First, there has been extensive research on the asymptotic behavior of finite mixtures models.~\citet{chen2016consistency} gives a recent, comprehensive review. Much of this literature, however, is about the problem of testing the order of the finite mixture~\citep[see][]{mclachlan2004finite}. There are several recent papers that instead address estimation.~\citet{Chen:2014eh} focuses on estimating the mixing proportion when components are only weakly separated.~\citet{Ho-Nguyen-Ann-16} gives results for the over-specified location-scale Gaussian mixtures.~\citet{gadat2016parameter} study the convergence rate of $\mathbb{L}^2$-norm estimators for a few settings of two component models.
Finally,~\citet{Anima-moment, hardt2015tight, Wu-denoise} explore the asymptotic properties of method of moments estimators in rather general settings of Gaussian mixtures.

Second, the problem of weak separation is a special case of the \textit{weak identification} problem especially common in econometrics. There are many examples of weak identification in other settings, including the \textit{weak instruments problem}~\citep{Staiger:1997} and the \textit{moving average unit root problem}, which is the source of the term \textit{pile up}~\citep{shephard1990probability,Andrews:2012jv}. See also~\citet{Chen:2014eh}.

Finally, although the technical discussion focuses narrowly on finite mixtures, our motivation remains the broader question of inference for causal effects within principal strata. To date, only a handful of papers have directly addressed the finite sample properties of mixtures for causal inference.~\citet{Griffin:2008bc} conduct extensive simulations and conclude that principal stratification models are generally impractical in social science settings.~\citet{Mattei:2013fb} caution that univariate mixture models often yield poor results and suggest jointly estimating effects for multiple outcomes, such as by assuming multivariate Normality.~\citet{Mercatanti:2013jm} proposes an approach for inference with a multimodal likelihood in the principal stratification setting. \citet{Frumento:2016tk} explore methods for quantifying uncertainty in principal stratification problems when the likelihood is non-ellipsoidal. See also~\citet{Chung:2004di},~\citet{zhang2008evaluating},~\citet{Richardson:2011wy}, and~\citet{Frumento:2012gz}.

%%%
%%% ASYMPTOTICS
%%% 
\section{Asymptotic properties of the MLE: Phase transition}
\label{sec:asymptotics}
In this section, we study the asymptotic behavior of the MLE under two distinct but representative settings of model~\eqref{eq: mix}:
first, when the variances $\sigma_{0}$ and $
\sigma_{1}$ are assumed known and equal; 
second, when the variances $\sigma_{0}$ and $\sigma_{1}$ are unknown 
but assumed to be equal. 
Overall, we demonstrate that worst-case convergence when the components are close together is generally slow.

\subsection{Known variances setting} 
\label{sec:known_variance}
Motivated by the illustrative simulations in Figure~\ref{fig:mle_examples}, we now explore the properties of the MLE, $\widehat{\Delta}^{\mle}$, when $\Delta$ is small but non-zero. In the classical asymptotic regime, where $\Delta$ is fixed as in Equation~\eqref{eq: mix}, it is immediate that $\widehat{\Delta}^{\mle}$ has a parametric rate of convergence in this simple example \citep[see][]{Redner:1984,Chen}. However, as shown in Figure~\ref{fig:jobs_mle_example_50SD}, this asymptotic regime can be a poor approximation to reality when components only have moderate separation. We therefore consider an asymptotic regime in which $\Delta_n$ shrinks as $n$ increases. 
Our core finding is that, under this regime in which the two components are only slightly separated and the variance is known, the convergence rate of the MLE for the difference in means is quite poor.

Under the assumption that variances are known, we re-parametrize Equation~\eqref{eq: mix} and assume that $Y_{i}, i \in \{1,\ldots,n\}$, are i.i.d. samples from the model:
\begin{equation}
\label{eq: block}
Y_{i} \stackrel{\text{iid}}{\sim} \pi N\left( \mu - \delta_n, \sigma \right) + (1-\pi)N\left(\mu + c\delta_n, \sigma \right),
\end{equation}
where $c := \frac{\pi}{1-\pi}$ and $\delta_n \in \Theta$ is a free parameter that varies with $n$.
We assume the equal variance case of $\sigma_0 = \sigma_1 = \sigma$ for a known $\sigma$.
Relative to Equation~\eqref{eq: mix}, $\mu_0 = \mu-\delta_n$, $\mu_1 = \mu+c\delta_n$, $\Delta = (1 + c)\delta_n$, and $\mu$ is the overall mean, $\mathbb{E}Y_i = \mu$. 
For simplicity, we set $\mu = 0$; all of the results in this section are applicable for any $\mu \in \mathbb{R}$. 
When $\mu = 0$ then the $\delta_{n}$ parameter is both the (negative) location of the first component as well as scaling of the separation of components $\Delta$; it thus corresponds to both a location and a separation parameter.
We focus on this separation parameter $\delta_{n}$ for ease of mathematical derivations; because $\Delta$ from Equation \eqref{eq: mix} is a constant re-scaling of $\delta$, all the asymptotic results equally apply. 
We further assume that $\delta_{n} \in \Theta$ where $\Theta$ is a compact subset of $\mathbb{R}$ and $0 \in \Theta$.
Finally, define $\widehat{\delta}_{n}^{\mle}$ as the MLE for $\delta_{n}$ for the model in \eqref{eq: block}. 

The following result shows the convergence rates of MLE for \eqref{eq: block} where the variances are assumed to be known:
\begin{theorem} \label{theorem:convergence_rate}
For the model~\eqref{eq: block}, the following holds for any $\epsilon > 0$
\begin{itemize}
\item[(a)] (Asymmetric regime) When $\pi \in (0,1/2)$, then 
\begin{eqnarray}
C_{1}(\epsilon) \left( \frac{1}{n}\right)^{1/6} \leq \sup \limits_{\delta_{n} \in \Theta_{1,n}(\epsilon)} \mathbb{E}_{\delta_{n}} \left( |\widehat{\delta}_{n}^{\mle} - \delta_{n}|\right) \leq C_{2}(\epsilon) \left(\frac{\log n}{n} \right)^{1/6}, \nonumber
\end{eqnarray}
where $\Theta_{1, n}(\epsilon) = \left\{ \delta: |\delta| \leq n^{- 1/6 + \epsilon} \right\}$.
\item[(b)] (Symmetric regime) When $\pi = 1/2$, then 
\begin{eqnarray}
C_{1}(\epsilon) \left( \frac{1}{n} \right)^{1/4} \leq \sup \limits_{\delta_{n} \in \Theta_{2,n}(\epsilon)} \mathbb{E}_{\delta_{n}} \left( \abss{|\widehat{\delta}_{n}^{\mle}| - |\delta_{n}|} \right) \leq C_{2}(\epsilon) \left(\frac{\log n}{n} \right)^{1/4}, \nonumber 
\end{eqnarray}
where $\Theta_{2, n} (\epsilon) = \left\{ \delta: |\delta| \leq n^{- 1/4 + \epsilon} \right\}$.
\end{itemize}
Here, $\mathbb{E}_{\delta_{n}}$ denotes the expectation taken with respect to the product measure with mixture density of $Y_{1},\ldots,Y_{n}$ under the model~\eqref{eq: block}. Furthermore, $C_{1}( \epsilon)$ and $C_{2} (\epsilon)$ are two positive constants depending only on $\epsilon$. 
Symmetry gives an analogous result for $\pi \in (1/2,1)$. 
\end{theorem}
\noindent
The proof of Theorem~\ref{theorem:convergence_rate} is provided in Appendix~\ref{subsec:proof:theorem:convergence_rate}. 
The variance parameter, $\sigma$ is subsumed in the constants and does not impact the rates.

Prior work \citep{Chen} has shown that when $\delta_n = 0$ the rate is of order $n^{-1/4}$ for the asymmetric case; the above therefore shows that there exists some $\delta_n \neq 0$ in a neighborhood of 0 where convergence is even worse than this degenerate case.
In particular, an immediate consequence of this theorem is that, for $\pi \neq 1/2$, there exists a sequence of $\delta_{n}$ going to 0 at no more than a $n^{-1/6}$ rate such that the error of the MLE is also of order $n^{-1/6}$.

For the symmetric regime we are simply looking at difference in magnitude, not sign.
This is because when $\pi = 1/2$ the sign of $\delta_{n}$ is not identifiable, and we find that
\begin{align*}
\sup \limits_{\delta_{n} \in \Theta} \mathbb{E}_{\delta_{n}} |\widehat{\delta}_{n}^{\mle} - \delta_{n}| \gtrsim n^{-1/r},
\end{align*}
for any $r \geq 2$ and for any fixed parameter space $\Theta$. Here, $\mathbb{E}_{\delta_{n}}$ denotes the expectation taken with respect to product measure with mixture density of $Y_{1},\ldots,Y_{n}$ under the model~\eqref{eq: block}; see the Appendix~\ref{subsec:extra_results} for the proof.

\paragraph{Connections to the Wasserstein metric.} 
The above connects to the Wasserstein metric, which has recently been used to study parameter estimation in mixture models \citep{Nguyen-13, Ho-Nguyen-Ann-16, Jonas-2017}, for additional interpretation of the results in Theorem \ref{theorem:convergence_rate}.
In particular, let $\widehat{G}_{n}^{\mle}$ denote a probability measure (or equivalently mixing measure) with two atoms $(-\widehat{\delta}_{n}^{\mle},c\widehat{\delta}_{n}^{\mle})$ whose weights are $(\pi,1-\pi)$ and $G_{n}$ a probability measure with two atoms $(-\delta_{n},c\delta_{n})$ whose weights are $(\pi,1-\pi)$, then we can verify that the results of Theorem \ref{theorem:convergence_rate} are equivalent to
\begin{eqnarray}
C_{1}(\epsilon) n^{-1/6} \leq \sup \limits_{\delta_{n} \in \Theta_{1,n}(\epsilon)} \mathbb{E}_{\delta_{n}} \left( W_{3}(\widehat{G}_{n}^{\mle},G_{n})\right) \asymp \sup \limits_{\delta_{n} \in \Theta_{1,n}(\epsilon)} \mathbb{E}_{\delta_{n}} \left(|\widehat{\delta}_{n}^{\mle} - \delta_{n}| \right) \leq C_{2}(\epsilon) \left(\frac{\log n}{n} \right)^{1/6}  \nonumber
\end{eqnarray}
under the asymmetric regime and 
\begin{eqnarray}
C_{1}(\epsilon) n^{-1/4} \leq \sup \limits_{\delta_{n} \in \Theta_{2,n}(\epsilon)} \mathbb{E}_{\delta_{n}} \left( W_{2}(\widehat{G}_{n}^{\mle},G_{n}) \right) \asymp \sup \limits_{\delta_{n} \in \Theta_{2,n}(\epsilon)} \mathbb{E}_{\delta_{n}} \left( \abss{\abss{\widehat{\delta}_{n}^{\mle}} - |\delta_{n}|} \right) \leq C_{2}(\epsilon) \left(\frac{\log n}{n} \right)^{1/4} \nonumber
\end{eqnarray}
under the symmetric regime.

\subsection{Unknown equal variances setting} 
\label{Section:MLE_equal_variances}

We now show that our previous results still generally hold when we relax the restriction that the variances are known. 
For the unknown equal variances setting, we assume that $Y_{1},\ldots,Y_{n}$ are i.i.d. samples from a two component location-scale Gaussian mixture with density
\begin{align}
\label{eqn:general_model}
Y_{i} \stackrel{\text{iid}}{\sim} \pi N\left(\mu - \delta_n, \sigma_n\right) + (1-\pi)N\left(\mu + c\delta_n, \sigma_n\right) .
\end{align}
Here, $\delta_{n}$ and $\sigma_{n}$ change with the sample size $n$ and converge to some limit points. 
We assume $\sigma_{n} \in \Omega$, a compact subset of $\mathbb{R}_{+}$. 
We set the overall mean of $\mu=0$ for convenience as before; $\delta_{n}$ is again a scaling of the gap between the two mixture means.
We define $(\widehat{\delta}_{n}^{\mle}, \widehat{\sigma}_{n}^{\mle})$ as the MLE for the separation and scale parameters for the model in~\eqref{eqn:general_model}. 
Unlike the previous convergence results with $\widehat{\delta}_{n}$ in the case with known variance, the convergence rates of $\widehat{\delta}_{n}$ and $\widehat{\sigma}_{n}$ are much harder to establish due to the strong dependence between the seperation parameter $\delta$ and scale parameter $\sigma$, which is determined by the following partial differential equation (PDE):
\begin{eqnarray}
\dfrac{\partial^{2}{f}}{\partial{\delta^{2}}}(x,\delta,\sigma) = 2 \dfrac{\partial{f}}{\partial{\sigma^{2}}}(x,\delta,\sigma), \label{eq:pde_Gaussian}
\end{eqnarray}
for all $x, \delta, \sigma$ and Normal density $f$. This dependence leads to worse convergence rates for parameter estimation for over-fit location-scale Gaussian mixtures~\citep{Ho-Nguyen-Ann-16} and for hypothesis testing for the number of components of location-scale Gaussian mixtures~\citep{Chen-2003}. 
Under the specific setting that we consider, this dependence leads to a new characterization of the asymptotic behavior of $\widehat{\delta}_{n}^{\mle}$, $|\widehat{\delta}_{n}^{\mle}|$, and $\widehat{\sigma}_{n}^{\mle}$ under the two regimes $\pi \in (0,1/2)$ and $\pi = 1/2$. 
To the best of our knowledge, these have not been previously addressed in the literature.
\begin{theorem} \label{theorem:convergence_rate_location_scale_Gaussian}
Take $\pi \in (0,1/2]$. Under the unknown equal variances setting~\eqref{eqn:general_model}, the following holds
\begin{itemize}
\item[(a)] (Asymmetric regime) When $\pi \in (0,1/2)$, then 
\begin{eqnarray}
C_{1}(\epsilon) \left( \frac{1}{n} \right)^{1/3} \leq \sup \limits_{(\delta_{n},\sigma_{n}) \in \mathcal{S}_{1,n}(\epsilon)} \mathbb{E}_{(\delta_{n},\sigma_{n})}\biggr(|\widehat{\delta}_{n}^{\mle} - \delta_{n}|^{2} + |(\widehat{\sigma}_{n}^{\mle})^{2} - \sigma_{n}^{2}| \biggr) \leq C_{2}( \epsilon) \left(\frac{\log n}{n} \right)^{1/3}, \nonumber
\end{eqnarray}
where $\mathcal{S}_{1,n}(\epsilon) = \left\{(\delta_{n}, \sigma_{n}): |\delta_{n}|^2 + |(\sigma_{n})^2 - (\overline{\sigma})^2| \leq n^{-1/3 + \epsilon}\right\}$ for any $\epsilon > 0$ and some positive constant $\overline{\sigma}$.
\item[(b)] (Symmetric regime) When $\pi = 1/2$, then
\begin{eqnarray}
C_{1}(\epsilon) \left( \frac{1}{n} \right)^{1/4} \leq \sup \limits_{(\delta_{n},\sigma_{n}) \in \mathcal{S}_{2,n}(\epsilon)} \mathbb{E}_{(\delta_{n},\sigma_{n})} \biggr(\biggr||\widehat{\delta}_{n}^{\mle}| - |\delta_{n}|\biggr|^{2} + |(\widehat{\sigma}_{n}^{\mle})^{2} - \sigma_{n}^{2}| \biggr) \leq C_{2}( \epsilon) \left(\frac{\log n}{n} \right)^{1/4}, \nonumber 
\end{eqnarray}
where $\mathcal{S}_{2,n}(\epsilon) = \left\{(\delta_{n}, \sigma_{n}): |\delta_{n}|^2 + |(\sigma_{n})^2 - (\overline{\sigma})^2| \leq n^{-1/4 + \epsilon}\right\}$ for any $\epsilon > 0$ and some positive constant $\overline{\sigma}$.
\end{itemize}
Here, $\mathbb{E}_{(\delta_{n},\sigma_{n})}$ denotes the expectation taken with respect to a product measure with a mixture density of $Y_{1},\ldots,Y_{n}$ under the unknown equal variances setting~\eqref{eqn:general_model}. Furthermore, $C_{1}( \epsilon)$ and $C_{2} (\epsilon)$ are two positive constants depending only on $\epsilon$.
\end{theorem}
\noindent
The proof of Theorem~\ref{theorem:convergence_rate} is provided in Appendix~\ref{subsec:proof:theorem:convergence_rate_location_scale_Gaussian}. 

A few comments are in order. First, under the asymmetric regime, the convergence rate of the separation parameter $\widehat{\delta}_{n}^{\mle}$ to $\delta_{n}$ is of an order no more than $n^{-1/ 6}$ (due to the squared term within the expectation) while that of scale parameter $(\widehat{\sigma}_{n}^{\mle})^{2}$ to $(\sigma_{n})^2$ is no more than order $n^{-1/ 3}$, as long as the true parameters $\delta_{n}$ and $\sigma_{n}$ belong to $S_{1,n}( \epsilon)$.
The PDE of the distribution in~\eqref{eq:pde_Gaussian} suggests the faster apparent convergence rate of the scale parameter relative to the separation parameter.
%The faster apparent convergence rate of scale parameter comparing to that of separation parameter can be explained via the PDE of the distribution in~\eqref{eq:pde_Gaussian}.

Second, under the symmetric regime, the worse-case convergence rate of $|\widehat{\delta}_{n}^{\mle}|$ to $|\delta_{n}|$ is $n^{-1/ 8}$, which is slower than the worst-case rate $n^{-1 / 4}$ of $(\widehat{\sigma}_{n}^{\mle})^{2}$ to $(\sigma_{n})^2$, when the true parameters $\delta_{n}$ and $\sigma_{n}$ belong to $S_{2,n}(\epsilon)$.
Here, we consider the absolute value of the separation parameter for the convergence as the sign of separation parameter is not identifiable under the symmetric setting. Furthermore, in contrast to the know variance setting~\eqref{eq: block}, the worse-case convergence rate of separation parameter under the symmetric regime is slower than that of separation parameter under the asymmetric regime. 
That fundamental difference can be again explained by the PDE of the location-scale Gaussian distribution.

%%%
%%% PILE UP
%%%
\section{Non-asymptotic properties of the MLE: Pile Up}
\label{sec:weak_separation}
Thus far, we have established rigorous asymptotic (minimax) behaviors of MLE under the asymmetric and symmetric cases of model~\eqref{eq: block} and model~\eqref{eqn:general_model}. The goal of this section is to shed some light on the non-asymptotic sample properties of the MLE. 
To facilitate the discussion, we focus solely on the known variances setting~\eqref{eq: block}, \textit{i.e.}, we want to analyze the non-asymptotic behavior of MLE when $\delta_n$ is near zero. 
%\avi{awk}
%We again set $\mu=0$.
We work with the likelihood function of our re-parameterized model (again, setting $\mu=0$). This allows us to directly obtain statements regarding the points of the maximum likelihood which in turn allows for the characterization of the MLE's behavior.
In particular, we first show that under our parameterization, zero (corresponding to no separation) will always be an inflection point if not a local mode.
Finally, we show that, in general, the local mode is in fact the MLE when the estimated overall variance is less than $\sigma$, the assumed component variance.
% which we set to $\sigma=1$ here.  
%We then characterize when zero is in fact the MLE. \avi{more here?}
%\avi{Do we set $\sigma=1$?}

\subsection{Zero as a local mode of the likelihood} 
Given an observation $Y=y$ from the mixture model \eqref{eq: block}, the log-likelihood for $\delta_n$ is 
\begin{equation}
\ell(\delta_n|Y=y) = \log\left(\pi e^{-0.5(y-\delta_n)^2} + (1-\pi) e^{-0.5(y-c\delta_n)^2}\right),
\end{equation}
where we set $\sigma=1$, though these results immediately extend to arbitrary $\sigma$.
The score function is then
\begin{equation}\label{eqn:score}
\ell'(\delta_n|Y=y) = 
 -\frac{ \pi e^{-0.5(y+\delta_n)^2}(y-\mu +\delta_n) - c(1-\pi)e^{-0.5(y-c\delta_n)^2}(y - c\delta_n) }{\pi e^{-0.5(y+ \delta_n)^2} + (1-\pi) e^{-0.5(y-c\delta_n)^2}}.
\end{equation}
Since $c= \frac{\pi}{1-\pi}$ with $\pi \in (0,1/2]$, it follows from \eqref{eqn:score} that 
\begin{equation}\label{eqn:scorezero}
\ell'(0|Y=y) = 0, ~\mathrm{for\, all}\, ~y \in \mathbb{R}.
\end{equation}

Given the samples $\boldsymbol{Y}_n = (Y_{1},Y_{2},\dots Y_{n})$ from model \eqref{eq: block}, Equation~\eqref{eqn:scorezero} yields the following approximation of the log-likelihood given samples $\boldsymbol{Y}_{n}$:
\begin{equation}
\ell(\delta_n| \boldsymbol{Y}_n) = \ell(0|\boldsymbol{Y}_n) + \frac{1}{2} \ell''(0|\boldsymbol{Y}_n) \delta_n^2 + O(\delta_n^2).
\end{equation}

In the event  that $\ell''(0|\boldsymbol{Y}_n) < 0$, zero is a local mode for the 
log-likelihood function $\ell(\delta_n|\boldsymbol{Y}_n)$; we call this event 
\begin{equation} \label{eq:E}
\mathcal{E} \equiv \{\ell''(0|\boldsymbol{Y}_n) < 0 \}.
\end{equation}
Direct calculation yields that
\begin{equation} \label{eqn:secder} 
\ell''(0|\boldsymbol{Y}_n) = c\, \left(\sum_{i=1}^nY_{i}^2 - n\right),
\end{equation}
and thus $\ell''(0|\boldsymbol{Y}_n) < 0$ when $\sum_{i=1}^nY_{i}^2 < n$. Equivalently, $\ell''(0|\boldsymbol{Y}_n) < 0$ when $\widehat{m}_2 < 1$, where $\widehat{m}_2 \equiv \frac{1}{n}\sum_{i=1}^nY_{i}^2$ is the observed second moment of the mixture distribution, and the assumed within-component variance is 1. We return to this connection to higher-order moments below.

\subsection{Zero as the global mode of the likelihood} 
After establishing that zero is a local mode of the likelihood when $\ell''(0|Y_n) < 0$, an important question is whether zero is also a global mode in this case. Let $\mathcal{F} \equiv \{\widehat{\delta}_{n}^{\mle} = 0\}$
 be the event that zero is also the global mode for the likelihood function $\ell(\delta|Y)$, where $\widehat{\delta}_{n}^{\mle}$ is the MLE under the setting of model \eqref{eq: block}. 
 We refer to the event $\mathcal{F}$ as \textit{pile up} throughout the paper. While it is clear that $\mathcal{F} \subset \mathcal{E}$, the reverse implication is not trivial. We divide our analysis into two cases: $\pi=1/2$ and  $\pi \in (0,1/2)$. We again denote $\widehat{m}_2 : = \frac{1}{n} \sum_{i = 1}^{n} Y_i^2$. 
 
\paragraph{Symmetric case.} When $\pi = \frac{1}{2}$, conditioning on the event $\mathcal{E}$ (equivalently $\widehat{m}_2 < 1$), we can check that
\begin{align}
\ell''(\delta|\boldsymbol{Y}_{n}) = \dfrac{4}{n}\sum \limits_{i=1}^{n}\dfrac{Y_{i}^2}{\parenth{\exp(-\delta Y_{i})+\exp(\delta Y_{i})}^2} - 1 \leq \widehat{m}_2 - 1 < 0 \nonumber
\end{align}
where the inequality is due to applying Cauchy-Schwarz $\exp(-\delta Y_{i})+\exp(\delta Y_{i}) \geq 2$ for all $i \in \left\{1,\ldots,n\right\}$. The above inequality implies that the log-likehood function $\ell(\delta|\boldsymbol{Y}_{n})$ is strictly concave under the event $\mathcal{E}$.
Therefore, zero is the global maximum of the log-likelihood function under the event $\mathcal{E}$. This leads to the following result regarding pile up.
\begin{proposition} \label{proposition:pile_up_symmetric_fix_scale}
Under the symmetric setting of location-scale Gaussian mixtures with known variances, $\mathcal{E} \equiv \mathcal{F}$, i.e., pile up occurs as long as $0$ is a local maxima of the log-likelihood function. 
\end{proposition}
The result of Proposition \ref{proposition:pile_up_symmetric_fix_scale} suggests that we can rewrite the representation of MLE under symmetric setting with known variances as
\begin{align}
\widehat{\delta}_{n}^{\mle} = \begin{cases} 0, & \text{if} \ \widehat{m}_2 < 1 \\
O_{p}(n^{-1/4}), & \text{if} \ \widehat{m}_2 \geq 1 \end{cases}. \nonumber
\end{align}
Thus, at least in the symmetric case, the MLE behaves like a threshold estimator analogous to the classic Hodges estimator~\citep[see][]{van2000asymptotic}.

\paragraph{Asymmetric case.}
Unlike the symmetric case, we can see via simulations that there are instances for which $\mathcal{E} \neq \mathcal{F}$ in relatively small samples. Nonetheless, these counter-examples are fairly rare; for $\Delta = (1+c)\delta_{n}= 0.25$, $\{\mathcal{E} \cap \mathcal{F}^c\}$ occurs in fewer than 3 percent of simulation draws with sample sizes less than $N = 500$, decreasing to below 1 percent with samples sizes of $N = 1000$ or more. Extensive simulation studies seem to imply that $\mathbb{P}_n(\mathcal{F}) \nearrow \mathbb{P}_n(\mathcal{E})$.\footnote{The index $n$ denotes the fact that the sampling distribution in \eqref{eq: block} changes with $n$.} We do not have a rigorous proof of this and therefore state it as a conjecture: 

\begin{conj}\label{eqn:conjenfn}
Under the asymmetric setting of location-scale Gaussian mixtures with known variances, if $\delta_{n} = O_{p}(n^{-1/6})$, then $\lim_{n\rightarrow \infty}\mathbb{P}_n(\mathcal{E} \cap \mathcal{F}) =  1$.
\end{conj}

Thus Conjecture \ref{eqn:conjenfn}, if true, implies that, for the asymmetric setting of location-scale Gaussian mixtures with known variances, the probability that pile up occurs, i.e., $\widehat{\delta}_{n} = 0$, can be well approximated by 
the event $\{\ell''(0|\boldsymbol{Y}_{n}) < 0 \}$. In other words, we can safely ignore the case in which zero is a local but not a global mode of the likelihood.

\begin{figure}[btp]
\centering
		 \begin{subfigure}[b]{0.55\textwidth}
				\includegraphics[width = \textwidth]{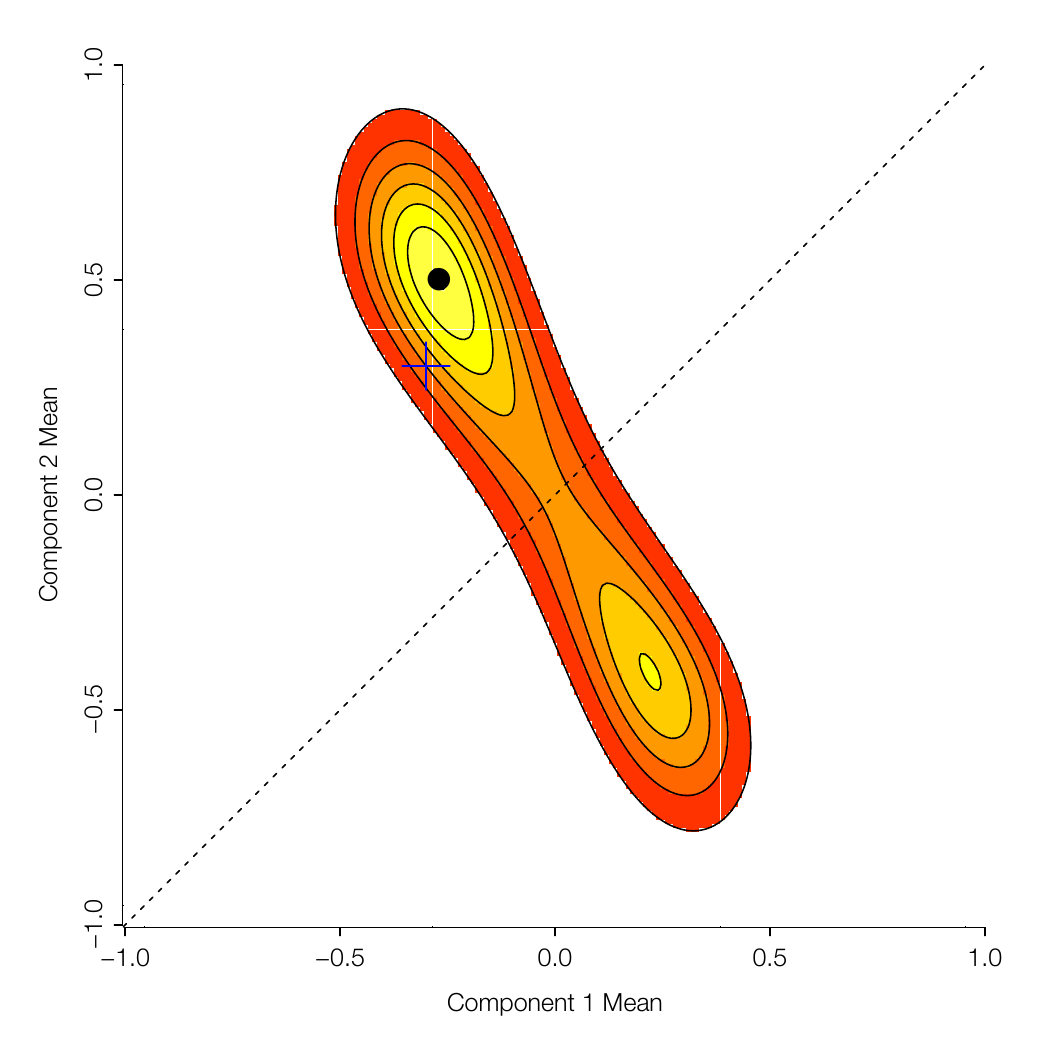}
				\caption{Example bimodal likelihood}
				\label{fig: example_ll_correct_mode}
		\end{subfigure}%
		\;\begin{subfigure}[b]{0.55\textwidth}
				\includegraphics[width = \textwidth]{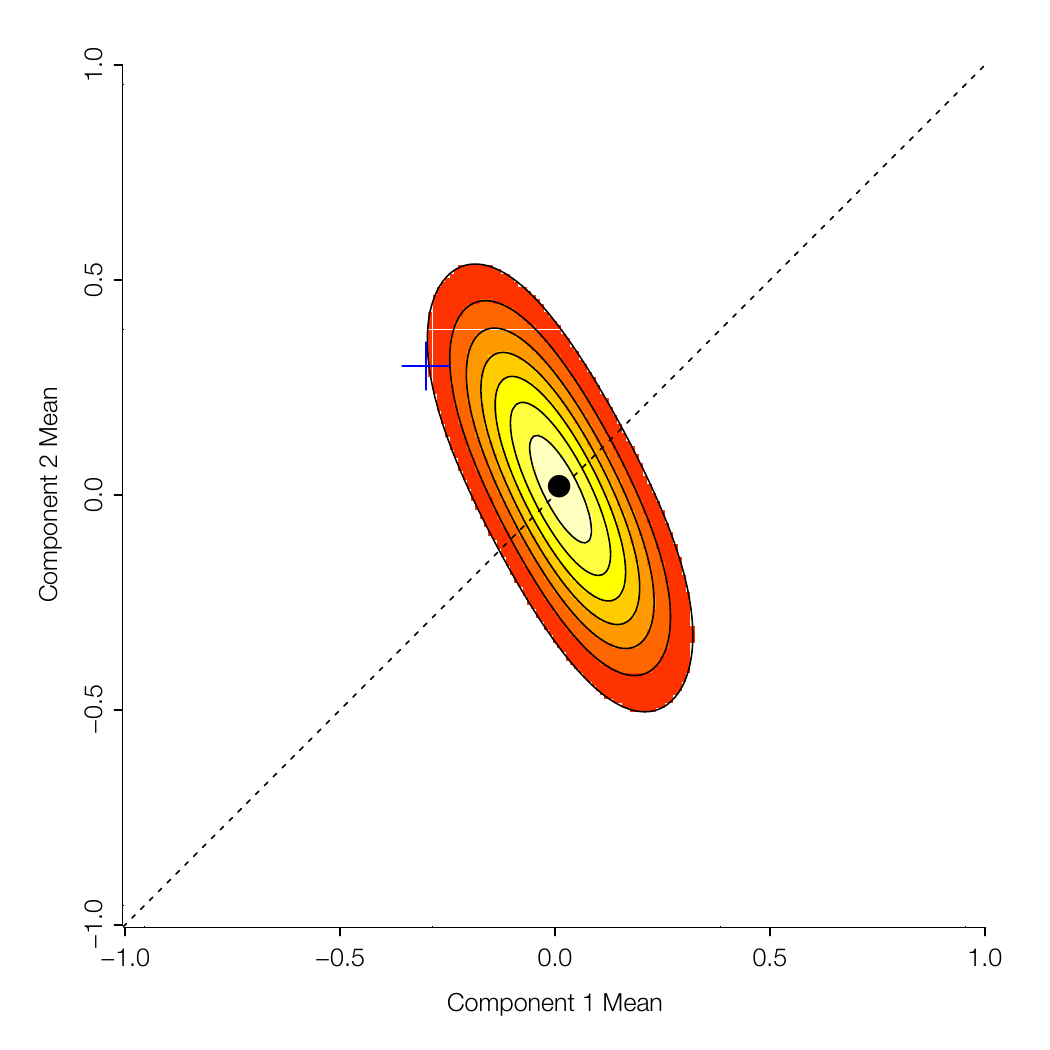} 
				\caption{Example unimodal likelihood}
				\label{fig: example_ll_unimodal}
		\end{subfigure}
\caption{Two example likelihoods for component means, with data generated via Equation~\eqref{eq: mix} with parameters $N = 200$, $\pi = 0.35$, and $\Delta = 0.6$. The `+' denotes the true component means.}
\label{fig:example_likelihoods}
\end{figure}

Figure~\ref{fig:example_likelihoods} shows this pile up phenomenon in practice. Specifically, Figures~\ref{fig: example_ll_correct_mode} and~\ref{fig: example_ll_unimodal} show the likelihood surfaces for two data sets generated via Equation~\eqref{eq: mix}, with $N = 200$, $\pi = 0.35$, and $\Delta = (1+c)\delta = 0.6$. In Figure~\ref{fig: example_ll_correct_mode}, the likelihood is bimodal and the global mode is close to the truth, albeit more extreme.\footnote{\label{fn:bias}The characterization of $\widehat{\delta}^{\text{mle}}_n$ as a Hodges-like estimator suggests that the MLE will be biased away from zero when $\widehat{\delta}^{\text{mle}}_n \neq 0$. This is closely related to the bias induced by introducing identifiability constraints, such as $\delta > 0$~\citep{Jasra:2005kl,fruhwirth2006finite}. In both cases, the MLE is the maximum of a truncated likelihood surface, truncated at the line $\delta = 0$.}
In Figure~\ref{fig: example_ll_unimodal}, the likelihood is unimodal and centered at zero, which is far from the truth. 

%%%
%%% METHOD OF MOMENTS
%%%
\section{Diagnostics for MLE pathologies}
\label{sec:mom_main}

The results above suggest that the higher-order moments of the mixture distribution play an important role in the finite sample properties of the MLE. 
We now construct diagnostics for the MLE using these moments.
First, we use these higher-order moments to construct diagnostics for pile up for the MLE, specifically the probability that pile up will occur given a set of moments, either observed moments or assumed moments.
We then construct similar diagnostics for the relative order of the components, as captured by the sign of $\Delta$.
Throughout, we consider the setting with known variances, since the corresponding moment equations are tractable in this case.

%%%
%%% PROBABILITY OF PILE UP
%%%

\subsection{Probability of pile up\label{sec:prob_pile_up}} 

The probability of pile up can be characterized by using the sampling distribution of the second moment, $Y^2$. 
In particular, we can determine $\mathbb{P}\{\widehat{m}_2 < 1\}$ using the first three moments of $Y^2$:
\begin{align}
m_2 &= \mathbb{E}[Y^2] = 1 + c\delta_n^2 \label{eq: BE_m2}\\
v_2 &= \mathbb{V}[Y^2] = 3 + 3(\pi + c^4(1-\pi))\delta_n^4 - m_2^2 \label{eq: BE_v2}\\
\Gamma_2 &= {1 \over v_2^{3/2}} \mathbb{E}|Y^2 - m_2|^3,\label{eq: BE_gamma2}
\end{align}
where we can obtain $\Gamma_2$ via Monte Carlo methods.
Using the Berry-Essen theorem for the convergence rates of a CLT, and assuming Conjecture \ref{eqn:conjenfn}, we can obtain the following bound for the probability of pile up:
\begin{align}\label{eqn:pileupprob}
|\mathbb{P}_n(\mathcal{E}) - \Phi(b_n)| \leq  0.7915 \frac{\Gamma_2}{\sqrt{n}}.
\end{align}
As we show in simulations, $\Phi(b_n)$ appears to be an excellent approximation to the empirical pile up probability, even though the bound, which depends on the sixth mixture moment, can be wide in practice. See supplementary materials.

\begin{figure}[bt]
\centering
		 \begin{subfigure}[b]{0.55\textwidth}
				\includegraphics[width = \textwidth]{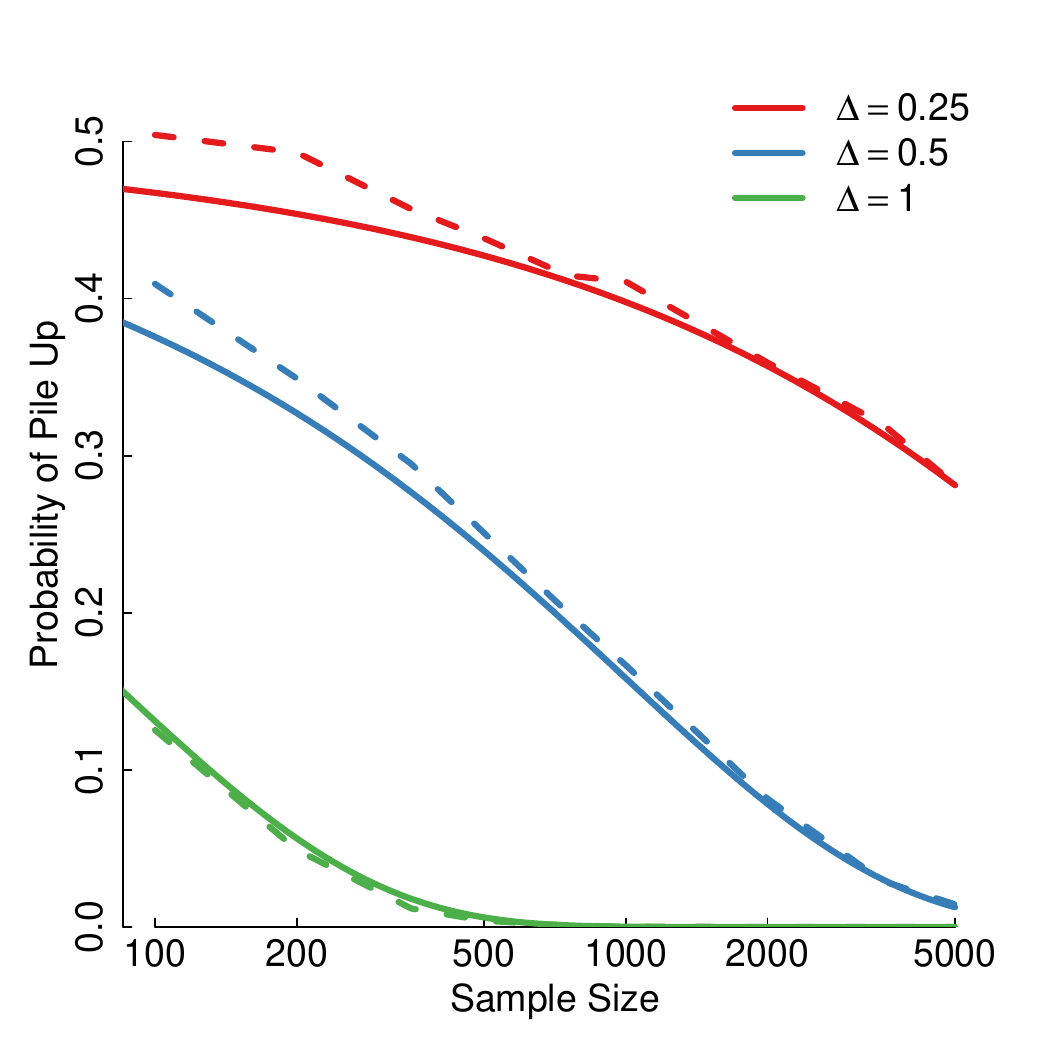}
				\caption{$\pi = 0.25$}
				\label{fig: norm_approx_vary_N_first}
		\end{subfigure}%
		\;\begin{subfigure}[b]{0.55\textwidth}
				\includegraphics[width = \textwidth]{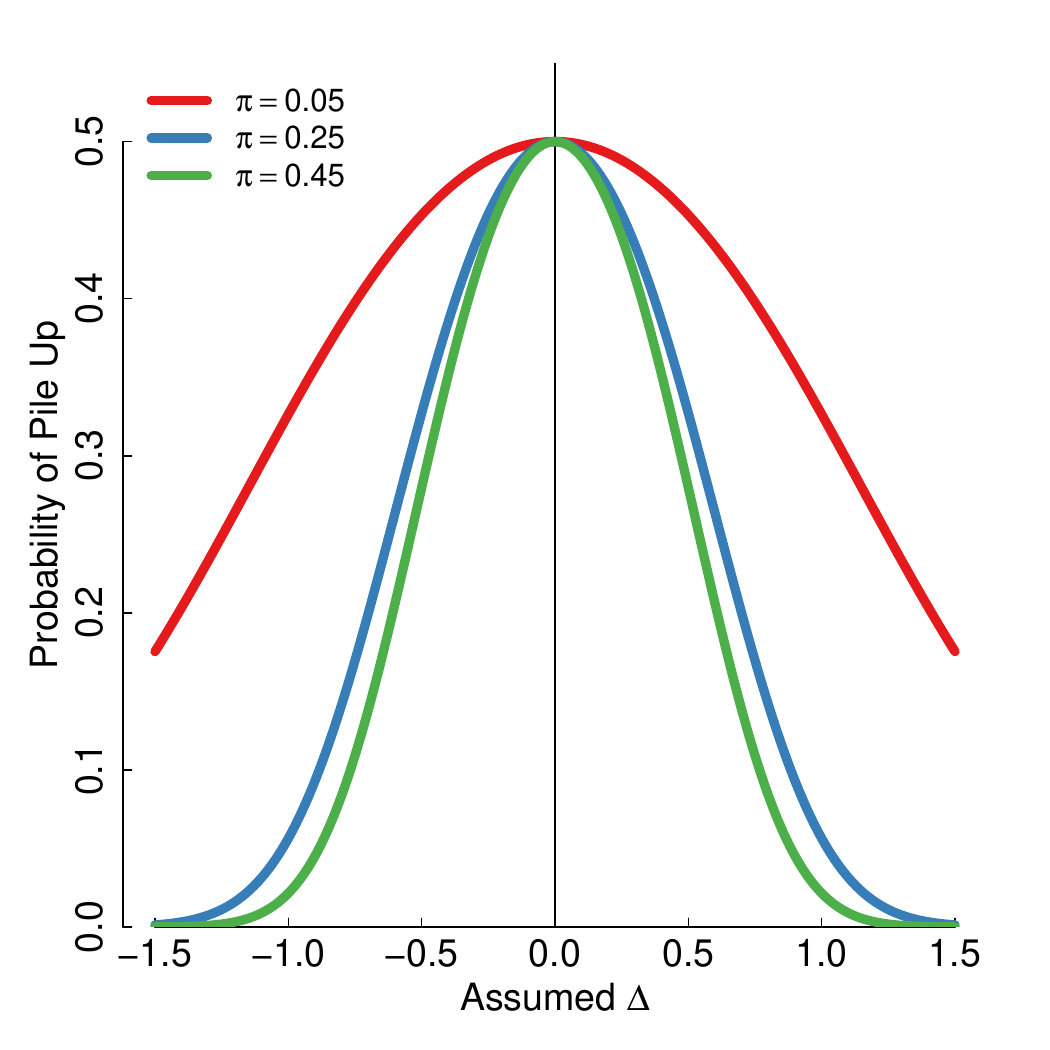} 
				\caption{$N = 200$}
				\label{fig: norm_approx_vary_Delta_first}
		\end{subfigure}
\caption{Probability of pileup given sample size and separation of means. Dotted lines are simulated values across 5,000 simulations; solid lines use the Normal approximation, $\Phi(b_n)$.}
\label{fig: norm_approx}
\end{figure}

We can use this result for practical diagnostics, both for planning a future analysis and for assessing a particular data set.
Figure~\ref{fig: norm_approx_vary_N_first} shows the pile up probability computed via simulation and via Equation~\eqref{eqn:pileupprob}, with $\pi = 0.325$, $\Delta = (1+c)\delta_{n} = 0.25$, and varying $n$. First, there is excellent agreement between the simulations and the Normal approximation, though $\Phi(b_n)$ slightly under-states the probabilities obtained via simulation. Second, while the probability of pile up is decreasing in both $n$ and $\Delta$, it is hardly a ``small sample'' issue. For $\Delta = 0.25$, which would be quite large in many social science applications, pile up remains a meaningful possibility even with sample sizes in the thousands. For $\Delta = 1.0$, which would be an implausibly large separation in many settings, the probability of pile up is still greater than 1 in 4 for $n = 5,000$. Finally, Figure~\ref{fig: norm_approx_vary_Delta_first} shows similar results for a moderate sample size of $N = 200$ but varying mixing proportions. In this case, the probability of pile up decreases as $\pi$ approaches $0.5$. We believe that figures such as these are useful diagnostics before observing the mixing distribution itself.

We can also incorporate information from the observed mixture distribution. First, we can plug in the observed empirical moments, $\widehat{m}_2$ and $\widehat{v}_2$, to calculate $\widehat{b} = \frac{1-\widehat{m}_2}{\sqrt{\widehat{v}_2/n}}$ and $\Phi(\widehat{b})$. This relies on the Normal approximation for the sampling distribution as well as precisely estimating $\widehat{v}_2$, which is the fourth moment of the observed mixture distribution and might be noisy in practice. Alternatively, we could use a case-resampling bootstrap to estimate $\mathbb{P}\{\widehat{m}_2 < 1\}$. Note that this is not the same as using the case-resampling bootstrap to estimate standard errors, which we advise against (see supplementary materials). Rather, this is analogous to the use of the bootstrap as a diagnostic tool in finite mixtures; see, for example,~\citet{Grun:2004}.
Finally, we note that an estimated MLE of zero still provides some information about the unknown parameter. For instance, if $\widehat{\Delta}^{\text{mle}} = (c+1)\widehat{\delta}_{n} = 0$, $\Delta = 0.2$ is a much more plausible value than $\Delta = 2.0$. We discuss this in the supplementary materials.

\subsection{Probability of a sign error}
\label{sec:sign_error}

We now turn to the sign of $\widehat{\Delta}^{\text{mle}}$ when $\pi \neq 1/2$ (the sign is not estimable when $\pi = 1/2$). Specifically, we define a \emph{sign error} as $\text{sgn}\left(\widehat{\Delta}^{\text{mle}}\right) \neq \text{sgn}\left(\Delta\right)$.
This is a well-studied issue in mixture modeling; for example, choosing the true mode in a multimodal likelihood is a classic problem~\citep[see][]{Gan:1999hh, Biernacki:2005eq}.~\citet{Redner:1984} give a foundational review of \textit{asymptotic} versus \textit{local} identifiability in mixtures. For a more recent perspective, see~\citet{kim2015empirical}, who introduce the concept of \textit{empirical} identifiability.

As with pile up, we use higher order moments for diagnosis. This is slightly more complicated than for pile up because $\text{sgn}\left(\widehat{\Delta}\right)$ is undefined when $\widehat{\Delta} = 0$. Thus, we need to consider the joint sampling distribution of both the second and third moments. In the setting with known, equal variances in Equation~\eqref{eq: block}, we have the following moment equations:
\begin{align*}
	m_2 &= \E[Y^2] = 1 + \pi(1-\pi)\Delta^2 \\
	m_3 &= \E[Y^3] = \pi(1-\pi)(1-2\pi)\Delta^3.
\end{align*}
Following~\citet{Tan}, the corresponding sample moments have the following distribution:
\begin{equation}
\label{eq: joint}
\left(\begin{array}{c} 
	\widehat{m}_2 \\
	\widehat{m}_3 
\end{array}\right)
\stackrel{\cdot}{\sim}
\mathcal{N}\left(\left(\begin{array}{c} 
m_2 \\
m_3
\end{array}\right)
,
\frac{1}{n}\left(\begin{array}{ccc}
\kappa_{11}\Delta^4+2m_2^2 & \kappa_{12}\Delta^5 + 6m_2m_3 \\
& \kappa_{22a}\Delta^6 + \kappa_{22b}m_2\Delta^4 + 6m_2^3\end{array}\right)
\right),
\end{equation}
with constants $\kappa_{11} = \pi(1-\pi)(1 - 6\pi(1-\pi))$; $\kappa_{12} = \pi(1-\pi)(1-2\pi)(1-12\pi(1-\pi))$; $\kappa_{22a} = \pi(1-\pi)(1 - 30\pi(1 -\pi) + 120 \pi^2(1 - \pi)^2) + 9\pi^2(1- \pi)^2(1 - 2\pi)^2$; and $\kappa_{22b} = 9\pi(1-\pi)(1 - 6\pi(1-\pi))$. Thus, we can approximate the joint probability of pile up, sign error, or neither for a given $\Delta$, $n$, and $\pi$, where we set $\Delta > 0$ for illustration:
\begin{equation}
\begin{aligned} \label{eq: prob_pathology}
&\mathbb{P}\left(\{\text{pile up};\;\text{sign error};\; \text{neither}\}\right) \approx \\
&\quad \mathbb{P}\left( \{\widehat{m}_2 < 1;\; \widehat{m}_2 > 1 \cap \widehat{m}_3 < 0;\; \widehat{m}_2 > 1 \cap \widehat{m}_3 > 0 \} \right)
\end{aligned}
\end{equation}
If desired, we could apply a similar Berry-Essen bound for these probabilities, as in Equation \eqref{eqn:pileupprob}.
  Instead, we simply invoke the Central Limit Theorem and use the Normal approximation in Equation~\eqref{eq: joint}.

\begin{figure}[!bt]
\centering
	\includegraphics[scale = 0.45]{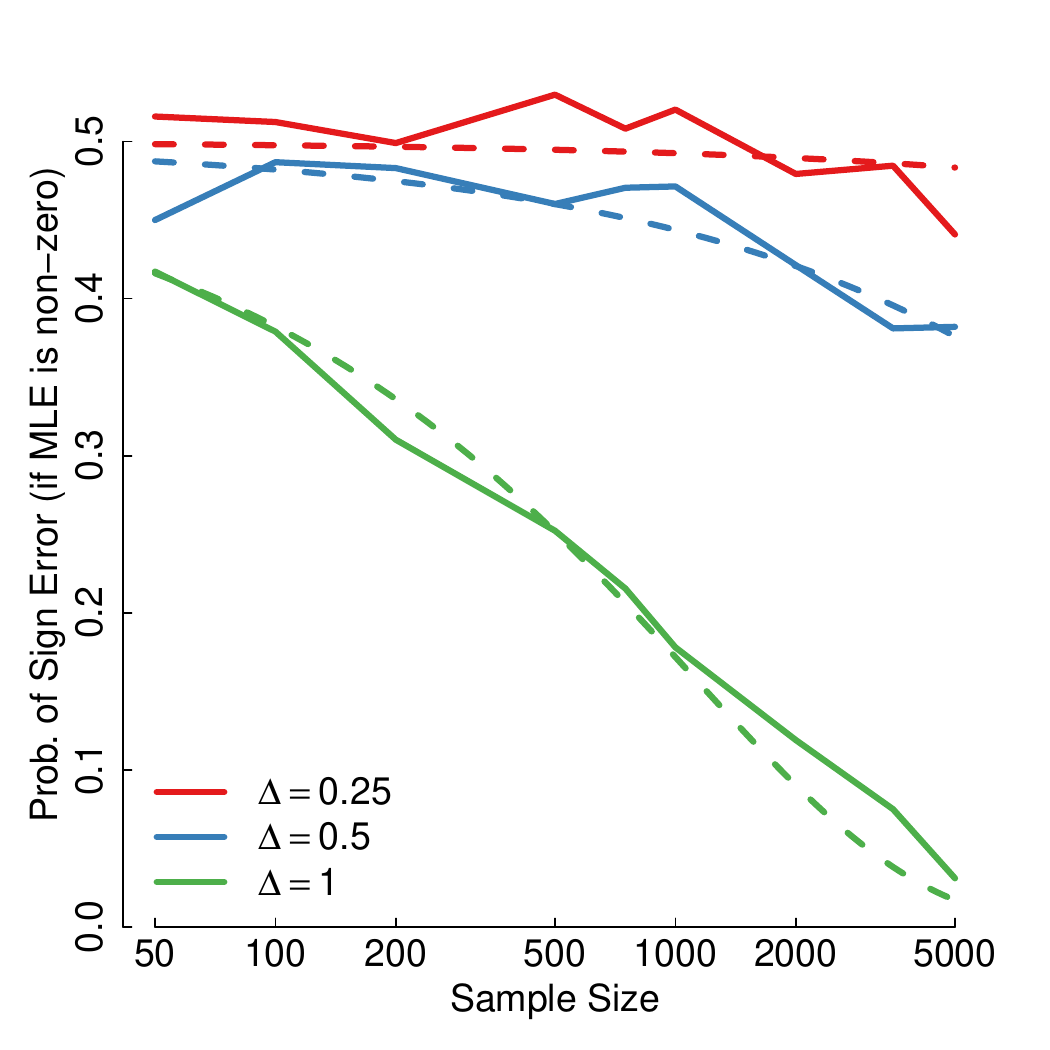}
	\caption{Probability that $\text{sgn}\left(\widehat{\Delta}\right) \neq \text{sgn}(\Delta)$ based on simulations (solid line) and the method of moments approximation in Equation~\eqref{eq: joint} (dotted line); based on $\pi = 0.25$ and 1000 simulations at each set of parameter values}
	\label{fig:prob_sign_error}
\end{figure}

Figure~\ref{fig:prob_sign_error} shows the conditional probability of sign error given no pile up across values of $N$ and $\Delta$ found by two methods: (1) direct simulation (simulations are restricted to draws in which $\widehat{\Delta}^{\text{mle}} \neq 0$); and (2) the tail probability of Equation \eqref{eq: prob_pathology} based on the Normal approximation in Equation~\eqref{eq: joint}. 
While the probability of a sign error decreases in both $n$ and $\Delta$, it remains remarkably high over plausible parameter values. Indeed, for $\Delta = 0.25$ the sign of $\Delta$ is essentially a coin flip, even with a sample size of 5,000. Importantly, conventional approaches for standard errors in the MLE~\citep{mclachlan2004finite} typically ignore this uncertainty. For additional discussion, see~\citet{kim2015empirical}. 

As in Section~\ref{sec:prob_pile_up}, we can assess the probability of sign error in practice. Based only on the sample size and mixing proportion, we can re-create Figure~\ref{fig:prob_sign_error} across plausible parameter values. We can also plug observed values into Equation~\eqref{eq: joint}. Alternatively, we can count the proportion of bootstrap replicates in which the sign of the bootstrapped third moment differs from the observed sign and $\widehat{m}_2 > 1$.

%%%%
%%%% CAUSAL INFERENCE OVERVIEW
%%%%
\section{Application to principal stratification\label{sec:causal}} 
We now motivate the use of finite mixtures in principal stratification. For our primary running example, we re-analyze the Job Search Intervention Study (JOBS II), a randomized field experiment of a mental health and job training intervention among unemployed workers~\citep{vinokur1995impact} that has been extensively studied in the causal inference literature~\citep{Jo:2009hx,Mattei:2013fb}. This is an example of one-sided noncompliance and is a simple but non-trivial example of the principal stratification setup. In the supplementary materials, we also re-analyze a randomized evaluation of JobCorps, the largest job training program in the US~\citep{Schochet:2008kj}. We briefly discuss these results at the end of this section.

\subsection{Setup}
We begin with the canonical example of a randomized experiment with noncompliance, such as JOBS II, and set up the problem using the potential outcomes framework \citep{Neyman, Rubin}. We observe $N$ individuals who are randomly assigned to a treatment group, $T_i = 1$, or control group, $T_i = 0$, with observed outcome, $Y$. For JOBS II, the primary outcome is a measure of depression six months after randomization. As usual, we assume that randomization is valid and that the Stable Unit Treatment Value Assumption holds~\citep[SUTVA;][]{Rubin:1980dg,imbens_rubin_book}. This allows us to define potential outcomes for individual $i$, $Y_i(0)$ and $Y_i(1)$, under control and treatment respectively, with observed outcome,
$Y_i^{\text{obs}} = T_iY_i(1) + (1-T_i)Y_i(0).$
The fundamental problem of causal inference is that we observe only one potential outcome for each unit. 
Finally, we define the Intent-to-Treat (ITT) effect as the impact of randomization on the outcome,
$\textrm{ITT} = \mathbb{E}[Y_i(1) - Y_i(0)].$
Throughout, we take expectations and probabilities to be over a hypothetical super-population.

The main complication is that only 55\% of those individuals assigned to treatment actually enrolled in the program. Let $D_i$ be an indicator for whether individual $i$ receives the treatment, with corresponding compliance $D_i(0)$ and $D_i(1)$ for control and treatment respectively. For simplicity, we assume that only individuals assigned to treatment can receive the active intervention (\textit{i.e.}, there is one-sided noncompliance), which is the case in the JOBS II evaluation. Formally, $D_i(0) = 0$ for all $i$. This gives two subgroups of interest: Never Takers, $D_i(1) = 0$, and Compliers, $D_i(1) = 1$. Following~\citet{angrist1996identification}  and~\citet{Frangakis:2002wp}, we refer to these subgroups interchangeably as \textit{compliance types} or \textit{principal strata}, $U_i \in \{\text{c}, \text{n}\}$, with ``c'' denoting Compliers and ``n'' denoting Never Takers. Table~\ref{tbl:jobs_obs_means} shows the relationship between observed groups and principal strata.

\begin{table}[bt]
	\centering
	\caption{Summary statistics for observed groups in JOBS II}
	\label{tbl:jobs_obs_means}
	\begin{tabular}{ccccl}
	$Z$ & $D^{\text{obs}}$ & \textbf{Observed Mean} & \textbf{Observed SD} & \textbf{Possible Principal Strata}\\
	\hline
	1 & 1 & -0.16 & 1.03 & Compliers\\
	1 & 0 & 0.05 & 0.96 & Never Takers \\
	0 & 0 & 0.14 & 0.99 & Compliers and Never Takers\\
	\hline
	\end{tabular}	
\end{table}

The two main estimands are the ITT effects for Compliers and Never Takers:
\begin{align*}
\textrm{ITT}_\text{c} &= \mathbb{E}[Y_i(1) - Y_i(0) \mid U_i = \text{c}] = \mu_{\text{c}1} - \mu_{\text{c}0}, \\
\textrm{ITT}_\text{n} &= \mathbb{E}[Y_i(1) - Y_i(0) \mid U_i = \text{n}] = \mu_{\text{n}1} - \mu_{\text{n}0},
\end{align*}
\noindent in which $\mu_{ut}$ represents the outcome mean for $U_i = u$ and $T_i = t$. We are primarily interested in $\text{ITT}_{\text{c}}$, the impact of randomization on Compliers, which measures the impact of actually enrolling in JOBS II. Since we observe stratum membership for individuals assigned to treatment, we can immediately estimate $\mu_{\text{c}1}$ and $\mu_{\text{n}1}$. Moreover, due to randomization, the observed proportion of Compliers in the treatment group is, in expectation, equal to the overall proportion of Compliers in the population, $\pi \equiv \mathbb{P}\{U_i = \text{c}\}$. Thus, we treat $\pi$ as essentially known or, at least, directly estimable. The main inferential challenge is that we do not observe stratum membership in the control group. Rather we observe a mixture of Compliers and Never Takers assigned to control: 
\begin{equation}
\label{eq: control}
Y_i^{\text{obs}} \mid T_i = 0\sim \pi f_{\text{c}0}(y_i) + (1 - \pi)f_{\text{n}0}(y_i),
\end{equation}
where $f_{u0}(y)$ is the distribution of potential outcomes for individuals in stratum $u$ assigned to control.

The standard solution for this problem is to invoke the exclusion restriction for Never Takers, which states that $\textrm{ITT}_{\text{n}} = 0$, or equivalently, $\mu_{\text{n}1} = \mu_{\text{n}0}$. Substantively, this states that the only impact of randomization on the outcome is by changing the intermediate variable, $D$.  This is often a reasonable assumption, since actual program participation---rather than the randomization itself---is typically the important factor in practice.
With this assumption, we can then estimate $\textrm{ITT}_\text{c}$ with the usual instrumental variables approach~\citep{angrist1996identification}. In JOBS II, however, there is a concern that randomization has a negative impact on depression levels for Never Takers~\citep[see][]{Mattei:2013fb}. Thus, assuming that $\text{ITT}_{\text{n}} = 0$ could lead to biased estimates for $\text{ITT}_\text{c}$.

\subsection{Model-based estimation}
In a seminal paper,~\citet{Imbens} outlined a model-based instrumental variables framework, proposing a parametric model for the outcome distribution conditional on stratum membership and treatment assignment, such as $f_{ut}(y_i) = \mathcal{N}(\mu_{ut}, \sigma_{ut}^2)$. While the exclusion restriction can strengthen inference in this setting, it is not strictly necessary. Instead, identification is based entirely on standard results for mixture models. 

Since~\citet{Imbens}, dozens of papers have used finite mixtures for estimating causal effects.\footnote{Some examples of other relevant papers are~\citet{Little:1998fn, Hirano:2000vg, Barnard:2003je, TenHave:2004hc, Gallop:2009bz, Zhang:2009ku, Elliott:2010ii, Zigler:2011kn, Frumento:2012gz, Page:2012kp, Schochet:2013jm}.} 
For one-sided noncompliance, we can write the observed data likelihood with mean-shifted standard Normal component distributions as:
\begin{align*}
\mathcal{L}_{\text{obs}}(\theta) = 
&\prod_{i:~T_i = 1,~D_i^{\text{obs}} = 1} \pi \phi(y_i; \mu_{\text{c}1})\;\; \times 
\prod_{i:~T_i = 1,~D_i^{\text{obs}} = 0} (1 - \pi) \phi(y_i; \mu_{\text{n}1})\;\; \times \\
&\quad \;\; \prod_{i:~T_i = 0} \left[ \pi \phi(y_i; \mu_{\text{c}0}) + (1 - \pi)\phi(y_i; \mu_{\text{n}0}) \right],
\end{align*}
where $\theta$ represents the vector of parameters and $\phi(y_i; \mu)$ is the Normal density with mean $\mu$ and variance 1. In practice, we often relax the assumption of known, common variance. Since the observed data likelihood for individuals with $T_i = 1$ immediately factors into the likelihood for the Compliers and the likelihood for the Never Takers, we can directly estimate $\mu_{\text{c}1}$ and $\mu_{\text{n}1}$. With one-sided noncompliance, we can also directly estimate $\pi$ among individuals assigned to treatment. 

The challenge is therefore to estimate $\mu_{\text{c}0}$ and $\mu_{\text{n}0}$ via a two-component homoskedastic Gaussian mixture with known mixing proportion, $\pi$.\footnote{Note that there is a very small amount of information about $\pi$ from the mixture model among those assigned to the control group. Given the other complications that arise in mixture modeling, we ignore this and regard $\pi$ as if it were estimated directly from the treatment group.} 
See~\citet{Mattei:2013fb} for further discussion of parametric mixture modeling in this setting.

\subsection{Application to JOBS II} 
We now turn to using the non-asymptotic results in Section~\ref{sec:mom_main} for estimation and diagnostics for JOBS II. 
We focus on a subset of $N = 410$ high risk individuals, with $N_1 = 278$ randomly assigned to treatment and $N_0 = 132$ to control. The finite mixture consists of the $N_0 = 132$ individuals assigned to control with mixing proportion $\widehat{\pi} = 0.45$. 

Table~\ref{tbl:jobs_obs_means} shows summary statistics for the three observed groups. We standardize the outcome by subtracting off the grand mean and dividing by $\widehat{\sigma}_1 = \sqrt{ \pi\widehat{\sigma}^2_{\text{n}1} + (1-\pi)\widehat{\sigma}^2_{\text{c}1}}$, the estimated within-component standard deviation under treatment.  Based on the group means, it is clear that workers who are observed to enroll in the program have lower depression, on average, than those who do not. Note that the point estimates for $\widehat{\sigma}_{\text{c}1}$ and $\widehat{\sigma}_{\text{n}1}$ are quite close, which is consistent with the equal variance assumption.% \avi{more on this}. 

\begin{figure}[!bt]
	\begin{center}
		\begin{subfigure}[b]{0.5\textwidth}
				\begin{center}
				\includegraphics[width = \textwidth]{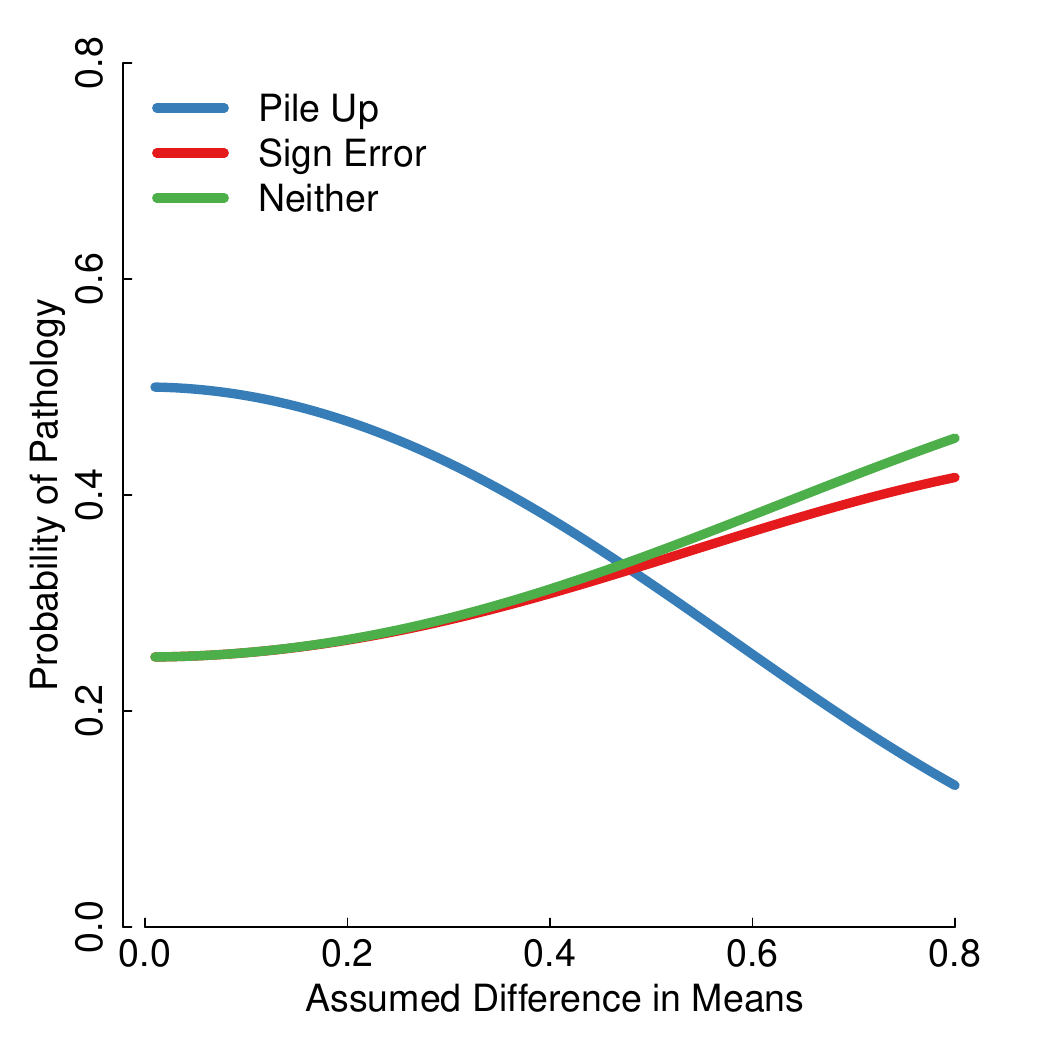} 
				\caption{Prob. of MLE Pathology}
				\label{fig:Jobs_Norm}
				\end{center}
		\end{subfigure}%
		~ \begin{subfigure}[b]{0.5\textwidth}
			\begin{center}
				\includegraphics[width = \textwidth]{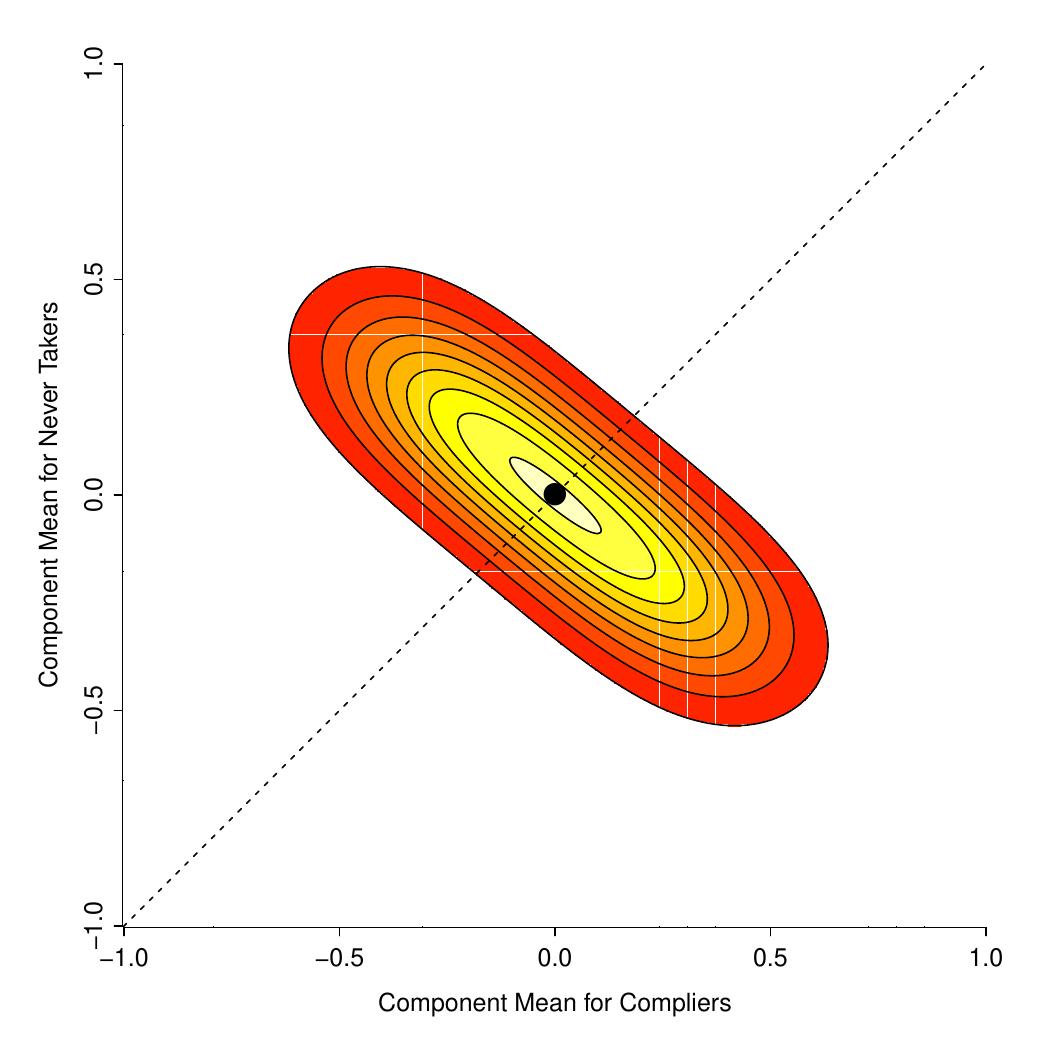}
				\caption{Observed likelihood}
				\label{fig:Jobs_ll}
			\end{center}
		\end{subfigure}\\[1em]
		
	\caption{Quality of Maximum Likelihood Estimation for the finite mixture model in JOBS II, with parameters $N = 132$ and $\pi = 0.45$. Panels (a) and (b) show the probability of MLE pathology and expected bias of the MLE if non-zero; Panel (c) shows the observed likelihood for the JOBS II mixture, with a maximum at $\mu_{\text{c}0} = \mu_{\text{n}0}$.}
	\label{fig:Jobs_Pathologies}
	\end{center}
	\end{figure}

First, we consider the expected performance of the mixture MLE based solely on the observed sample size and mixing proportion. Figure~\ref{fig:Jobs_Norm} gives the probability of pile up and sign error over a range of plausible values of $\Delta$ using the Normal approximation in Equation~\eqref{eq: joint} and the observed JOBS II values of $N = 132$ and $\widehat{\pi} = 0.45$. The pattern is striking. For values of $\Delta < 0.5$, the most likely estimate of the MLE is zero, regardless of the true value of $\Delta$. If the MLE is non-zero, the probability of correctly estimating the sign of $\Delta$ is only slightly better than a coin flip. 

Second, we incorporate information from the mixture distribution itself. First, the observed second and third moments are $\widehat{m}_2 = 0.96$ and $\widehat{m}_3 = 0.17$ (after centering the mixture distribution). When we plug the observed values into the Normal approximations in Equation~\eqref{eq: joint}, the probability of pile up is 0.63 and the probability of a sign error is 0.31. The corresponding probabilities based on the case-resampling bootstrap are nearly identical, 0.64 and 0.29 respectively. Thus, prior to any estimation, we believe that the probability of a pathological MLE is high.

Figure~\ref{fig:Jobs_ll} shows the observed likelihood surface for Equation~\eqref{eq: mix} fit to the JOBS II data. The likelihood is unimodal and centered at zero, which is consistent with the univariate results in~\citet{Mattei:2013fb}.\footnote{We can see this using the summary statistics in~\citet{Mattei:2013fb}. For the univariate model without the exclusion restriction, their Table 1 gives point estimates $\widehat{\mu}_{\text{c}1} = 1.96$ and $\widehat{\mu}_{\text{n}1} = 2.08$ on the depression scale. The treatment effect point estimates are $\widehat{\textrm{ITT}}_{\text{c}} = -0.206$ and $\widehat{\textrm{ITT}}_{\text{n}} = -0.084$, which imply $\widehat{\mu}_{\text{c}0} = 1.96 + 0.206 = 2.166$ and $\widehat{\mu}_{\text{n}0} = 2.08 + 0.084 = 2.164$. Therefore, $\widehat{\Delta} \approx 0$. By contrast, the implied estimate for $\Delta$ from their bivariate model is $\widehat{\Delta} = 0.261$, which is roughly three-quarters of a standard deviation on the depression scale. Finally, note that the model in~\citet{Mattei:2013fb} assumes unknown, unequal variances.}  %The standard 95\% confidence interval for the MLE is $[-0.9, 0.9]$. 
	Given the high probability of pile up \textit{ex ante}, our analysis suggests that we should interpret the MLE of $\widehat{\Delta}^{\text{mle}} = 0$ with caution.

\subsection{Application to Job Corps} 
In the supplementary materials, we provide a detailed re-analysis of a randomized evaluation of JobCorps, the largest job training program in the US~\citep{Schochet:2008kj}. Following~\citet{Lee:2009eg} and~\citet{Zhang:2009ku}, we are interested in the impact of Job Corps on (log) hourly wages, which is a measure of job quality. This quantity, however, is only well defined for a certain sub-population, known as {\em always employed} individuals. This is a principal causal effect and is sometimes referred to as the {\em Survivor Average Causal Effect} (SACE). While more complicated than non-compliance in JOBS II, we can again formulate the question as estimating the component means in a Normal finite mixture model. We focus on a mixture of $N = 3,371$ individuals with $\pi = 0.06$. Thus, while the mixing proportion is relatively extreme, the sample size is considerable.

Despite the large sample size, we continue to find pathological estimates from the Normal mixture model. First, based on the diagnostics we propose above, the probability of pile up is around one-third, which is surprising given the large sample size. Rather than find that $\widehat{\Delta}^{\mle} = 0$, however, we estimate an implausibly large $\widehat{\Delta}^{\mle} = -4.5$ standard deviations. This estimate is well outside outside the minimax bounds, $\Delta \in [-2.4, 2.2]$, suggesting that bias might be substantial.\footnote{Following~\citet{Lee:2009eg}, we calculate minimax bounds via trimmed means of the mixture distribution. Specifically, we bound $\mu_{\text{NE}1}$ via the mean of the $\pi = 0.06$ individuals with, respectively, the lowest and highest values of hourly wages, with similar bounds for $\mu_{\text{EE}1}$.} See the supplementary materials for additional analysis. In practice, the simplest explanation for these results is that the simple Normal mixture model in Equation~\eqref{eq: mix} is a poor fit to the data. At the same time, it is difficult to imagine a different parametric mixture model that would be a better fit. This suggests that parametric finite mixtures might not be an effective strategy here.

%%%
%%% DISCUSSION
%%%
\section{Discussion\label{sec:discussion}}

We find that maximum likelihood estimates for component-specific means in finite mixtures can yield pathological results in a range of practical settings. These pathologies are particularly relevant for estimating causal effects in principal stratification models, which are often based on estimates of component means. Echoing previous work~\citep[\textit{e.g.},][]{Griffin:2008bc}, we therefore caution researchers on the use and interpretation of model-based estimates of component-specific parameters, especially for causal inference.

First, we suggest that, whenever possible, researchers consider alternative approaches to inference that do not rely on model-based estimation. In the context of principal stratification, these alternatives often rely on constant treatment effect assumptions or on conditional independence across multiple outcomes~\citep[e.g.,][]{Jo:2002cw, Jo:2009hx, Ding:2011cw}. When such restrictions are not possible, we recommend that researchers first compute nonparametric bounds~\citep[see][]{Zhang:2003kk, Grilli:2008ih, Lee:2009eg, miratrix2018bounding}. 

Second, researchers might nonetheless be interested in leveraging parametric assumptions for estimation. In this case, we suggest that researchers use our results to assess the probability of pathological results for different parameter values. Similar to design analysis, these calculations can provide practical guidance on whether mixture modeling will yield useful inference. One possibility is to incorporate multiple outcomes, such as in~\citet{Mattei:2013fb}. This can greatly improve inference; intuitively, the distance between components will be greater in multivariate space, in effect, giving larger $\Delta$ and easier separation~\citep[see also][]{mercatanti2015improving}. 

Third, we have focused on maximum likelihood rather than Bayesian methods \citep{fruhwirth2006finite}. The Bayesian approach offers some distinct advantages over likelihood-based inference.\footnote{The Bayesian approach also introduces some unique challenges that we do not address here, namely the label-switching problem~\citep{Celeux:2000uw,Jasra:2005kl} and the difficulty of specifying vague prior distributions for finite mixtures~\citep{Grazian:2015uf}.} For example, the Bayesian can incorporate informative prior information, which can be especially important in finite mixture modeling; see, for example,~\citet{Aitkin:1985we,Hirano:2000vg, Chung:2004di,Lee:2008wb,Gelman:2010bk}. Moreover, our concern about sign error is trivial in the Bayesian setting: the global mode is simply a poor summary of a multi-modal posterior. More broadly, the weak identification issues we highlight in this paper are not necessarily relevant to a strict Bayesian.~\citet{Imbens} and~\citet{Mattei:2013fb}, for example, characterize weak identification as substantial regions of flatness in the posterior, which increases uncertainty but does not lead to any fundamental challenges.\footnote{\citet{Imbens} note that ``issues of identification [in the Bayesian perspective] are quite different from those in the frequentist perspective because with proper prior distributions, posterior distributions are always proper. The effect of adding or dropping assumptions is directly addressed in the phenomenological Bayesian approach by examining how the posterior predictive distributions for causal estimands change.''} Nonetheless, we argue that our results are highly relevant for Bayesians who are also interested in good frequency properties~\citep{Rubin:1984wm}. In the supplementary materials, we offer evidence that the pathological behaviors we document for the MLE also hold for the posterior mean and median with some ``default'' prior values. In this sense, we conduct a Frequentist evaluation of a Bayesian procedure~\citep[\textit{e.g.},][]{rubin2004multiple} and find poor frequency properties overall. More generally, we agree that informative prior information can be a powerful tool for improving inference in this setting. Finding suitable priors for finite mixture models is a topic for future research.

Going forward, we hope that the approach outlined here can serve as a useful template for studying the behavior of mixture model estimates in finite samples. Moreover, we considered only a very simple case in this paper; in the future, we plan to assess inference for much richer models, especially those common in principal stratification. Finally, we are actively exploring alternative estimation strategies, particularly those that more directly leverage Bayesian methods and that can give sensible point estimates. In the end, inference in the Twilight Zone is possible. But we must proceed with caution.

%%%%%%%%%%%%%%%%%%%%%%%%%%%%%%%%%%%%%%%%%%%%%%%%%%%%%%%%%%%%%%%%%%%%%%%%%%%%%%%%%%%%%%%%%%%%%%%
%%%
%%% BIBLIOGRAPHY
%%%

\clearpage
\singlespacing
\begin{small}
\bibliographystyle{apalike}
\bibliography{combined_mix_ref}
\end{small}

%%%%%%%%%%%%%%%%%%%%%%%%%%%%%%%%%%%%%%%%%%%%%%%%%%%%%%%%%%%%%%%%%%%%%%%%%%%%%%%%%%%%%%%%%%%%%%%
%%%
%%% SUPPLEMENTARY MATERIALS
%%%

\clearpage
\begin{center}
\textbf{\Large{Supplementary Materials for ``Weak separation in mixture models and implications for principal stratification''}}
\end{center}

%%%%%%%%%%%%%%%%%%%%%%%%%%%%%%%%%%%%%%%%%%%%%%%%%%%%%%%%%%%%%%%%%%%%%%%%%%%%%%%%%%%%%%%%%%%%%%%
%%%
%%% APPENDIX
%%%
\appendix
\renewcommand\thefigure{\thesection.\arabic{figure}}

%%%
%%% METHOD OF MOMENTS
%%%
\section{Robust estimation via method of moments}

%\avi{magnitude of $\delta$}
Rather than use higher order moments as diagnostics, we can instead use the method of moments directly for estimation. 
Several recent papers have highlighted the attractive properties of method of moment estimators for general mixture models~\cite{Anima-moment, Wu-denoise}.
Applying these results, we show that the method of moments approach has similar asymptotic properties to the MLE but better finite sample properties; in particular, the method of moments is not susceptible to pile up.

First, in the setting with known, equal variances in Equation~\eqref{eq: block}, we have the following moment equations:
\begin{align}
\label{eq:mix_moments}
	m_1 &= \E[Y] \;\,= \mu \nonumber \\
	m_2 &= \E[Y^2] = 1 + c\delta^2 \\
	m_3 &= \E[Y^3] = \frac{1-2\pi}{1-\pi}c\delta^3, \nonumber
\end{align}
where $\Delta = (1+c)\delta$. Since there is no information in the first moment about $\delta$, we consider two estimators based on the second and third moments:\footnote{In principle, we could also consider a generalized method of moments estimator based on both the second and third moments, though this is less transparent than the estimators we discuss below. See~\cite{Anima-moment, hardt2015tight, Wu-denoise}.}
$$| \widehat{\delta}_{m_2}| : = \biggr| \dfrac{\widehat{m}_{2} - 1}{c} \biggr|^{1/ 2} \qquad\qquad\qquad 
\widehat{\delta}_{m_3} : =  \biggr[ \dfrac{(1 - \pi)}{c\,(1 - 2 \pi)}~  \widehat{m}_{3}\biggr]^{1/ 3},$$
where $\hat{m}_2$ and $\hat{m}_3$ are the sample second and third (non-central) moments, respectively. First, the absolute value for $| \widehat{\delta}_{m_2}|$ is necessary because there is no information about sign of $\delta$ in the second moment. Thus, $\widehat{\delta}_{m_2}$ is a natural estimator when $\pi = 1/2$. By contrast, when $\pi \in (0, 1/2)$, $\hat{\delta}_{m_3}$ will estimate both the magnitude and sign of $\delta$. 

The following result establishes that these estimators have asymptotic behavior similar to the MLE, as described in Theorem~\ref{theorem:convergence_rate}.
\begin{proposition}
\label{prop:rates_moments}
Given the formulations of estimators 
%$\widehat{\delta}_{\text{sym}}$ and $\widehat{\delta}_{\text{asym}}$, 
$\widehat{\delta}_{m_2}$ and $\widehat{\delta}_{m_3}$, for the setting of known equal variances~\eqref{eq: block}, the following holds
\begin{itemize}
\item[(a)] (Asymmetric regime) When $\pi \in (0,1/2)$, then 
\begin{eqnarray}
	\sup \limits_{\delta_{n} \in \Theta} \abss{ \abss{\widehat{\delta}_{m_2}} - \abss{\delta_{n}}} &=& \; O_{p}\parenth{n^{- 1/4}}, \\[0.5em]
	\sup \limits_{\delta_{n} \in \Theta} \abss{\widehat{\delta}_{m_3} - \delta_{n}} &=& \; O_{p}\parenth{n^{- 1/ 6}}.
\end{eqnarray}
\item[(b)] (Symmetric regime) When $\pi = 1/2$, then 
\begin{eqnarray}
	\sup \limits_{\delta_{n} \in \Theta} \abss{ \abss{\widehat{\delta}_{m_2}} - \abss{\delta_{n}}} &=& \; O_{p}\parenth{n^{- 1/4}},
\end{eqnarray}
\end{itemize}
where $\widehat{\delta}_{m_3}$ is undefined when $\pi = 1/2$.
%\begin{align*}
%	\sup \limits_{\delta_{n} \in \Theta} \abss{\widehat{\delta}_{m_3} - \delta_{n}} &= \; O_{p}\parenth{n^{- 1/ 6}} \\[0.5em]
%	\sup \limits_{\delta_{n} \in \Theta} \abss{ \abss{\widehat{\delta}_{m_2}} - \abss{\delta_{n}}} &= \; O_{p}\parenth{n^{- 1/4}}.
%\end{align*}
\end{proposition}
While these simple estimators have the same asymptotic behavior as the MLE, neither $\widehat{\delta}_{m_2}$ nor $\widehat{\delta}_{m_3}$ are susceptible to pile up. It suggests that the moment estimators under the simple setting of known equal variances are more robust than the MLE. 
%\avi{more}

%%%
%%% JOB CORPS
%%%
\section{Analysis of Job Corps}
\label{sec:jobcorps}

\subsection{Setup.} 
Following~\cite{Zhang:2009ku}, we use the principal stratification framework to define the impact of Job Corps on hourly wages. Let $S$ be an indicator for employment, with corresponding potential outcomes $S_i(0)$ and $S_i(1)$ and observed employment status $S_i^{\text{obs}}$ for individual $i$. We then define principal strata, $U$, based on the joint distribution, $\{S_i(0), S_i(1)\}$:
\begin{equation*}
U_i = \begin{cases}
 EE & \mbox{ if } S_i(1) = 1, S_i(0) = 1 \\
 EN & \mbox{ if } S_i(1) = 1, S_i(0) = 0 \\
 NE & \mbox{ if } S_i(1) = 0, S_i(0) = 1 \\
 NN & \mbox{ if } S_i(1) = 0, S_i(0) = 0
 \end{cases}.
\end{equation*}
We are interested in the impact of randomization on the \textit{always employed} strata, $EE$.  This is sometimes known as a \textit{Survival Average Causal Effect} and is closely related to the idea of ``truncation due to death''~\cite[see][]{Zhang:2009ku}. Finally, following~\cite{Lee:2009eg}, we invoke the \textit{monotonicity} assumption, which states that random encouragement to enroll in a job training program can only increase employment, $S_i(1) \geq S_i(0)$; thus the $NE$ group does not exist.\footnote{While this simplifies the analysis and allows us to highlight the role of finite mixture modeling,~\cite{Zhang:2009ku} argue against this assumption. In particular, they argue that enrolling in a job training program might raise an individual's \textit{reservation wage} and, as a result, make that individual less likely to accept a lower paying job. We merely note that relaxing this assumption further complicates the analysis, since the mixing proportions are no longer identified non-parametrically.}

\begin{table}[bt]
	\centering
	\caption{Summary statistics for observed groups in Job Corps}
	\label{tbl:jobcorps_obs_means}
	\begin{tabular}{ccccl}
	$Z$ & $S^{\text{obs}}$ & \textbf{Observed Mean} & \textbf{Observed SD} & \textbf{Possible Principal Strata}\\
	\hline
	1 & 1 & 0.03 & 1.013 & $EE$ and $EN$ \\
	1 & 0 & --- & --- & $NN$ \\
	0 & 1 & -0.05 & 1 & $EE$ \\
	0 & 0 & --- & --- & $NN$ and $EN$ \\
	\hline
	\end{tabular}	
\end{table}

Table~\ref{tbl:jobcorps_obs_means} shows the relationship between principal strata and the observed groups, based on $Z$ and $S^{\text{obs}}$. Under monotonicity, we directly observe \textit{always employed} individuals ($EE$) assigned to the control group. We can therefore directly estimate the average outcome for this group, $\mu_{\text{EE}0}$. We can also directly estimate the proportion of $EE$ individuals via $\widehat{\pi}_{EE} = \mathbb{P}[S \mid Z_i = 0]$, the proportion of \textit{never employed} individuals ($NN$) via $\widehat{\pi}_{\text{NN}} = 1 - \mathbb{P}[S \mid Z_i = 1]$, and the proportion of the induced to employment individuals ($EN$) via $\widehat{\pi}_{\text{EN}} = 1 - \widehat{\pi}_{\text{NN}} - \widehat{\pi}_{\text{EE}}$. Without additional assumptions, however, we cannot estimate $\mu_{\text{EE}1}$, instead observing a mixture of $EE$ and $EN$ individuals. Consistent with~\cite{Zhang:2009ku} and~\cite{Frumento:2012gz}, we therefore assume that log-hourly wages follow a mixture of Gaussians with known mixing proportion, as in Equation~\eqref{eq: mix} in the main text. Note that this mixture is much simpler than the full model considered in~\cite{Zhang:2009ku}, which accounts for some important additional complications.

\subsection{Diagnostics.} We focus on a complete case subset used by~\cite{Lee:2009eg} of $N = 9,145$ individuals, with $N_1 = 5,546$ randomly assigned to treatment and $N_0 = 3,599$ to control. The mixture model consists of the $N_{11} = 3,371$ individuals assigned to treatment who are employed, with mixing proportion $\widehat{\pi} = 0.06$. 

Table~\ref{tbl:jobcorps_obs_means} shows summary statistics for observable groups. We standardize the outcome by subtracting off the grand mean and dividing by $\widehat{\sigma}_0$, the estimated standard deviation for individuals assigned to control who are employed. This is also the standard deviation for $EE$ individuals assigned to control. Since hourly wage is only defined for employed workers, the rows with $S^{\text{obs}} = 0$ have undefined outcomes.

	\begin{figure}[hbtp]
	\begin{center}
		\begin{subfigure}[b]{0.4\textwidth}
				\begin{center}
				\includegraphics[width = \textwidth]{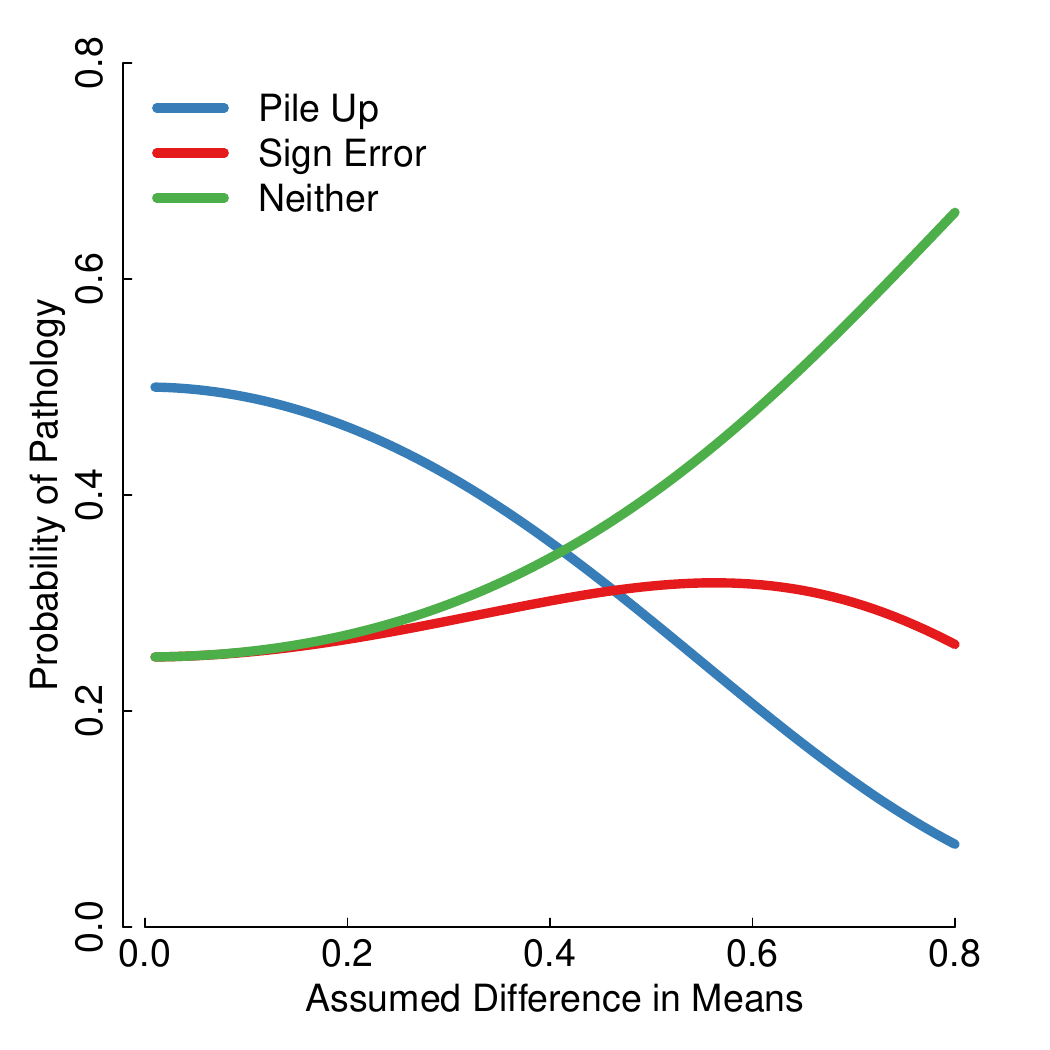} 
				\caption{Prob. of MLE Pathology}
				\label{fig:JobCorps_Norm}
				\end{center}
		\end{subfigure}%
		~ \begin{subfigure}[b]{0.4\textwidth}
			\begin{center}
				\includegraphics[width = \textwidth]{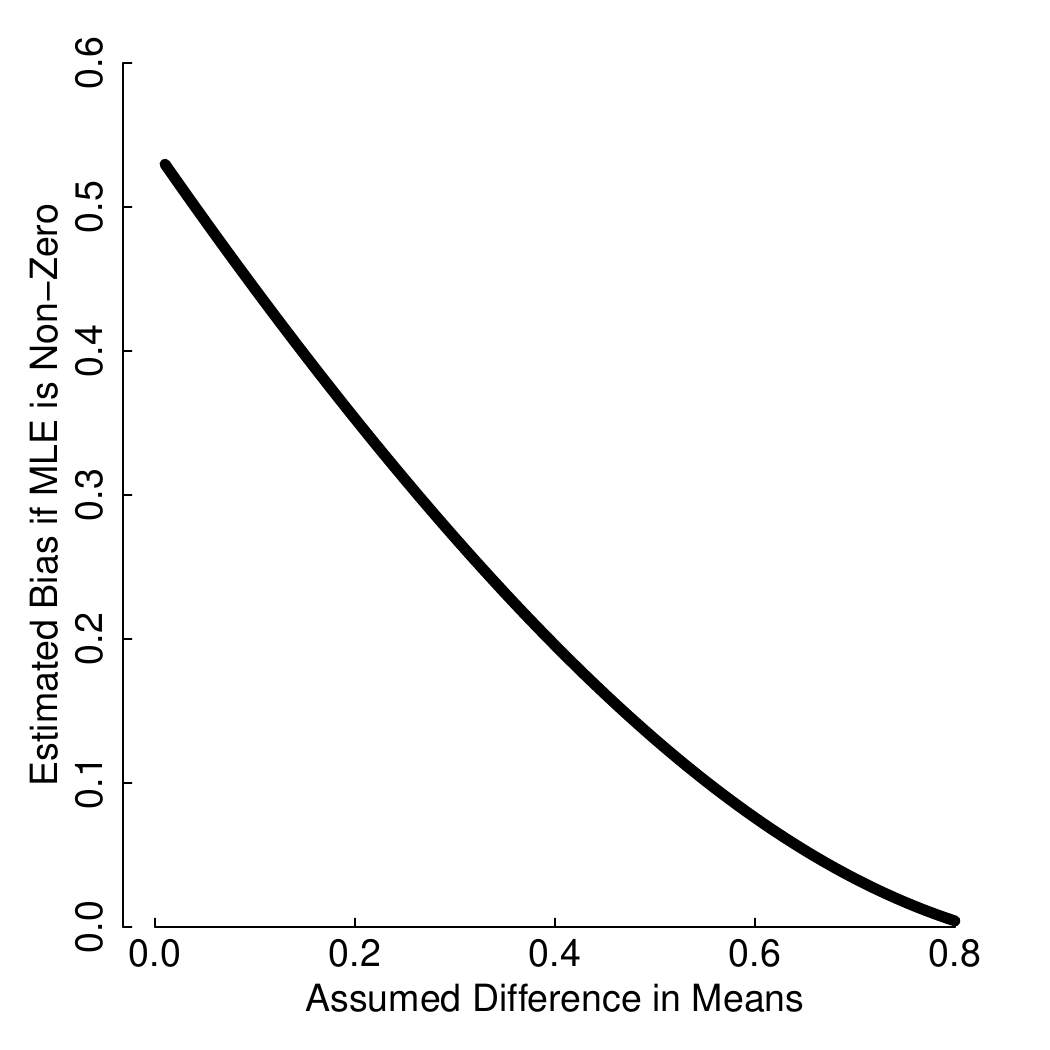}
				\caption{Expected Bias if MLE is non-zero}
				\label{fig:JobCorps_Bias}
			\end{center}
		\end{subfigure}\\
		~ \begin{subfigure}[b]{0.5\textwidth}
			\begin{center}
				\includegraphics[width = \textwidth]{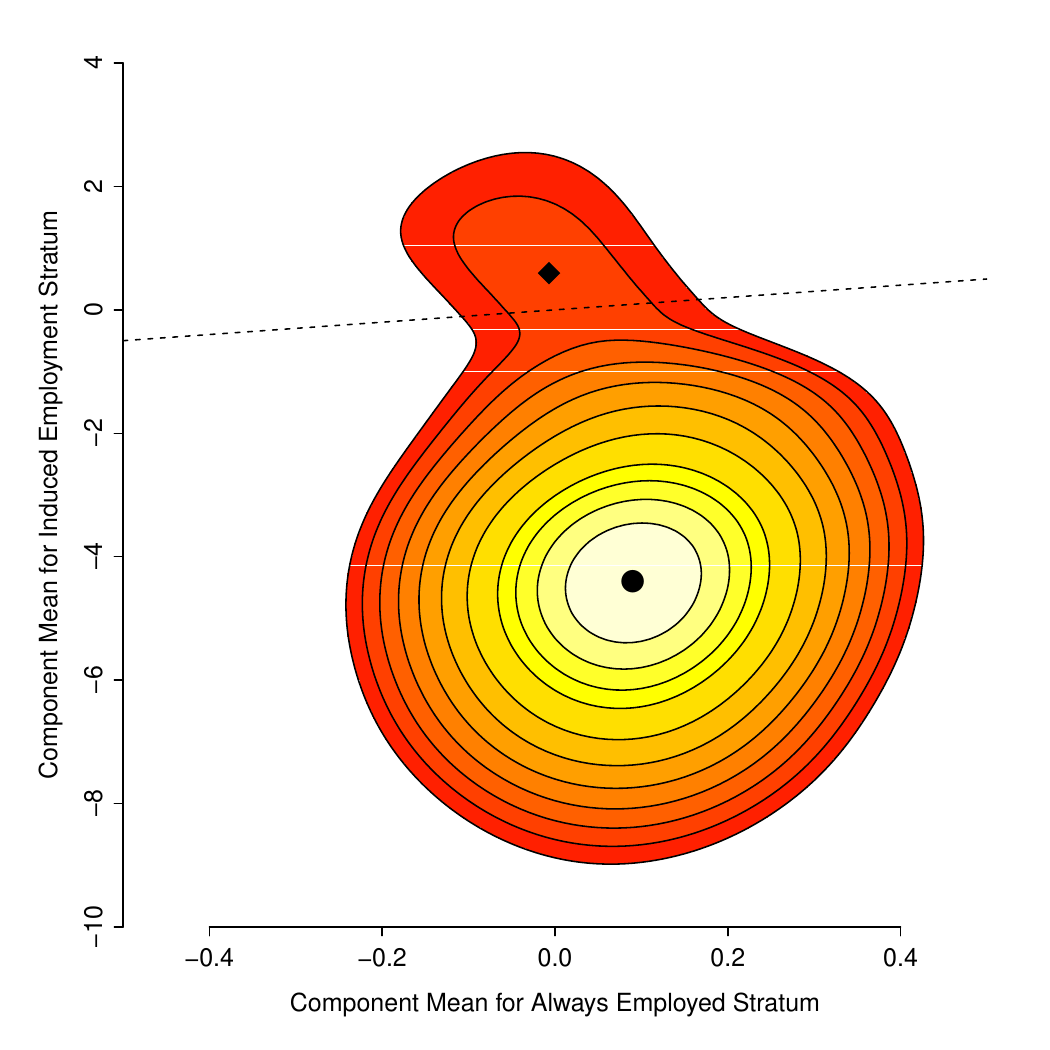}
				\caption{Observed likelihood}
				\label{fig:Jobcorps_ll}
			\end{center}
			\end{subfigure}
		\caption{Quality of Maximum Likelihood Estimation for the finite mixture model in Job Corps, with parameters  $N = 3,371$ and $\pi = 0.06$. Panels (a) and (b) show the probability of MLE pathology and expected bias of the MLE if non-zero; Panel (c) shows the observed likelihood for the Job Corps mixture, with a global mode and a local mode. The dotted line denotes equal component means.}
	\label{fig:JobCorps_Pathologies}
	\end{center}
	\end{figure}
	
Figure~\ref{fig:JobCorps_Norm} gives the probability of pile up and sign error over a range of plausible values of $\Delta$ using the Normal approximation in Equation~\eqref{eq: joint} and the observed Job Corps mixtures parameters of $N = 3,371$ and $\widehat{\pi} = 0.06$. As in Figure~\ref{fig:Jobs_Norm}, pile up is a major concern, though the probability of a sign error is somewhat less \textit{ex ante}, in part because the mixing proportion is much closer to 0. Figure~\ref{fig:JobCorps_Bias} shows the bias of the MLE if the MLE is non-zero and the sign is correct. As with JOBS II, the bias can be severe.

We can also incorporate the higher order moments of the mixture distribution. In this case, the observed second and third moments are $\widehat{m}_2 = 1.03$ and $\widehat{m}_3 = -0.87$, respectively (after centering the mixture distribution). Plugging the observed values into the Normal approximations in Equation~\eqref{eq: joint}, the pile up probability of 0.34 and the sign error probability is 0.03. The corresponding probabilities based on the case-resampling bootstrap are nearly identical, 0.34 and 0.04 respectively. 

Figure~\ref{fig:Jobcorps_ll} shows the observed likelihood for the mixture model. The MLE is at $\widehat{\mu}_{\text{EE} 1}^{\mle} = 0.09$ and $\widehat{\mu}_{\text{NE} 1}^{\mle} = -4.40$, which implies $\widehat{\Delta}^{\mle} = -4.49$ standard deviations. This is clearly an extreme estimate. Transforming these estimates to \$ per hour shows that $\widehat{\mu}_{\text{EE} 1}^{\mle} = \$8.24$ per hour and $\widehat{\mu}_{\text{NE} 1}^{\mle} = \$0.09$ per hour, which is far below feasible hourly wages in this sample. This estimate is also outside the minimax bounds, $\Delta \in [-2.4, 2.2]$.\footnote{Following~\cite{Lee:2009eg}, we calculate minimax bounds via trimmed means of the mixture distribution. Specifically, we bound $\mu_{\text{NE}1}$ via the mean of the $\pi = 0.06$ individuals with, respectively, the lowest and highest values of hourly wages, with similar bounds for $\mu_{\text{EE}1}$.}
There is also a local mode centered at $\widehat{\mu}_{\text{EE} 1}^{\mle} = -0.01$ and $\widehat{\mu}_{\text{NE} 1}^{\mle} = 0.59$, which implies  $\widehat{\Delta}^{\mle} = 0.60$ standard deviations. In units of \$ per hour, this is $\widehat{\mu}_{\text{EE} 1}^{\mle} = \$7.47$ per hour and $\widehat{\mu}_{\text{NE} 1}^{\mle} = \$13.64$ per hour. While far more feasible than the global mode, these estimates are still worrisome, since it is unlikely that the group induced to employment by Job Corps would have hourly wages nearly twice those of the always employed group; see Figure~\ref{fig:JobCorps_Bias}. Regardless, the likelihood at the MLE is considerably higher than at the local mode, with $-2*(\ell(+0.60|Y) - \ell(-4.49|Y)) = 296$.  Taken together, these results suggest that maximum likelihood does not give practically useful results in this example. 
% $\widehat{mu}_{\text{EE} 0}^{\mle} = \$7.38$ per hour.

In practice, the simplest explanation for these results is that the simple Normal mixture model in Equation~\eqref{eq: mix} in the main text is a poor fit to the data. At the same time, however, it is difficult to imagine a more plausible parametric mixture model in this setting. Thus parametric finite mixtures might not be an effective strategy in this example.

\section{Validating the Normal approximations}
\label{subsec: val}

\begin{figure}
	\centering
	\includegraphics[scale = 0.9]{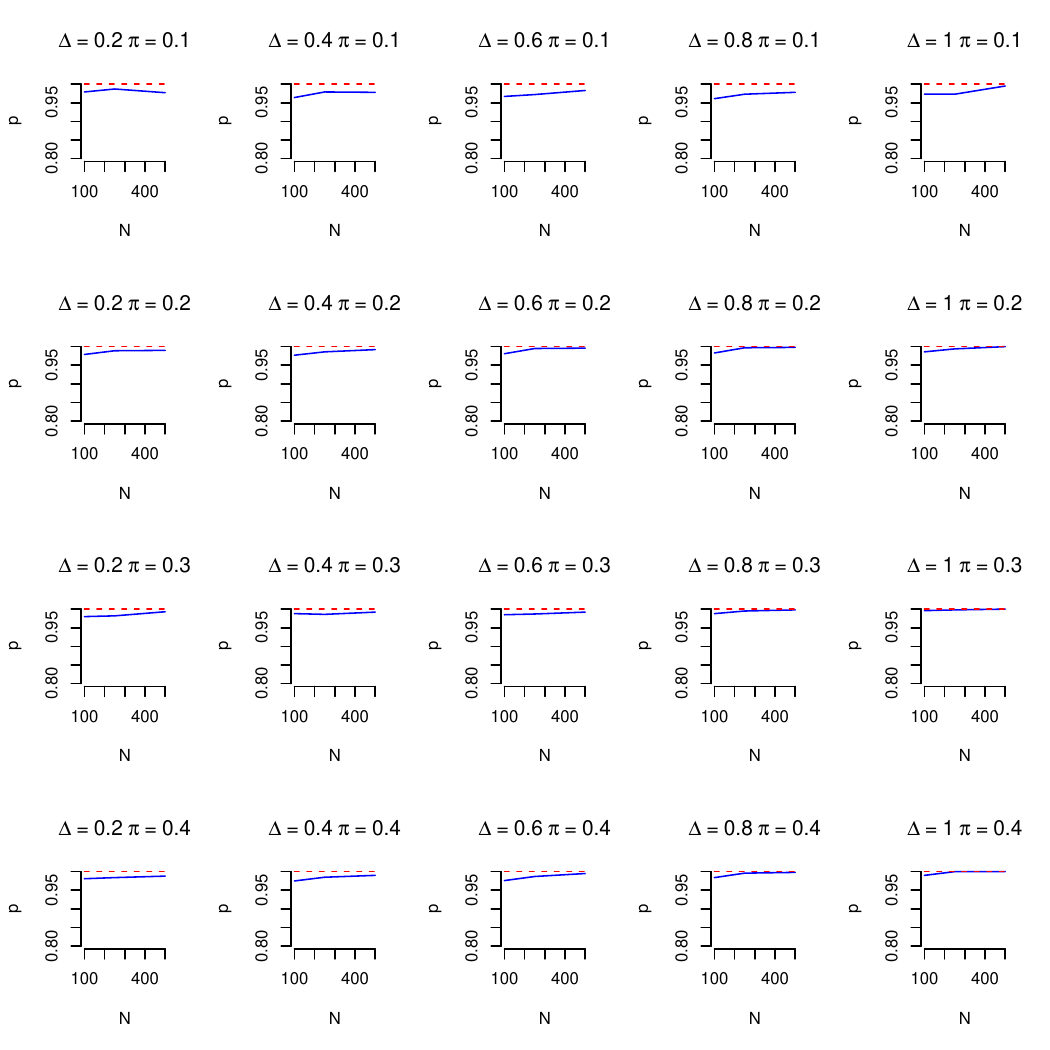}
	\caption{Probability that the diagnostic based on the second moment ($1\{\widehat{m}_2 < 1\}$) agrees with whether or not pile up was observed in simulation. The dotted red line perfect correspondence at each tested $N$. The blue line is the average agreement probability over 1000 simulated data sets.
}
	\label{fig: pile_comp}
\end{figure}

We present figures testing the agreement of the moment-based Normal approximations with their corresponding pathologies assessed via simulation. Figure~\ref{fig: pile_comp} compares the incidence of pile up and $\widehat{m}_2 < 1$ for a range of values of $\pi,\Delta,$ and $N$. The blue line indicates the probability the method of moments estimator indicator of pile up ($1\{\widehat{m}_2 < 1\}$) agrees with whether or not pile up was observed in simulation. The results are averaged over 1000 simulated data sets. Unsurprisingly, the correspondence improves as $N$ increases and is worst when $\pi = 0.1$, the case in which the mixture is its most asymmetric. Overall, however, the Normal approximation provides an excellent estimator for whether pile up has occurred in the sample.

\begin{figure}
	\centering
	\includegraphics[scale = 0.9]{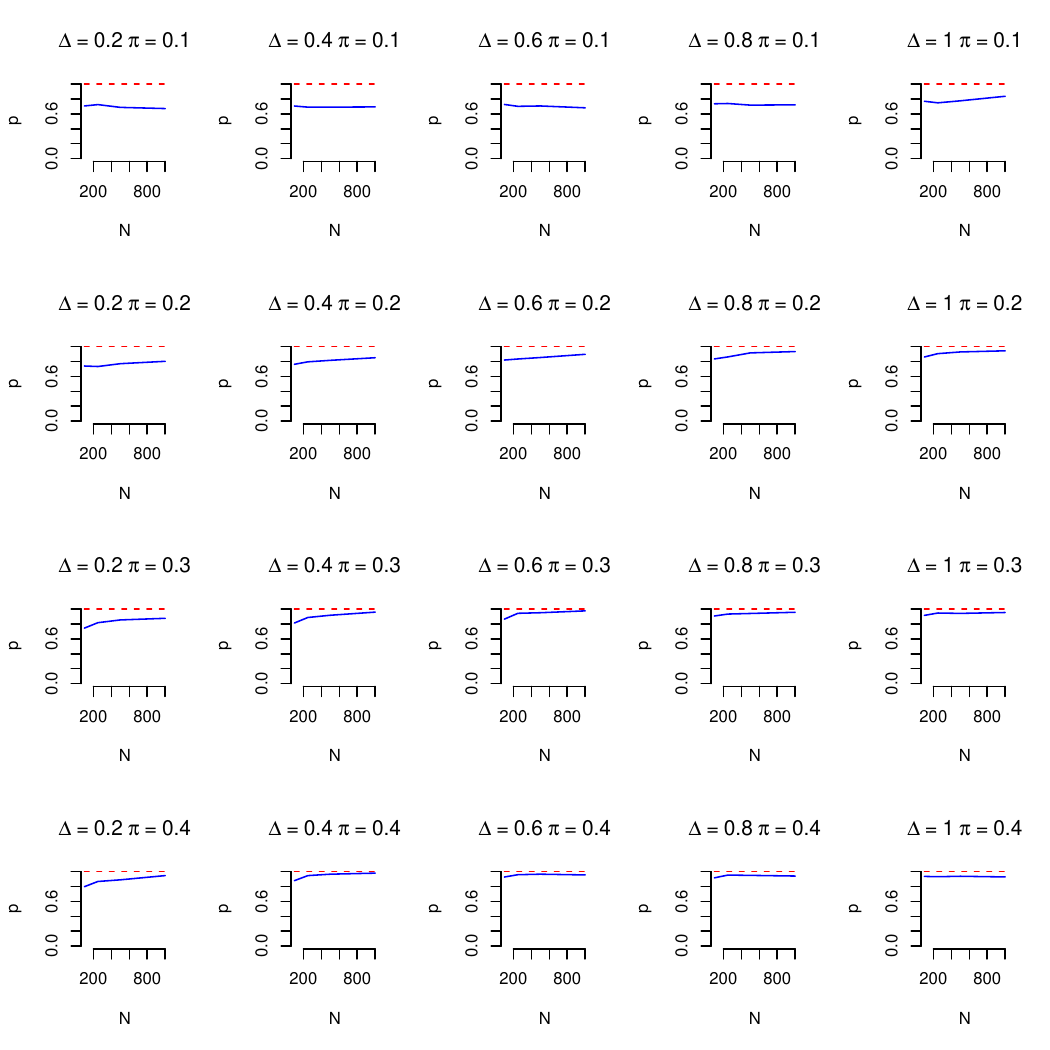}
	\caption{Probability that the diagnostic based on the third moment agrees with whether or not the wrong sign pathology was observed in simulation. The dotted red line perfect correspondence at each tested $N$. The blue line is the average agreement probability over 1000 simulated data sets.}
	\label{fig: sign_comp}
\end{figure}

Figure~\ref{fig: sign_comp} shows the corresponding plots for assessing the sign of $\Delta$. Here, due to the extra noise in $m_3$, the correspondence is much less sharp. The discrepancies are most noticeable when $\pi$ is close to $0$ and $\Delta$ is small. 

\begin{figure}[hbt]
	\centering
	\includegraphics[scale = 0.5]{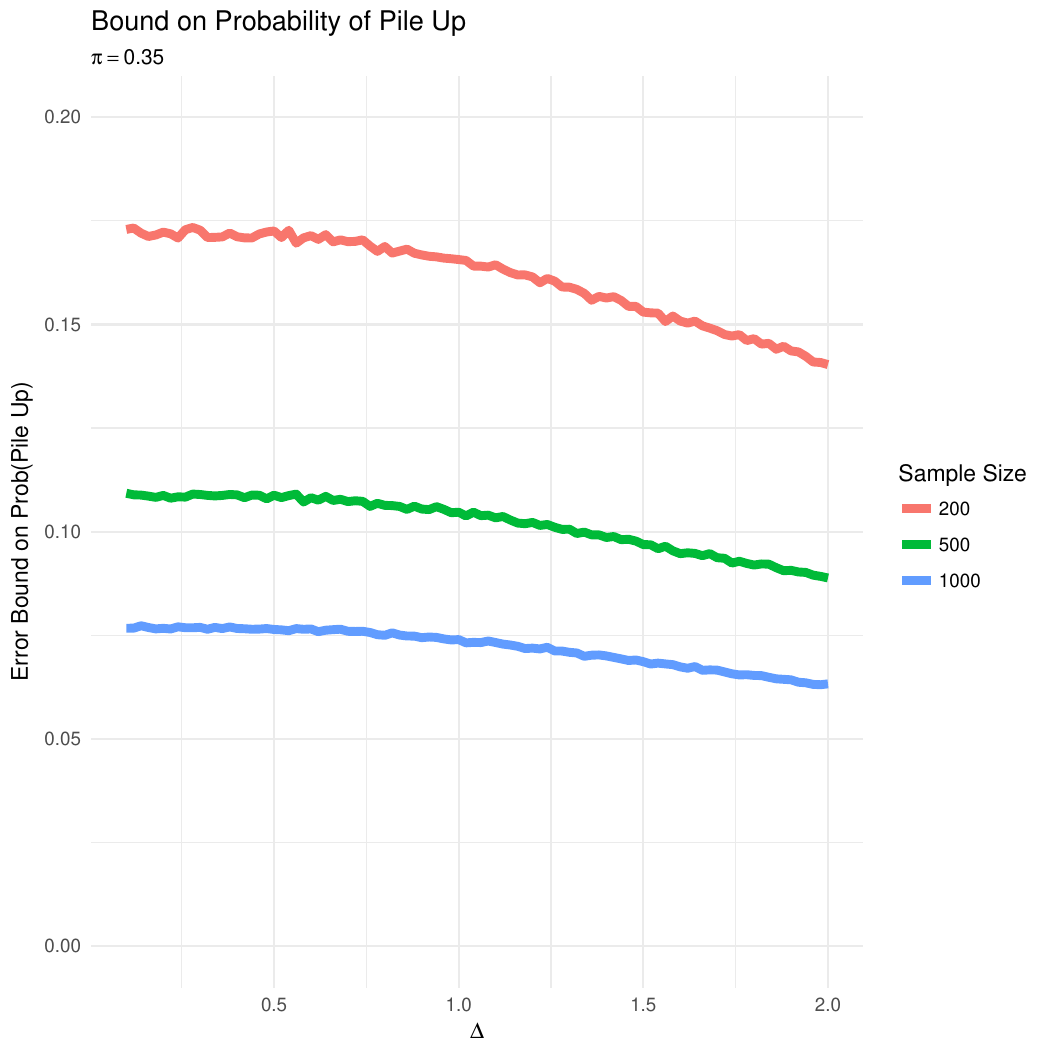}
	\caption{Berry-Essen bound for probability of pile up for $\pi = 0.35$ and a range of values of $N$ and $\Delta$.}
	\label{fig: BE_Bound}
\end{figure}

%%%
%%% ROBUST INFERENCE
%%%
\section{Confidence sets via inverting tests}
\label{sec: ri}

Given the poor performance of the MLE, we are interested in methods that perform well even when $\Delta$ is small. Based on the large literature on weak identification in other settings, we presume that many such methods are possible. As a starting point, we suggest an approach to construct confidence intervals based on inverting a sequence of tests. This approach is widely used in other weak identification settings, namely weak instruments~\cite{Staiger:1997, Kang:2015te} and the unit root moving average problem~\cite{Mikusheva:2007wl}. It is also closely related to the method of constructing confidence intervals for causal effects by inverting a sequence of Fisher Randomization Tests~\cite{rosenbaum2002observational}. 

At the same time, this approach has its drawbacks. First, while test inversion yields confidence sets with good coverage properties, it does not necessarily yield good point estimates. In particular, it is possible to construct a Hodges-Lehmann-style estimator via the point on the grid with the highest $p$-value~\cite{Hodges:1963}. But since pile up and sign error remain issues, any point estimator in this case should be interpreted with caution. Second, the coverage guarantees hold only when the model is correctly specified; under even moderate mis-specification, the resulting estimator can cease to exist~\cite{Gelman:Blog}. Importantly, the MLE performs poorly \textit{even when the model is correctly specified}. Alternatively, researchers uninterested in test inversion for confidence intervals might nonetheless be interested in using this approach to assess model fit. If the proposed procedure rejects everywhere, this is evidence that the Normal mixture model is a poor fit.

We discuss two basic approaches here. Our first approach is a version of the grid bootstrap of~\cite{andrews1993exactly} and~\cite{hansen1999grid}, which generates Monte Carlo $p$-values by simulating fake data sets from the null hypothesis. While the grid bootstrap is conceptually straightforward and enjoys theoretical guarantees~\cite{Mikusheva:2007wl}, it is also computationally intensive. Our second approach is therefore a fast approximation that directly uses the Normal sampling distribution in Equation~\eqref{eq: joint} of the main text to derive a $\chi^2$ test at each grid point. To demonstrate these methods, we first outline inference for $\Delta$ alone and then extend this to inference for the component-specific means, $\mu_0$ and $\mu_1$.

\subsection{Overview of grid bootstrap}
To conduct a grid bootstrap, we first need a grid. Define \textbf{$\Delta$} $= \{\Delta_0,\Delta_1,\ldots,\Delta_n\}$ with $\Delta_i > \Delta_j$ for $i > j$. 
The immediate goal is then to obtain a $p$-value for the following null hypotheses for each value $\Delta_j \in$ \textbf{$\Delta$}: 
\begin{equation}
\label{eq: pile_hyp}
H_0: \Delta = \Delta_j \hspace{1mm}\text{vs.}\hspace{1mm}H_1: \Delta \neq \Delta_j.
\end{equation}
For convenience we first center the data (i.e., we set $\mu = 0$ as in the main text). Next, we need a test statistic, $t(\mathbf{y}, \Delta_j)$, that is a function of the observed (or simulated) data and the value of $\Delta$ under the null hypothesis, $\Delta = \Delta_j$. For a given $N$, and initially assuming $\pi$ and $\sigma^2$ are known, we then obtain exact $p$-values through simulation with the following procedure:

\begin{itemize}
	\item For each $\Delta_j \in \textbf{$\Delta$}$
	\begin{itemize}
		\item Calculate the observed test statistic, $t^{\obs}_{j} = t(\mathbf{y}^{\obs}, \Delta_j)$.
		\item Generate $B$ data sets of size $N$ from the model
	$$\mathbf{y}_{j}^\ast \stackrel{\text{iid}}{\sim} \pi \mathcal{N}\left(\frac{\Delta_{j}}{2}, \sigma^2\right) + (1 - \pi) \mathcal{N}\left(-\frac{\Delta_{j}}{2}, \sigma^2\right).$$
		\item For each simulated $\mathbf{y}_{j}^\ast$, compute $t^\ast_{j} = t(\mathbf{y}_{j}^\ast, \Delta_j)$.
		\item Calculate the empirical $p$-value of $t^{\obs}_{j}$ as a function of the null distribution, $t^\ast_{j}$.	
		\end{itemize}
	
	\item Calculate the confidence set, $\textrm{CS}_{\alpha}(\Delta) = \{ \Delta_j : p(\Delta_j) > 1-\alpha \}$
	for a specified significance level $\alpha$, where $p(\Delta_j)$ is the empirical $p$-value of $\widehat{\Delta}^{\text{mle}}$ assuming that $\Delta = \Delta_j$. 

\end{itemize}
\noindent Note that the resulting confidence set might not be continuous, which could occur if the sampling distribution is strongly bimodal.

\subsection{Constructing a test statistic}
So long as the model is correctly specified, this approach yields an exact $p$-value for any valid test statistic, up to Monte Carlo error~\cite{Mikusheva:2007wl}. We propose a test statistic based on the joint distribution of $\widehat{m}_2$ and $\widehat{m}_3$.\footnote{There are many possible alternatives. For example,~\cite{Frumento:2016tk} suggest test statistics based on scaled log-likelihood ratios. Another option is to use univariate test statistics based on $\widehat{m}_2$ or $\widehat{m}_3$.} Equation~\ref{eq: joint} suggests a natural combination of the estimated cumulants:

\begin{equation}\label{eq: cumulant_test_stat}
t_{m}(\mathbf{y}, \Delta_j) = (d_2, d_3) \text{Var}(m_2,m_3)^{-1} (d_2,d_3)^T, % \sim \chi^2_2
\end{equation}
where $d_k = \widehat{m}_k - m_k$, and we use the assumed null of $\Delta = \Delta_j$ to obtain $(m_2,m_3)$ and $\text{Var}(m_2, m_3)$. %Note that this statistic takes into account information about how much $\hat{\kappa}_2$ differs from $0$. 
As we saw, the Normal approximation in Equation~\eqref{eq: joint} in the main text is excellent, even for modest sample sizes (say $N > 100$). This implies:
$$t_{m}(\mathbf{y}, \Delta_j) \stackrel{a}{\sim} \chi^2_2.$$
We can therefore obtain a $p$-value via a Wald test, rather than via simulation, at each grid point, which is much faster computationally. 

Finally, to use these approaches to estimate component means, we need to (1) expand the grid, and (2) expand the test statistic. A natural choice for a grid of points is the two-dimensional grid over $\mu_0$ and $\mu_1$. To expand the test statistic, we directly use the first three cumulants from Equation~\eqref{eq: joint} from the main text and from~\cite{Tan} to obtain a joint test statistic as in Equation~\eqref{eq: cumulant_test_stat}:

\begin{equation}\label{eq: cumulant_test_stat3}
t_{m}(\mathbf{y}, \Delta_j) = (d_1, d_2, d_3) \text{Var}(\kappa_1, \kappa_2,\kappa_3)^{-1} (d_1, d_2, d_3)^T \sim \chi^2_3.
\end{equation}

\begin{figure}[btp]
\centering
		 \begin{subfigure}[b]{0.3\textwidth}
				\includegraphics[width = \textwidth]{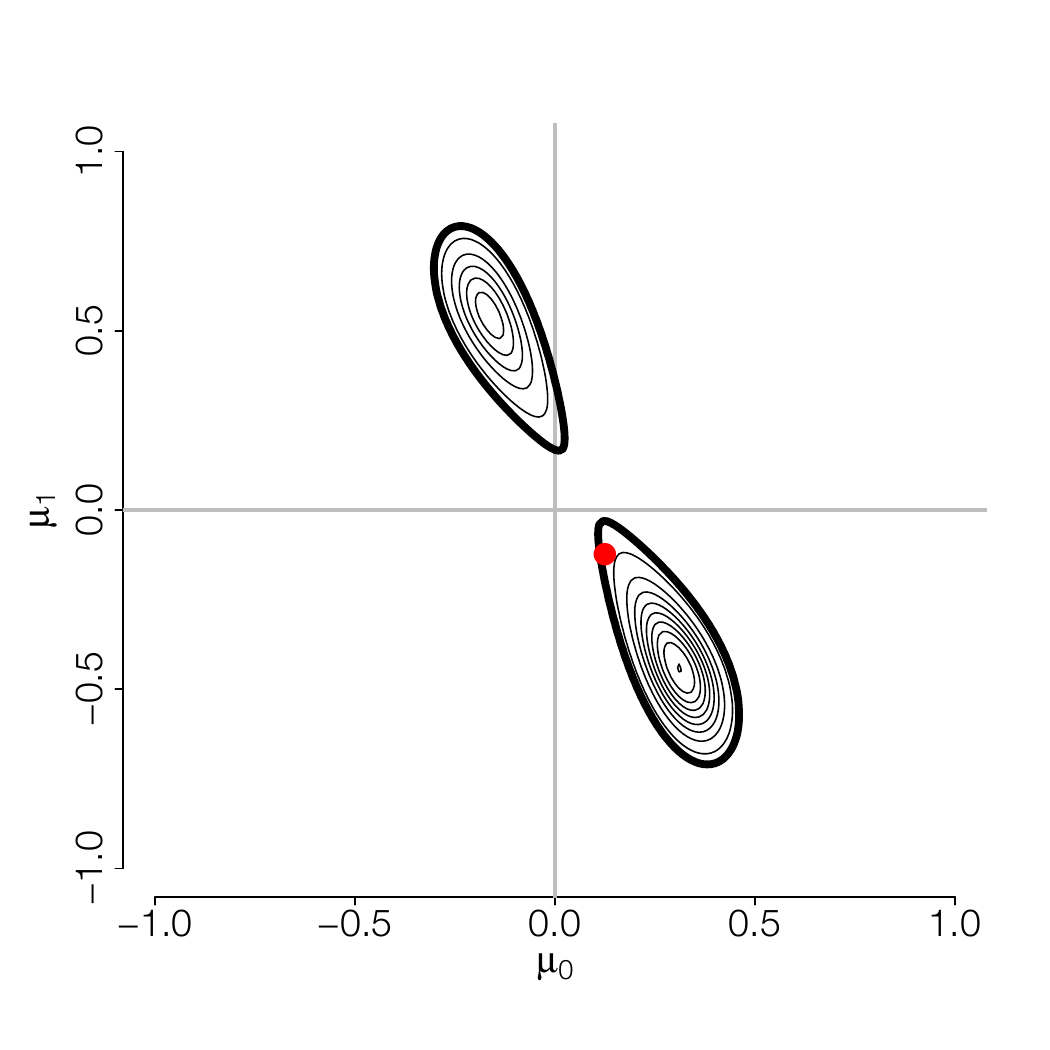}
		\end{subfigure}%
		~\begin{subfigure}[b]{0.3\textwidth}
				\includegraphics[width = \textwidth]{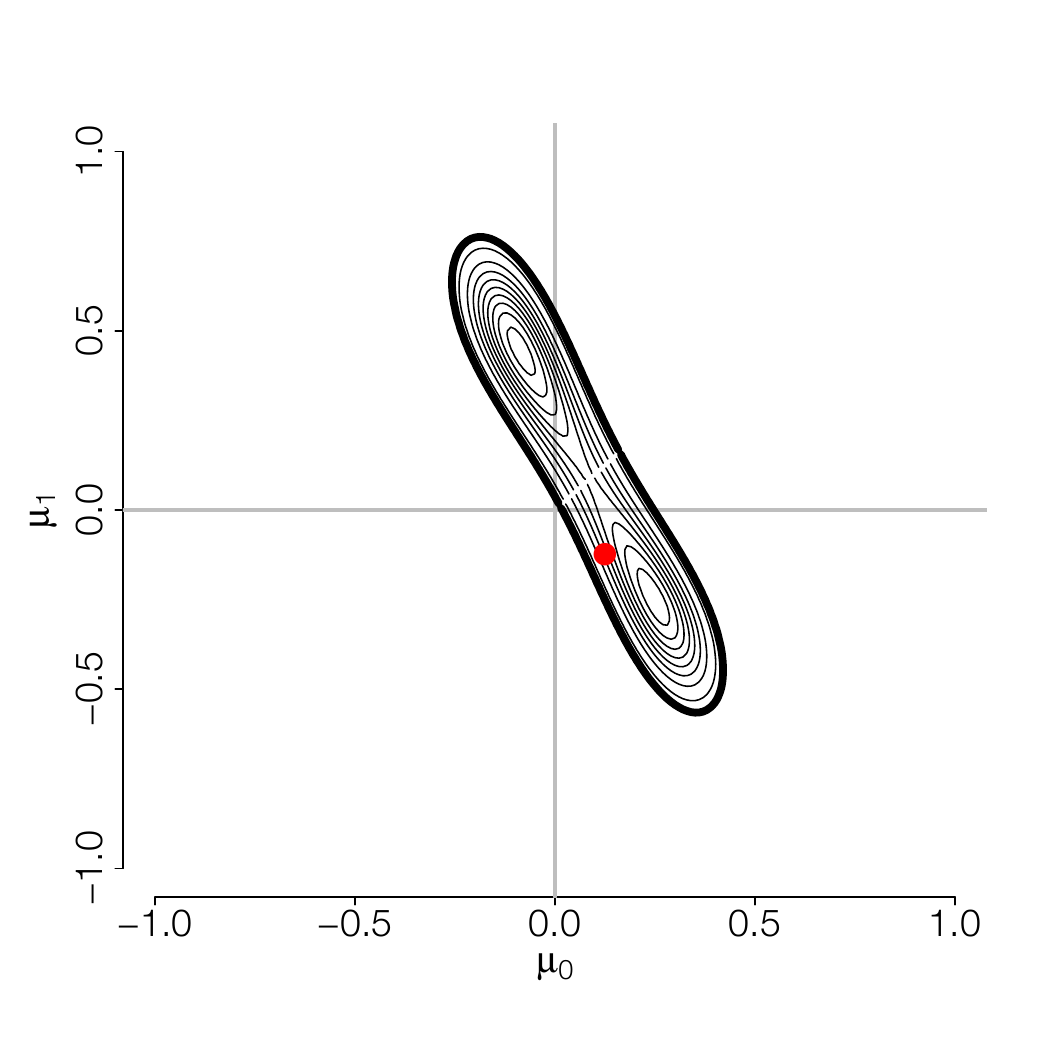} 
		\end{subfigure}
		~\begin{subfigure}[b]{0.3\textwidth}
				\includegraphics[width = \textwidth]{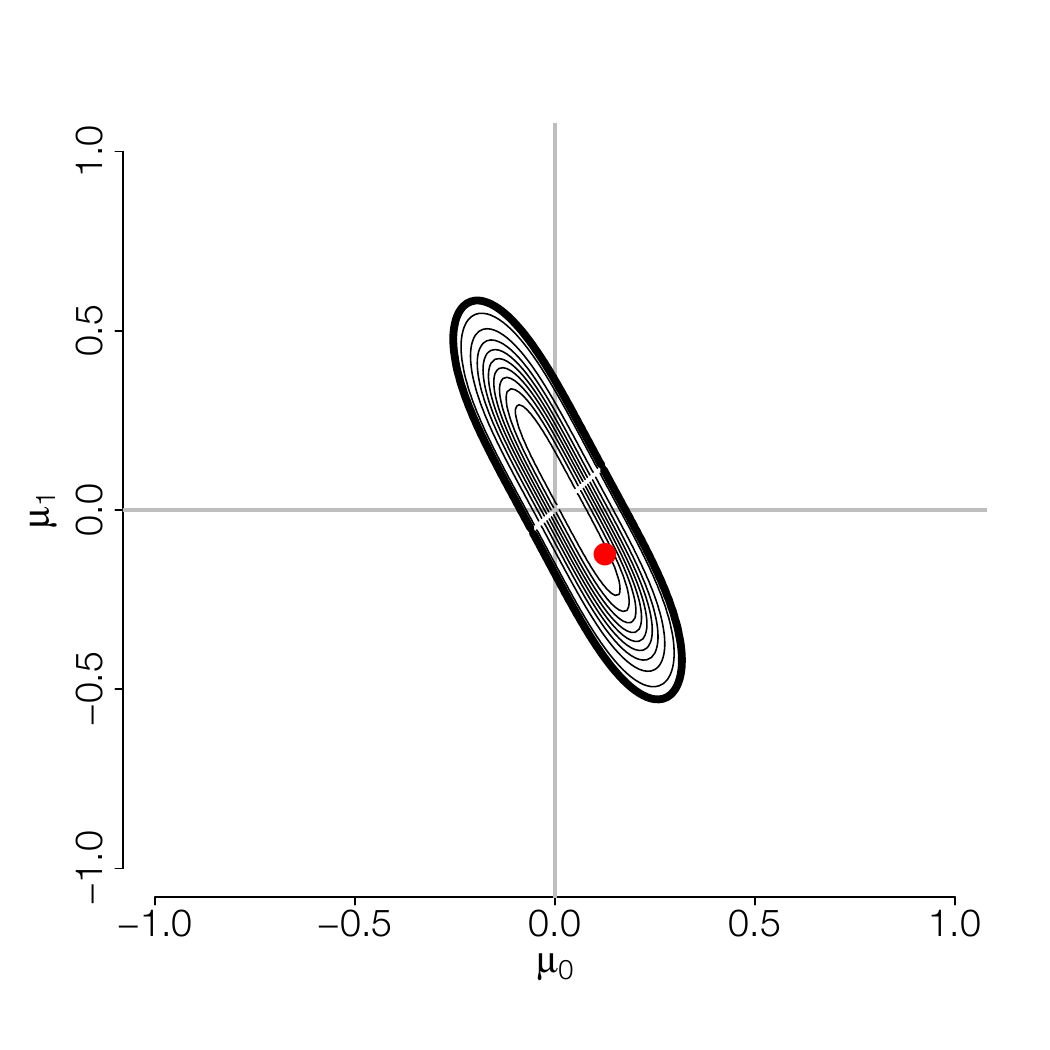} 
		\end{subfigure}
\caption{Three examples of the grid of Wald test $p$-values from Equation~\ref{eq: cumulant_test_stat3}. The three simulated data sets were drawn from Equation~\eqref{eq: mix} in the main text with $N = 1000$, $\pi = 0.325$, $\sigma^2 = 1$, $\mu_0 = \frac{1}{8}$, $\mu_1 = -\frac{1}{8}$. The dark line shows the cutoff for $p = 0.05$. The red dot shows the true value. Note that the Wald test is undefined when $\mu_0 = \mu_1$.}
\label{fig: MOM_for_means_examples}
\end{figure}

As above, we can obtain $p$-values via the grid bootstrap rather than via the $\chi^2$ distribution. Figure~\ref{fig: MOM_for_means_examples} shows the distribution of $p$-values for three different examples from the same data generating process, with  $N = 1000$, $\pi = 0.325$, $\sigma^2 = 1$, $\mu_0 = +\frac{1}{8}$, $\mu_1 = -\frac{1}{8}$.\footnote{Note that the $\chi^2$ distribution no longer holds when $\mu_0 = \mu_1$. While we can use a univariate Normal distribution to obtain a valid $p$-value in this case, this additional complication is generally unnecessary in practice.}

\begin{figure}[hbtp]
	\centering
	\includegraphics[scale = 0.6]{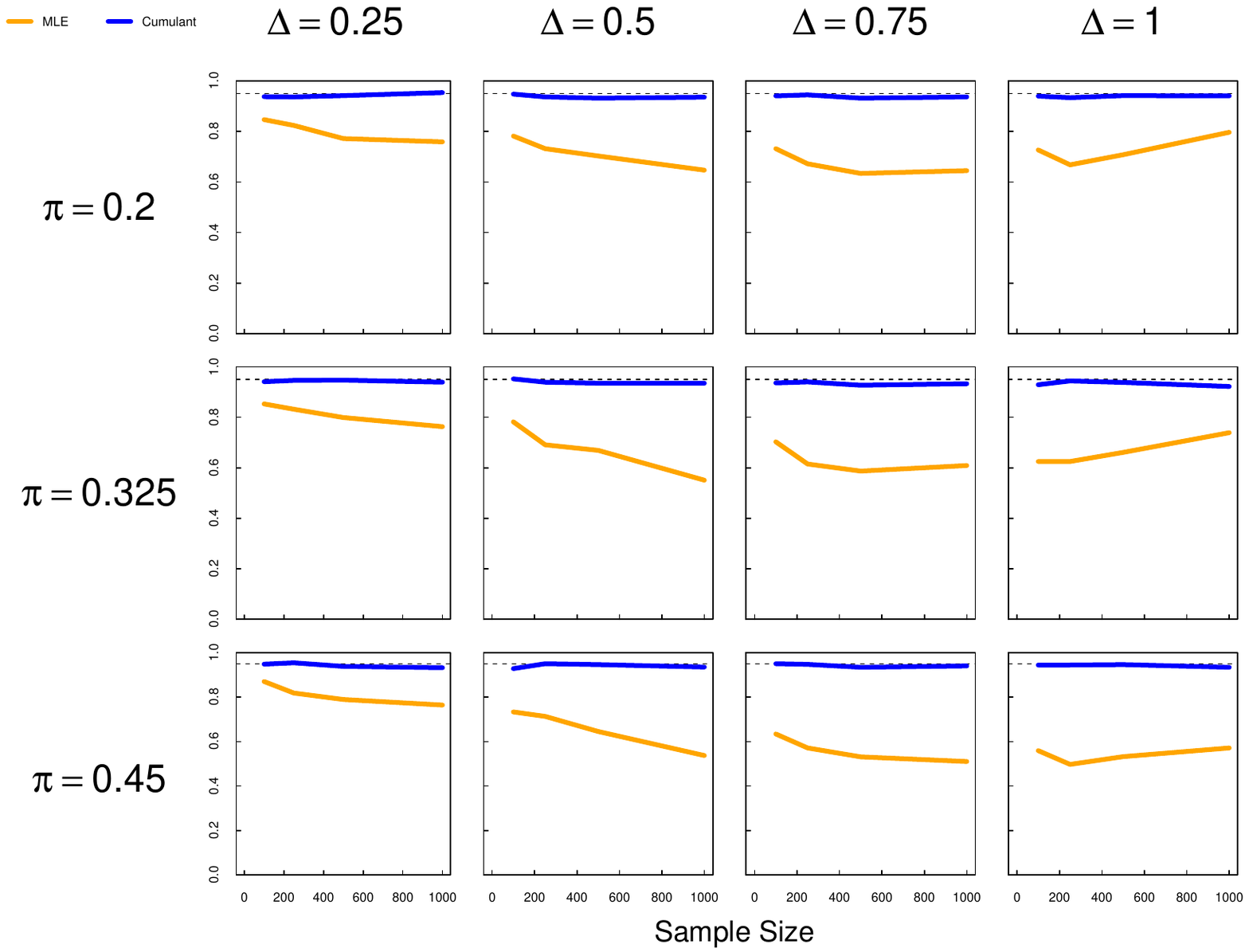}
	\caption{Coverage for 95\% confidence sets based on the test inversion algorithm described in Section~\ref{sec: ri}. The results for the MLE are for the standard finite mixtures estimator.}
	\label{fig:CS_coverage}
\end{figure}

Figure~\ref{fig:CS_coverage} shows the 95\% coverage for the confidence sets obtained through this fast approximation. As expected, the coverage is essentially exact. In particular, 95\% coverage for this procedure is far better than the corresponding coverage based on the MLE.

%\subsection{Coverage for confidence sets via test inversion}
%
%Figure~\ref{fig:CS_coverage} shows the coverage probabilities for $95\%$ confidence sets based on the test inversion algorithm described in Section~\ref{sec: ri}. Note that the coverage of the grid bootstrap intervals are practically exact and dramatically outperform the corresponding ``naive'' confidence intervals for the MLE.

\subsection{Grid bootstrap for principal stratification model}

In the full principal stratification model, we directly estimate the outcome means for Compliers and Never Takers assigned to treatment, $\widehat{\mu}_{\text{c}1}$ and $\widehat{\mu}_{\text{n}1}$, and use the finite mixture model to estimate corresponding outcome means for Compliers and Never Takers assigned to control, $\widehat{\mu}_{\text{c}0}$ and $\widehat{\mu}_{\text{n}0}$. Our goal is inference for $\text{ITT}_\text{c} = \widehat{\mu}_{\text{c}1} - \widehat{\mu}_{\text{c}0}$ and $\text{ITT}_\text{n} = \widehat{\mu}_{\text{n}1} - \widehat{\mu}_{\text{n}0}$. While this is straightforward given estimates for $\mu_{\text{c}0}$ and $\mu_{\text{n}0}$, we only have confidence sets for these means.

We therefore propose the following approach to obtaining $(1-\alpha) 100\%$ confidence sets for $\text{ITT}_{\text{c}}$ and $\text{ITT}_{\text{n}}$:

\begin{itemize}
	\item Use a grid bootstrap or test inversion to obtain a joint $(1-\alpha/2) 100\%$ confidence set for $\mu_{\text{c}0}$ and $\mu_{\text{n}0}$, which we can project into univariate confidence sets, $\text{CS}_{\alpha/2}(\mu_{\text{c}0})$ and $\text{CS}_{\alpha/2}(\mu_{\text{n}0})$
	\item Directly obtain $(1-\alpha/2) 100\%$ confidence intervals via the Normal distribution for $\mu_{\text{c}1}$ and $\mu_{\text{n}1}$, $\text{CS}_{\alpha/2}(\mu_{\text{c}1})$ and $\text{CS}_{\alpha/2}(\mu_{\text{n}1})$
	\item For $\text{ITT}_{\text{c}}$ (repeat for $\text{ITT}_{\text{n}}$):
	\begin{itemize}
		\item If $\text{CS}_{\alpha/2}(\mu_{\text{c}0})$ is not disjoint, obtain a $(1-\alpha) 100\%$ confidence interval for $\text{ITT}_{\text{c}}$:
			\begin{align*}
			\text{CS}_{\alpha}^{UB}(\text{ITT}_{\text{c}}) &= \text{CS}_{\alpha/2}^{UB}(\mu_{\text{c}1}) - \text{CS}_{\alpha/2}^{LB}(\mu_{\text{c}0}) \\
			\text{CS}_{\alpha}^{LB}(\text{ITT}_{\text{c}}) &= \text{CS}_{\alpha/2}^{LB}(\mu_{\text{c}1}) - \text{CS}_{\alpha/2}^{UB}(\mu_{\text{c}0})
			\end{align*}
		\item If  $\text{CS}_{\alpha/2}(\mu_{\text{c}0})$ is disjoint, repeat the above calculations for each separate segment and then take the union
	\end{itemize}
\end{itemize}

This yields valid confidence sets for both treatment effects of interest. If desired, we could incorporate an additional Bonferroni correction to account for the two separate intervals.

Finally, if desired, we can extend this procedure to account for uncertainty in $\pi$ and $\sigma$, which are nuisance parameters for the desired hypothesis tests. We can therefore use results from~\cite{Berger:1994vh} to obtain valid $p$-values in this context. First, we obtain a $(1 - \gamma)$-level joint confidence set for $CS_{\gamma}(\pi, \sigma^2)$, such as via case-resampling bootstrap, with $\gamma$ very small, such as $\gamma = 0.001$. We obtain a valid $p$-value for, say, $\Delta$, by taking the maximum $p$-value over $CS_{\gamma}(\pi, \sigma^2)$ plus a correction for the added uncertainty:
\begin{eqnarray*}
p_\gamma(\Delta_0) = \sup_{(\pi, \sigma^2) \in CS_\gamma(\pi, \sigma^2)} p(\Delta_0) + \gamma.\label{eq::p-ci}
\end{eqnarray*} 
See~\cite{Nolen:2011fk} and~\cite{ding2016randomization} for further discussion of the validity of this approach.

\section{Failure of resampling methods}\label{sec:resampling}
\begin{figure}[hbtp]
	\centering
	\includegraphics[scale = 0.9]{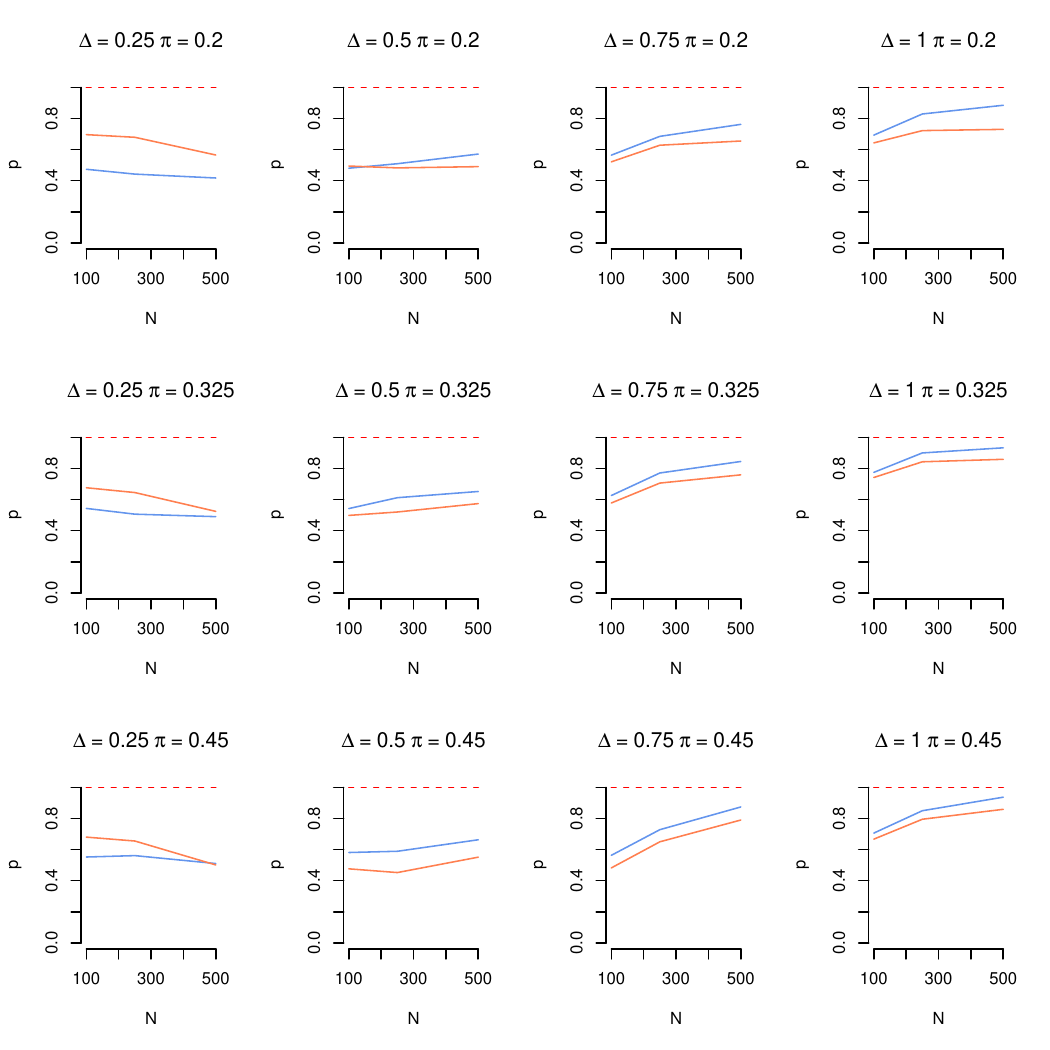}
	\caption{Coverage probabilities for 95\% confidence sets based on the case-resampling and subsampling intervals. The blue line represents the case-resampling coverage probability, while the blue line represents the subsampling coverage probability.}
	\label{fig: boot_sub}
\end{figure}

Resampling methods, such as the case-resampling bootstrap, are common in finite mixture model settings. For example,~\cite{mclachlan2004finite} recommend using the bootstrap to improve estimation of standard errors when the Fisher information yields a poor approximation~\cite{Grun:2004}. Others have suggested subsampling in similar settings~\cite{Andrews:2000}. Figure~\ref{fig: boot_sub} shows the coverage for 95\% confidence sets based on the case-resampling and subsampling intervals. Clearly, the coverage is far from nominal.

The form of $\widehat{\Delta}^{\text{mom}}$ shows why the performance of these methods is so poor. As the work~\cite{Bickel:1981} proved, for the bootstrap to be consistent in the iid context, the mapping from the underlying distribution of the data to the distribution of the statistic must be continuous~\cite{Andrews:2000}. Clearly, 
\begin{equation*}
\widehat{\Delta}^{\text{mom}} =  \text{sgn}(\widehat{m}_3) \sqrt{\frac{ \widehat{m}_2 - 1 }{\pi(1 - \pi)}}
\end{equation*}
is not a continuous mapping from the sample to $\widehat{\Delta}^{\text{mom}}$, with a boundary at $m_2 \geq 1$ and a discontinuity at $m_3 = 0$.\footnote{In some promising recent work,~\cite{Laber:2011jl} explore bootstrap-type methods with non-continuous mappings. We hope to explore this more in the future.} In the related case of the unit root problem,~\cite{Mikusheva:2007wl} shows that other resampling methods also fail, including subsampling and the $m$ of $n$ bootstrap. In the context of principal stratification,~\cite{Zhang:2009ku} note that confidence intervals based on the bootstrap often fail when the likelihood is multimodal.~\cite{Frumento:2016tk} offer additional discussion in this setting.

\section{Frequency Performance of the Posterior Mean and Median}\label{sec:bayes}

\begin{figure}[h]
\centering
		 \begin{subfigure}[b]{0.5\textwidth}
				\includegraphics[width = \textwidth]{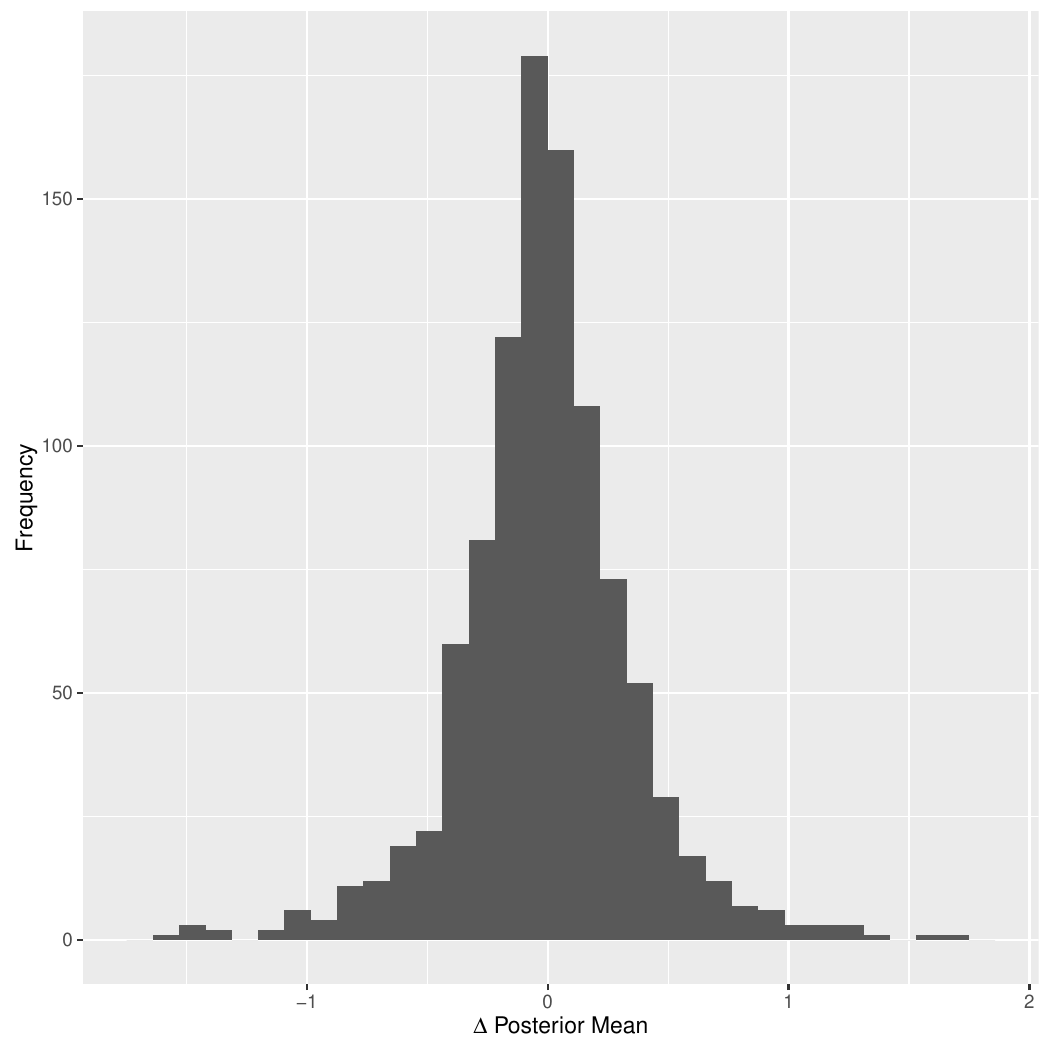}
				\caption{$\Delta = 0.5$}
				\label{fig: norm_approx_vary_N}
		\end{subfigure}%
		~\begin{subfigure}[b]{0.5\textwidth}
				\includegraphics[width = \textwidth]{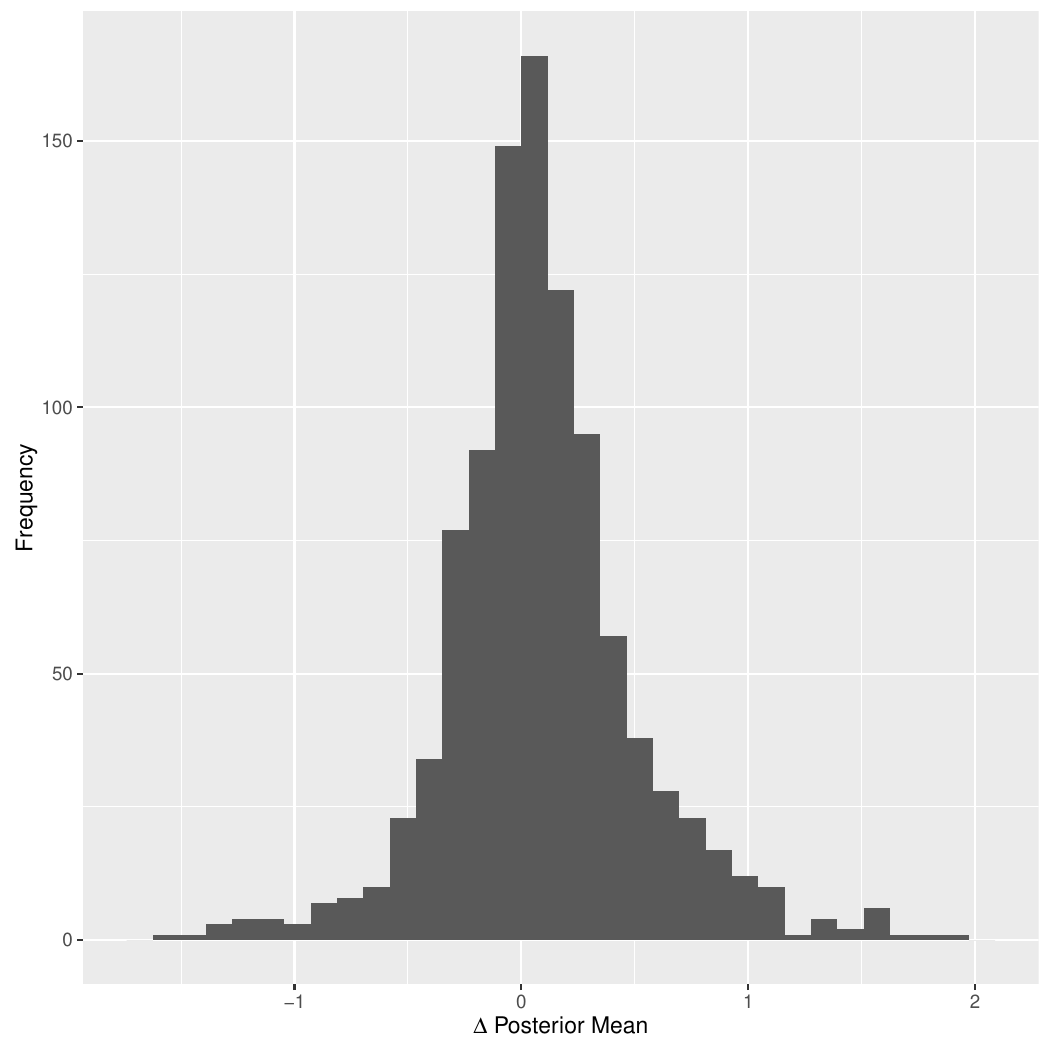} 
				\caption{$\Delta = 1$}
				\label{fig: norm_approx_vary_Delta}
		\end{subfigure}
\caption{Histograms of the posterior mean for $\Delta$ calculated via MCMC draws from \texttt{bayesm}. The histogram on the left
is for $\Delta = 0.5$, while the histogram on the right is for $\Delta = 1$. Both histograms have $N = 100$, $\pi = 0.3$,
and $\sigma = 1$. }
\label{fig: post_mean}
\end{figure}

Bayesian inference for finite mixtures introduces some unique challenges for specifying priors~\cite[e.g.,][]{Grazian:2015uf}. Nonetheless, inference for a posterior with a sufficiently vague prior should be broadly similar to inference based on the likelihood alone. Thus, without an informative
prior for $\{\mu_0,\mu_1\}$ in the two-component Gaussian mixture, the posterior mean and median should exhibit similar
pathologies to those exhibited by the MLE. We test this intuition using the \texttt{bayesm} package in R. Figure~\ref{fig: post_mean}
shows histograms of the posterior mean of $\Delta$ when the true $\Delta$ is $0.5$ and $1$, $\pi = 0.3$, and $N = 100$.
We use the default priors of the \texttt{bayesm} package except in the case of the Dirichlet parameter, which is set to reflect
that $\pi = 0.3$ is known (i.e., we assume a very informative prior). The histograms exhibit the same behavior as the MLE of $\Delta$. In particular, the estimator concentrates
around $0$ and seems unable to differentiate between $\Delta > 0$ and $\Delta < 0$. 

\begin{figure}[h]
\centering
		 \begin{subfigure}[b]{0.5\textwidth}
				\includegraphics[width = \textwidth]{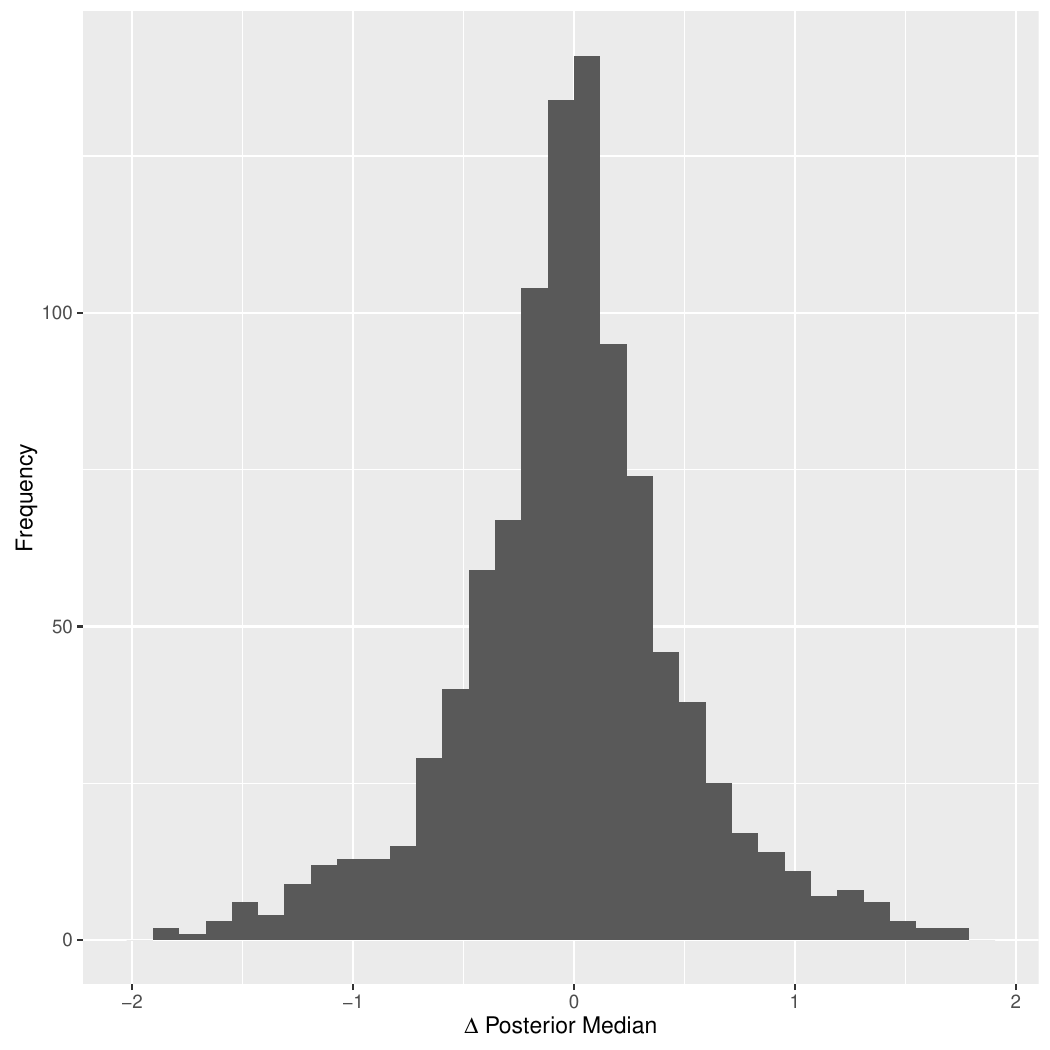}
				\caption{$\Delta = 0.5$}
				\label{fig: norm_approx_vary_N}
		\end{subfigure}%
		~\begin{subfigure}[b]{0.5\textwidth}
				\includegraphics[width = \textwidth]{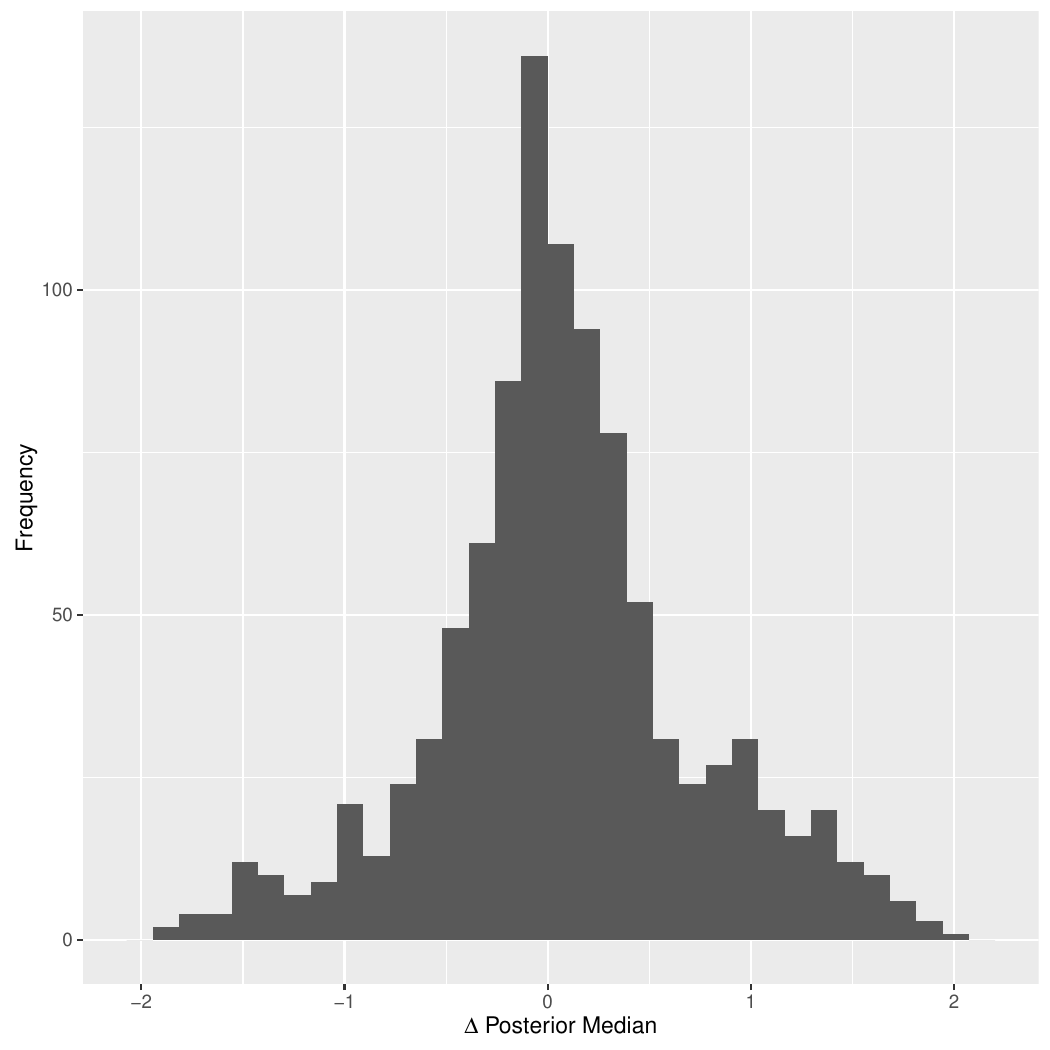} 
				\caption{$\Delta = 1$}
				\label{fig: norm_approx_vary_Delta}
		\end{subfigure}
\caption{Histograms of the posterior median for $\Delta$ calculated via MCMC draws from \texttt{bayesm}. The histogram on the left
is for $\Delta = 0.5$, while the histogram on the right is for $\Delta = 1$. Both histograms have $N = 100$, $\pi = 0.3$,
and $\sigma = 1$. }
\label{fig: post_median}
\end{figure}

Figure~\ref{fig: post_median} shows the corresponding plot for the distribution of the posterior median of $\Delta$.
As we can see, the median also concentrates about $0$ and appears unable to determine the sign of $\Delta$.

%%%
%%% PROOFS
%%%

\clearpage
\section{Proofs}
\label{sec:proof_append}
\setcounter{figure}{0}    
In this appendix, we provide detailed proofs for the key asymptotic results in Section \ref{sec:asymptotics}. We first start with the proof regarding convergence rates of $\widehat{\delta}_{n}^{\mle}$ and $\abss{\widehat{\delta}_{n}^{\mle}}$ under the asymmetric and symmetric setting of model \eqref{eq: block}. 
%%%%%%%%%%%%%%%%%%%%%%%%%%%%%%%%%%%%%%%%%%%
%%%%%%%%%%%%%
\subsection{PROOF OF THEOREM \ref{theorem:convergence_rate}}
\label{subsec:proof:theorem:convergence_rate}
%%%%%%%%%%%%%
%%%%%%%%%%%%%%%%%%%%%%%%%%%%%%%%%%%%%%%%%%%
Throughout this proof, for the ease of presentation, we denote 
\begin{align}
g(x,\delta) : = \pi\phi(x,-\delta) + (1-\pi)\phi(x,c\delta), \nonumber
\end{align}
for any $\delta  \in \Theta$ where $\left\{\phi(x,\delta)\right\}$ denotes the family of Gaussian distribution with location parameter $\delta$ and scale is fixed to be 1. Additionally, we also remind that $c = \pi/(1-\pi)$, with this quantity thus being a known constant.
To streamline the argument, we divide the proof into two parts. In Section~\ref{subsub:upper_bound_know_variance}, we provide the proof for the upper bounds of the convergence rate of MLE. Then, in Section~\ref{subsub:lower_bound_know_variance}, we present the proof for the lower bounds.

\subsubsection{Proof for upper bounds}
\label{subsub:upper_bound_know_variance}
The proof technique for the upper bounds utilizes the strategy of comparing the convergence rate of density estimation to that of parameter estimation in mixture models, which had been employed successfully in the previous work \cite{Chen, Nguyen-13, Ho-Nguyen-Ann-16, Jonas-2017}. 
\paragraph{Convergence rate of density estimation} The convergence rate of density estimation in Gaussian mixture models had been studied rigorously in the literature \cite{Ghosal-2001}. Regarding our model \eqref{eq: block}, we have the following result regarding the convergence rate of $g(x,\widehat{\delta}_{n}^{\mle})$ to $g(x,\delta_{n})$ under Hellinger metric.
\begin{proposition} \label{proposition:Hellinger_convergence_density_estimation}
Under the setting of model \eqref{eq: block}, the following holds
\begin{align}
\sup \limits_{\delta_{n} \in \Theta} \mathbb{E}_{\delta_{n}} \left(h\parenth{g(x,\widehat{\delta}_{n}^{\mle}),g(x,\delta_{n})}\right) \lesssim \left(\frac{\log n}{n} \right)^{1/2}, \nonumber
\end{align}
where $\Theta$ is a bounded (growing) parameter space. Here, $\mathbb{E}_{\delta_{n}}$ denotes the expectation taken with respect to product measure with mixture density of $Y_{1},\ldots,Y_{n}$ under the model~\eqref{eq: block}. 
\end{proposition} 
The proof of the above result follows from a standard application of Theorem 7.4 in \cite{Vandegeer-2000}; therefore, it is omitted.
\paragraph{From density estimation to parameter estimation}
Equipped with $(\log n/n)^{1/2}$ rate of density estimation in Proposition \eqref{proposition:Hellinger_convergence_density_estimation}, to achieve the convergence rates of $\widehat{\delta}_{n}^{\mle}$ and $\abss{\widehat{\delta}_{n}^{\mle}}$ under the asymmetric and symmetric setting of model \eqref{eq: block}, it is sufficient to demonstrate the following result:
\begin{lemma} \label{lemma:density_to_parameter}
Given $\pi \in (0,1/2]$ and $\Theta = [-1, 1]$, the following holds
\begin{itemize}
\item[(a)] (Asymmetric regime) When $\pi \in (0,1/2)$, then 
\begin{eqnarray}
\inf \limits_{\delta^{(1)},\delta^{(2)} \in \Theta} h\parenth{g(x,\delta^{(1)}),g(x,\delta^{(2)})}/\abss{\delta^{(1)} - \delta^{(2)}}^{3} > 0. \nonumber
\end{eqnarray}
\item[(b)] (Symmetric regime) When $\pi = 1/2$, then 
\begin{eqnarray}
\inf \limits_{\delta^{(1)},\delta^{(2)} \in \Theta} h\parenth{g(x,\delta^{(1)}),g(x,\delta^{(2)})}/\biggr||\delta^{(1)}| - |\delta^{(2)}|\biggr|^{2} > 0. \nonumber
\end{eqnarray}
\end{itemize}
\end{lemma}
\begin{proof} (a) Due to the basic inequality between total variational distance and Hellinger distance $h \geq V$, it suffices to prove that
\begin{eqnarray}
\inf \limits_{\delta^{(1)},\delta^{(2)} \in \Theta} V\parenth{g(x,\delta^{(1)}),g(x,\delta^{(2)})} /|\delta^{(1)} - \delta^{(2)}|^{3} > 0. \label{eqn:inequality_one}
\end{eqnarray}
Assume that the conclusion of \eqref{eqn:inequality_one} does not hold. It implies that we can find two sequences $\left\{\delta_{n}^{(1)}\right\}$ and $\left\{\delta_{n}^{(2)}\right\}$ such that $V(g(x,\delta_{n}^{(1)}),g(x,\delta_{n}^{(2)})) /|\delta_{n}^{(1)} - \delta_{n}^{(2)}|^{3} \to 0$ as $n \to \infty$. For the simplicity of the presentation, we only the consider the most challenging setting of sequences $\left\{\delta_{n}^{(1)}\right\}$ and $\left\{\delta_{n}^{(2)}\right\}$ when $\delta_{n}^{(1)} \to 0$, $\delta_{n}^{(2)} \to 0$ as $n \to \infty$. The proof for other possibilities of these sequences can be argued in the similar fashion. Now, we have two distinct cases regarding the convergence of $\delta_{n}^{(1)}$ and $\delta_{n}^{(2)}$.
\paragraph{Case a.1:} $\delta_{n}^{(1)}/\delta_{n}^{(2)} \not \to 1$ as $n \to \infty$ (Here, the limit can be thought as that of some subsequence of $\delta_{n}^{(1)}/\delta_{n}^{(2)}$. However, we replace this subsequence by the whole sequence of $\delta_{n}^{(1)}/\delta_{n}^{(2)}$ for the simplicity of the presentation). Under this case, we divide our argument into several steps.
\paragraph{Step 1 - Taylor expansion} Now, the following equality holds
\begin{eqnarray}
\dfrac{g(x,\delta_{n}^{(1)}) - g(x, \delta_{n}^{(2)})}{|\delta_{n}^{(1)} - \delta_{n}^{(2)}|^{3}} & = & \dfrac{\pi(\phi(x,-\delta_{n}^{(1)}) - \phi(x,-\delta_{n}^{(2)}))}{|\delta_{n}^{(1)} - \delta_{n}^{(2)}|^{3}} \nonumber \\
& & \hspace{8 em} +\dfrac{(1-\pi)(\phi(x,c\delta_{n}^{(1)}) - \phi(x,c\delta_{n}^{(2)}))}{|\delta_{n}^{(1)} - \delta_{n}^{(2)}|^{3}}. \nonumber 
\end{eqnarray}
Invoking Taylor expansion up to the third order, we obtain that
\begin{eqnarray}
\phi(x,-\delta_{n}^{(1)}) - \phi(x,-\delta_{n}^{(2)}) & = & \sum \limits_{\alpha=1}^{3} \dfrac{(\delta_{n}^{(2)} - \delta_{n}^{(1)})^{\alpha}}{\alpha!}\dfrac{\partial^{\alpha}{\phi}}{\partial{\delta^{\alpha}}}(x,-\delta_{n}^{(2)})+ R_{1}(x), \nonumber \\
\phi(x,c\delta_{n}^{(1)}) - \phi(x,c\delta_{n}^{(2)}) & = &  \sum \limits_{\alpha=1}^{3} \dfrac{c^{\alpha}(\delta_{n}^{(1)} - \delta_{n}^{(2)})^{\alpha}}{\alpha!}\dfrac{\partial^{\alpha}{\phi}}{\partial{\delta^{\alpha}}}(x,c\delta_{n}^{(2)})+R_{2}(x) \nonumber \\
& = & \sum \limits_{\alpha=1}^{3} \dfrac{c^{\alpha}(\delta_{n}^{(1)} - \delta_{n}^{(2)})^{\alpha}}{\alpha!}\biggr(\sum \limits_{\tau = 0}^{3-\alpha} \dfrac{(c+1)^{\tau}(\delta_{n}^{(2)})^{\tau}}{\tau!} \dfrac{\partial^{\alpha+\tau}{\phi}}{\partial{\delta^{\alpha+\tau}}}(x,-\delta_{n}^{(2)}) \nonumber \\
& & \hspace{16 em} +R_{2,\alpha}(x) \biggr)+R_{2}(x), \nonumber
\end{eqnarray}
where $R_{1}(x)$, $R_{2}(x)$ are respectively the Taylor remainders up to the third order from performing Taylor expansion around $-\delta_{n}^{(2)}$ and $c\delta_{n}^{(2)}$ while $R_{2,\alpha}$ are Taylor remainders up to the order $3-\alpha$ from performing Taylor expansion around $-\delta_{n}^{2}$ in $\dfrac{\partial^{\alpha}{\phi}}{\partial{\delta^{\alpha}}}(x,c\delta_{n}^{(2)})$ as $1 \leq \alpha \leq 3$. Here, the Taylor remainders $R_{1}(x)$ and $R_{2}(x)$ satisfy 
\begin{align}
\max \{\|R_{1}(x)\|_{\infty}, \|R_{2}(x)\|_{\infty}\} = O\parenth{|\delta_{n}^{(1)} - \delta_{n}^{(2)}|^{3+\gamma}}, \label{eqn:Taylor_first_bound}
\end{align} 
where $\gamma>0$ is some positive constant. It implies that $R_{1}(x)/ |\delta_{n}^{(1)} - \delta_{n}^{(2)}|^{3} \to 0$ and $R_{2}(x) / |\delta_{n}^{(1)} - \delta_{n}^{(2)}|^{3} \to 0$ for all $x \in \mathbb{R}$. Similarly, $\|R_{2,\alpha}(x)\|_{\infty} = O(|\delta_{n}^{(2)}|^{3-\alpha+\gamma})$  as $1 \leq \alpha \leq 3$. As $\delta_{n}^{(1)}/\delta_{n}^{(2)} \not \to 1$, we have $|\delta_{n}^{(2)}|/|\delta_{n}^{(1)} - \delta_{n}^{(2)}| \not \to + \infty$. Therefore, we have $|\delta_{n}^{(2)}|^{r-\alpha+\gamma}/|\delta_{n}^{(1)} - \delta_{n}^{(1)}|^{r-\alpha} \to 0$ as $n \to \infty$, which eventually leads to 
\begin{align}
(\delta_{n}^{(1)} - \delta_{n}^{(2)})^{\alpha}\|R_{2,\alpha}(x)\|_{\infty}/|\delta_{n}^{(1)} - \delta_{n}^{(2)}|^{3} \to 0 \label{eqn:Taylor_second_bound}
\end{align} 
for all $1 \leq \alpha \leq 3$. Governed by the previous results, the following representation holds
\begin{eqnarray}
\dfrac{g(x,\delta_{n}^{(1)}) - g(x, \delta_{n}^{(2)})}{|\delta_{n}^{(1)} - \delta_{n}^{(2)}|^{3}} & = & \dfrac{\pi\biggr(\sum \limits_{\alpha=1}^{3} \dfrac{(\delta_{n}^{(2)} - \delta_{n}^{(1)})^{\alpha}}{\alpha!}\dfrac{\partial^{\alpha}{\phi}}{\partial{\delta^{\alpha}}}(x,-\delta_{n}^{(2)})+R_{1}(x)\biggr)}{|\delta_{n}^{(1)} - \delta_{n}^{(2)}|^{3}} \nonumber \\
& & \hspace{ - 10 em} + \dfrac{(1-\pi)\biggr(\sum \limits_{\alpha=1}^{3} \dfrac{c^{\alpha}(\delta_{n}^{(1)} - \delta_{n}^{(2)})^{\alpha}}{\alpha!}\biggr(\sum \limits_{\tau = 0}^{3-\alpha} \dfrac{(c+1)^{\tau}(\delta_{n}^{(2)})^{\tau}}{\tau!} \dfrac{\partial^{\alpha+\tau}{\phi}}{\partial{\delta^{\alpha+\tau}}}(x,-\delta_{n}^{(2)})+R_{2,\alpha}(x) \biggr)+R_{2}(x)\biggr)}{|\delta_{n}^{(1)} - \delta_{n}^{(2)}|^{3}} \nonumber \\
& := & \dfrac{\sum \limits_{\alpha=1}^{3} A_{n,\alpha}\dfrac{\partial^{\alpha}{\phi}}{\partial{\delta^{\alpha}}}(x,-\delta_{n}^{(2)})+ R(x)}{|\delta_{n}^{(1)} - \delta_{n}^{(2)}|^{3}}, \label{eqn:Taylor_representation}
\end{eqnarray}
where $R(x) = \pi R_{1}(x) + (1-\pi)\sum \limits_{\alpha=1}^{3} \dfrac{c^{\alpha}(\delta_{n}^{(1)} - \delta_{n}^{(2)})^{\alpha}}{\alpha!}R_{2,\alpha}(x)+ (1-\pi)R_{2}(x)$ for all $x \in \mathbb{R}$.  Invoking the bounds with Taylor remainders $R_{1}(x), R_{2}(x)$, and $R_{2,\alpha}(x)$ in \eqref{eqn:Taylor_first_bound}, \eqref{eqn:Taylor_second_bound}, we have $\|R(x)\|_{\infty} / |\delta_{n}^{(1)} - \delta_{n}^{(2)}|^{3} \to 0$ as $n \to \infty$. 
\paragraph{Step 2 - Non-vanishing coefficients} Assume that the coefficients $A_{n,\alpha}/|\delta_{n}^{(1)} - \delta_{n}^{(2)}|^{3} \to 0$ as $n \to \infty$ for all $1 \leq \alpha \leq 3$. From the formulations of $A_{n,\alpha}$ in \eqref{eqn:Taylor_representation}, we can quickly compute that $A_{n,1} = 0$ while
\begin{eqnarray}
& & \hspace{ - 6 em} A_{n,2} = \dfrac{c}{2}(\delta_{n}^{(2)} - \delta_{n}^{(1)})(\delta_{n}^{(1)}+\delta_{n}^{(2)}), \nonumber \\
& & \hspace{ - 6 em} A_{n,3} = \dfrac{\pi(\delta_{n}^{(2)} - \delta_{n}^{(1)})^{3}}{3!} + (1-\pi)c(c+1)^{2}(\delta_{n}^{(2)})^{2}\dfrac{(\delta_{n}^{(1)} - \delta_{n}^{(2)})}{2!} \nonumber \\
& & + (1-\pi)(c+1)c^{2}\dfrac{(\delta_{n}^{(1)} - \delta_{n}^{(2)})^{2}}{2!}\delta_{n}^{(2)} + (1-\pi)c^{3}\dfrac{(\delta_{n}^{(1)} - \delta_{n}^{(2)})^{3}}{3!}. \nonumber
\end{eqnarray}
As $A_{n,2}/|\delta_{n}^{(1)} - \delta_{n}^{(2)}|^{3} \to 0$, it implies that $(\delta_{n}^{(1)} + \delta_{n}^{(2)}) / |\delta_{n}^{(1)} - \delta_{n}^{(2)}|^{2} \to 0$, which leads to $\delta_{n}^{(1)}/\delta_{n}^{(2)} \to -1$ as $n \to \infty$. Plugging this limit into $A_{n,3}/|\delta_{n}^{(1)} - \delta_{n}^{(2)}|^{3} \to 0$ yields the following equation
\begin{eqnarray}
\dfrac{8\pi}{3!} - (1-\pi)c(c+1)^{2} + 2(1-\pi)c^{2}(c+1) - \dfrac{8(1-\pi)c^{3}}{3!} = 0, \nonumber
\end{eqnarray}
which has only a unique solution $\pi =1/2$, a contradiction to the assumption of asymmetric setting, i.e., $\pi \in (0,1/2)$. Therefore, not all the coefficients $A_{n,\alpha}/|\delta_{n}^{(1)} - \delta_{n}^{(2)}|^{3} \to 0$ when $n \to \infty$ as $1 \leq \alpha \leq 3$.
\paragraph{Step 3 - Fatou's argument} Denote $m_{n} = |\delta_{n}^{(1)} - \delta_{n}^{(2)}|^{3}/\max \limits_{1 \leq \alpha \leq 3} |A_{n,\alpha}|$. Since not all the coefficients $A_{n,\alpha}/|\delta_{n}^{(1)} - \delta_{n}^{(2)}|^{3} \to 0$ as $1 \leq \alpha \leq 3$, we have $m_{n} \not \to \infty$. Therefore, we obtain that
\begin{eqnarray}
m_{n} \dfrac{g(x,\delta_{n}^{(1)}) - g(x, \delta_{n}^{(2)})}{|\delta_{n}^{(1)} - \delta_{n}^{(2)}|^{3}} = m_{n} \dfrac{\sum \limits_{\alpha=1}^{3} A_{n,\alpha}\dfrac{\partial^{\alpha}{\phi}}{\partial{\delta^{\alpha}}}(x,-\delta_{n}^{(2)})+ R(x)}{|\delta_{n}^{(1)} - \delta_{n}^{(2)}|^{3}} \to \sum \limits_{\alpha = 1}^{3} \beta_{\alpha}\dfrac{\partial^{\alpha}{\phi}}{\partial{\delta^{\alpha}}}(x,0), \nonumber
\end{eqnarray}
for all $x$ where $A_{n,\alpha}/\max \limits_{1 \leq \alpha \leq 3} |A_{n,\alpha}| \to \beta_{\alpha}$ as $1 \leq \alpha \leq 3$ such that at least one of $\beta_{\alpha}$ has absolute value to be 1. Invoking Fatou's lemma, the following holds
\begin{align}
0 = \lim \limits_{n \to \infty} \dfrac{m_{n}V\parenth{g(x,\delta_{n}^{(1)}),g(x,\delta_{n}^{(2)})}}{|\delta^{(1)} - \delta^{(2)}|^{3}} & \geq \int \mathop {\lim \inf}\limits_{n \to \infty} \dfrac{m_{n}\abss{g(x,\delta_{n}^{(1)}) - g(x, \delta_{n}^{(2)})}}{|\delta_{n}^{(1)} - \delta_{n}^{(2)}|^{3}}dx \nonumber \\
& = \int \sum \limits_{\alpha = 1}^{3} \beta_{\alpha}\dfrac{\partial^{\alpha}{\phi}}{\partial{\delta^{\alpha}}}(x,0) dx. \nonumber
\end{align}
The above inequality leads to $\sum \limits_{\alpha = 1}^{3} \beta_{\alpha}\dfrac{\partial^{\alpha}{\phi}}{\partial{\delta^{\alpha}}}(x,0) = 0$ for almost surely $x$. Nevertheless, due to the strong order identifiability of location Gaussian distribution \cite{Chen}, the above equation implies that $\beta_{\alpha} = 0$ for all $1 \leq \alpha \leq 3$, which is a contradiction. Therefore, Case a.1 cannot holds.
\paragraph{Case a.2:}  $\delta_{n}^{(1)}/\delta_{n}^{(2)} \to 1$ as $n \to \infty$. It implies that $|\delta_{n}^{(2)}|/|\delta_{n}^{(1)} - \delta_{n}^{(2)}| \to \infty$ as $n \to \infty$. As $V\parenth{g(x,\delta_{n}^{(1)}),g(x, \delta_{n}^{(2)})}/|\delta_{n}^{(1)} - \delta_{n}^{(2)}|^{3} \to 0$, it implies that 
\begin{align}
V\parenth{g(x,\delta_{n}^{(1)}),g(x, \delta_{n}^{(2)})}/|\delta_{n}^{(1)} - \delta_{n}^{(2)}|^{2} \to 0, \nonumber
\end{align}
as $n \to \infty$ for all $x \in \mathbb{R}$. Similar to the Taylor expansion argument in Step 1 in Case a.1, by means of Taylor expansion up to the second order, we obtain that
\begin{eqnarray}
\dfrac{g(x,\delta_{n}^{(1)}) - g(x, \delta_{n}^{(2)})}{|\delta_{n}^{(1)} - \delta_{n}^{(2)}|^{2}} & = & \dfrac{\pi(\phi(x,-\delta_{n}^{(1)}) - \phi(x,-\delta_{n}^{(2)}))+(1-\pi)(\phi(x,c\delta_{n}^{(1)}) - \phi(x,c\delta_{n}^{(2)}))}{|\delta_{n}^{(1)} - \delta_{n}^{(2)}|^{2}} \nonumber \\ 
& & \hspace{ - 9 em} = \dfrac{\pi\biggr(\sum \limits_{\alpha=1}^{2} \dfrac{(\delta_{n}^{(2)} - \delta_{n}^{(1)})^{\alpha}}{\alpha!}\dfrac{\partial^{\alpha}{\phi}}{\partial{\delta^{\alpha}}}(x,-\delta_{n}^{(2)})+R_{1}'(x)\biggr)}{|\delta_{n}^{(1)} - \delta_{n}^{(2)}|^{2}} \nonumber \\
& & \hspace{ - 9 em} + \dfrac{(1-\pi)\biggr(\sum \limits_{\alpha=1}^{2} \dfrac{c^{\alpha}(\delta_{n}^{(1)} - \delta_{n}^{(2)})^{\alpha}}{\alpha!}\biggr(\sum \limits_{\tau = 0}^{2-\alpha} \dfrac{(c+1)^{\tau}(\delta_{n}^{(2)})^{\tau}}{\tau!} \dfrac{\partial^{\alpha+\tau}{\phi}}{\partial{\delta^{\alpha+\tau}}}(x,-\delta_{n}^{(2)})+R_{2,\alpha}'(x) \biggr)+R_{2}'(x)\biggr)}{|\delta_{n}^{(1)} - \delta_{n}^{(2)}|^{2}} \nonumber \\
& & \hspace{ - 9 em} = \dfrac{\sum \limits_{\alpha=1}^{2} A_{n,\alpha}\dfrac{\partial^{\alpha}{\phi}}{\partial{\delta^{\alpha}}}(x,-\delta_{n}^{(2)})+ R'(x)}{|\delta_{n}^{(1)} - \delta_{n}^{(2)}|^{2}} \to 0, \nonumber
\end{eqnarray}
where $\|R'(x)\|_{\infty} = O\parenth{|\delta_{n}^{(2)}|^{1+\gamma}|\delta_{n}^{(1)} - \delta_{n}^{(2)}|}$ for some $\gamma>0$. By means of the calculations with $A_{n,\alpha}$ in Case a.1, we have 
\begin{align}
\dfrac{\|R'(x)\|_{\infty}}{|A_{n,2}|} = \dfrac{O\parenth{\abss{\delta_{n}^{(2)}}^{1+\gamma}\abss{\delta_{n}^{(1)} - \delta_{n}^{(2)}}}}{\abss{\delta_{n}^{(2)} - \delta_{n}^{(1)}}\abss{\delta_{n}^{(1)}+\delta_{n}^{(2)}}} \to 0. \nonumber
\end{align}
Now, if $A_{n,\alpha}/|\delta_{n}^{(1)} - \delta_{n}^{(2)}|^{2} \to 0$ for all $1 \leq \alpha \leq 2$, we have $|\delta_{n}^{(1)}+\delta_{n}^{(2)}|/|\delta_{n}^{(1)} - \delta_{n}^{(2)}| \to 0$, which  implies that $\delta_{n}^{(1)}/\delta_{n}^{(2)} \to -1$, a contradiction to the assumption of Case a.2. According to the argument in Step 3 in Case a.1, by denoting $m_{n}' = |\delta_{n}^{(1)} - \delta_{n}^{(2)}|^{2}/\max \limits_{1 \leq \alpha \leq 2} |A_{n,\alpha}|$, we have $m_{n}' \not \to \infty$. Therefore, we have
\begin{eqnarray}
m_{n}'\dfrac{g(x,\delta_{n}^{(1)}) - g(x, \delta_{n}^{(2)})}{|\delta_{n}^{(1)} - \delta_{n}^{(2)}|^{2}} \to \sum \limits_{\alpha=1}^{2} \tau_{\alpha}\dfrac{\partial^{\alpha}{\phi}}{\partial{\delta^{\alpha}}}(x,0), \nonumber
\end{eqnarray}
for all $x$ for some coefficients $\tau_{\alpha}$ such that at least one of them has absolute value to be 1. By virtue of Fatou's lemma in Step 3 in Case a.1 with $\lim \limits_{n \to \infty} V\parenth{g(x,\delta_{n}^{(1)}),g(x, \delta_{n}^{(2)})}/|\delta_{n}^{(1)} - \delta_{n}^{(2)}|^{2}$, we achieve that $\sum \limits_{\alpha=1}^{2} \tau_{\alpha}\dfrac{\partial^{\alpha}{\phi}}{\partial{\delta^{\alpha}}}(x,0)=0$ for almost surely $x$. However, the strong identifability of location Gaussian distribution implies that $\tau_{\alpha} = 0$ for all $1 \leq \alpha \leq 2$, which is a contradiction. Therefore, Case a.2 cannot happen. 

Combining the results from Case a.1 and Case a.2, we achieve the conclusion of \eqref{eqn:inequality_one}. As a consequence, the conclusion of part (a) of Lemma \ref{lemma:density_to_parameter} follows.

(b) Similar to the proof strategy of part (a), to obtain the conclusion of this result, it is sufficient to demonstrate that
\begin{eqnarray}
\inf \limits_{\delta^{(1)},\delta^{(2)} \in \delta} V\parenth{g(x,\delta^{(1)}),g(x,\delta^{(2)})}/\biggr||\delta^{(1)}| - |\delta^{(2)}|\biggr|^{2} > 0. \label{eqn:inequality_second}
\end{eqnarray}
Assume that the conclusion of \eqref{eqn:inequality_second} does not hold. It implies that we can find two sequences $\left\{\delta_{n}^{(1)}\right\}$ and $\left\{\delta_{n}^{(2)}\right\}$ such that 
\begin{eqnarray}
V\parenth{g(x,\delta_{n}^{(1)}),g(x,\delta_{n}^{(2)})}/\biggr||\delta_{n}^{(1)}| - |\delta_{n}^{(2)}|\biggr|^{2}\to 0 \nonumber
\end{eqnarray}
as $n \to \infty$. Similar to the proof argument of part (a), we only consider the possibility that $\delta_{n}^{(1)} \to 0$ and $\delta_{n}^{(2)} \to 0$ as $n \to \infty$. Now, we have two different settings of $\delta_{n}^{(1)}$ and $\delta_{n}^{(2)}$.
\paragraph{Case b.1:} $\delta_{n}^{(1)}/\delta_{n}^{(2)} \not \to 1$ as $n \to \infty$ and $\delta_{n}^{(1)}\delta_{n}^{(2)} \geq 0$ for all $n$ (Here, the limit and the inequality can be thought as those of some subsequence of $\delta_{n}^{(1)}$ and $\delta_{n}^{(2)}$. However, we replace this subsequence by the whole sequence of $\delta_{n}^{(1)}$ and $\delta_{n}^{(2)}$ for the simplicity of the presentation). Under that setting, we have
\begin{eqnarray}
\dfrac{V\parenth{g(x,\delta_{n}^{(1)}),g(x,\delta_{n}^{(2)})}}{\biggr||\delta_{n}^{(1)}| - |\delta_{n}^{(2)}|\biggr|^{2}} = \dfrac{V\parenth{g(x,\delta_{n}^{(1)}),g(x,\delta_{n}^{(2)})}}{|\delta_{n}^{(1)} - \delta_{n}^{(2)}|^{2}} \to 0. \nonumber
\end{eqnarray}
To ease the understanding, we divide our argument for Case b.1 into two separate steps.
\paragraph{Step 1 - Taylor expansion} By means of Taylor expansion up to the second order as that of Case a.2 in the proof of part (a), we obtain that
\begin{eqnarray}
\dfrac{g(x,\delta_{n}^{(1)}) - g(x,\delta_{n}^{(2)})}{|\delta_{n}^{(1)} - \delta_{n}^{(2)}|^{2}} = \dfrac{\sum \limits_{\alpha=1}^{2} A_{n,\alpha}\dfrac{\partial^{\alpha}{\phi}}{\partial{\delta^{\alpha}}}(x,-\delta_{n}^{(2)})+ R'(x)}{\abss{\delta_{n}^{(1)} - \delta_{n}^{(2)}}^{2}} \to 0, \nonumber
\end{eqnarray}
where $R'(x)$ is a combination of Taylor remainders such that 
\begin{align}
\|R'(x)\|_{\infty} = O\parenth{\abss{\delta_{n}^{(2)}}^{1+\gamma}\abss{\delta_{n}^{(1)} - \delta_{n}^{(2)}}}, \nonumber
\end{align}
for some positive constant $\gamma$ and $A_{n,\alpha}$ are defined as in that in Case a.2 when $\pi=1/2$. Since $\delta_{n}^{(1)}/\delta_{n}^{(2)} \not \to 1$, we have $|\delta_{n}^{(2)}|/|\delta_{n}^{(1)} - \delta_{n}^{(2)}| \not \to \infty$. Therefore, it leads to
\begin{align}
\|R(x)\|_{\infty}/|\delta_{n}^{(1)} - \delta_{n}^{(2)}| \to 0 \nonumber
\end{align}
as $n \to \infty$. 
\paragraph{Step 2 - Non-vanishing coefficients and Fatou's argument} Assume that $A_{n,\alpha}/|\delta_{n}^{(1)} - \delta_{n}^{(2)}|^{2} \to 0$ for all $1 \leq \alpha \leq 2$. From the formulation of $A_{n,2}$, we have 
\begin{align}
(\delta_{n}^{(1)}+\delta_{n}^{(2)})/|\delta_{n}^{(1)} - \delta_{n}^{(2)}| \to 0. \nonumber
\end{align} 
It implies that $\delta_{n}^{(1)}/\delta_{n}^{(2)} \to -1$ as $n \to \infty$, which is a contradiction to the condition that $\delta_{n}^{(1)}\delta_{n}^{(2)} \geq 0$. Therefore, not all of the coefficients of $A_{n,\alpha}/|\delta_{n}^{(1)} - \delta_{n}^{(2)}|^{2}$ go to 0. From here, by means of the Fatou's argument in Step 3 of Case a.1, we achieve the conclusion that Case b.1 cannot hold. 
\paragraph{Case b.2} $\delta_{n}^{(1)}/\delta_{n}^{(2)} \not \to 1$ and $\delta_{n}^{(1)}\delta_{n}^{(2)} <0$ for all $n$. Under that setting, we have
\begin{eqnarray}
\dfrac{V\parenth{g(x,\delta_{n}^{(1)}),g(x,\delta_{n}^{(2)})}}{\biggr||\delta_{n}^{(1)}| - |\delta_{n}^{(2)}|\biggr|^{2}} = \dfrac{V\parenth{g(x,\delta_{n}^{(1)}),g(x,\delta_{n}^{(2)})}}{\abss{\delta_{n}^{(1)} + \delta_{n}^{(2)}}^{2}} \to 0. \nonumber
\end{eqnarray}
We also divide the argument of Case b.2 into two main key steps.
\paragraph{Step 1 - Taylor expansion} By means of Taylor expansion up to the second order, we obtain
\begin{eqnarray}
\dfrac{g(x,\delta_{n}^{(1)}) - g(x,\delta_{n}^{(2)})}{|\delta_{n}^{(1)} + \delta_{n}^{(2)}|^{2}} & = & \dfrac{\dfrac{1}{2}(\phi(x,-\delta_{n}^{(1)}) - \phi(x,\delta_{n}^{(2)}))+\dfrac{1}{2}(\phi(x,\delta_{n}^{(1)}) - \phi(x,-\delta_{n}^{(2)}))}{|\delta_{n}^{(1)} + \delta_{n}^{(2)}|^{2}} \nonumber \\ 
& & \hspace{ - 8 em} = \dfrac{\dfrac{1}{2}\biggr(\sum \limits_{\alpha=1}^{2} \dfrac{(-\delta_{n}^{(2)} - \delta_{n}^{(1)})^{\alpha}}{\alpha!}\dfrac{\partial^{\alpha}{\phi}}{\partial{\delta^{\alpha}}}(x,\delta_{n}^{(2)})+R_{1}''(x)\biggr)}{|\delta_{n}^{(1)} + \delta_{n}^{(2)}|^{2}} \nonumber \\
& & \hspace{ - 8 em} + \dfrac{\dfrac{1}{2}\biggr(\sum \limits_{\alpha=1}^{2} \dfrac{(\delta_{n}^{(1)} + \delta_{n}^{(2)})^{\alpha}}{\alpha!}\biggr(\sum \limits_{\tau = 0}^{2-\alpha} \dfrac{2^{\tau}(-\delta_{n}^{(2)})^{\tau}}{\tau!} \dfrac{\partial^{\alpha+\tau}{\phi}}{\partial{\delta^{\alpha+\tau}}}(x,\delta_{n}^{(2)})+R_{2,\alpha}''(x) \biggr)+R_{2}''(x)\biggr)}{|\delta_{n}^{(1)} + \delta_{n}^{(2)}|^{2}} \nonumber \\
& & \hspace{ - 8 em} : = \dfrac{\sum \limits_{\alpha=1}^{2} B_{n,\alpha}\dfrac{\partial^{\alpha}{\phi}}{\partial{\delta^{\alpha}}}(x,\delta_{n}^{(2)})+ R''(x)}{|\delta_{n}^{(1)} + \delta_{n}^{(2)}|^{2}} \to 0, \nonumber
\end{eqnarray}
where $R''(x)$ is the combination of Taylor remainders such that 
\begin{align}
\|R''(x)\|_{\infty} = O\parenth{|\delta_{n}^{(2)}|^{1+\gamma}|\delta_{n}^{(1)} + \delta_{n}^{(2)}|}, \nonumber
\end{align}
which implies that $\|R''(x)\|_{\infty}/|\delta_{n}^{(1)} + \delta_{n}^{(2)}|^{2} \to 0$ as $n \to \infty$. 
\paragraph{Step 2 - Non-vanishing coefficients and Fatou's argument} Assume that $B_{n,\alpha}/|\delta_{n}^{(1)}+\delta_{n}^{(2)}|^{2} \to 0$ for all $1 \leq \alpha \leq 2$. Direct computation with $B_{n,2}$ implies that 
\begin{align}
(\delta_{n}^{(1)}-\delta_{n}^{(2)})/|\delta_{n}^{(1)}+\delta_{n}^{(2)}| \to 0 \nonumber
\end{align} 
as $n \to \infty$. It leads to $\delta_{n}^{(1)}/\delta_{n}^{(2)} \to 1$, which is a contradiction to the assumption that $\delta_{n}^{(1)}\delta_{n}^{(2)}<0$. From here, the Fatou's argument in Step 3 of Case a.1, we also obtain the conclusion that Case b.2 does not hold.
\paragraph{Case b.3} $\delta_{n}^{(1)}/\delta_{n}^{(2)} \to 1$ as $n \to \infty$. This implies that $\delta_{n}^{(1)}\delta_{n}^{(2)} > 0$ when $n$ is sufficiently large. From here, the proof argument of this case is similar to that of Case a.2 in part (a), which also yields the contradiction. 

As a consequence, we achieve the conclusion of part (b) of the lemma. 
\end{proof}
\subsubsection{Proof for lower bounds}
\label{subsub:lower_bound_know_variance}
(a) Based on the proof technique of Theorem 3.2 in~\cite{Jonas-2017}, to achieve the conclusion with the lower bound of part (a) of the theorem, it is sufficient to demonstrate that
\begin{align}
\inf \limits_{\delta^{(1)},\delta^{(2)} \in \Theta_{1,n}} h\parenth{g(x,\delta^{(1)}),g(x,\delta^{(2)})}/\biggr|\delta^{(1)} - \delta^{(2)}\biggr|^{r} = 0 \label{eqn:minimax_location_first}
\end{align}
for any $1 \leq r < 3$. We divide the proof argument for the above result into several key steps.
\paragraph{Step 1 - Constructing sequences} In fact, we construct two sequences $\left\{\delta_{n}^{(1)}\right\}$ and $\left\{\delta_{n}^{(2)}\right\}$ such that $\delta_{n}^{(1)} = -\delta_{n}^{(2)}$ for all $n \geq 1$ and $\delta_{n}^{(1)} \to 0$ as $n \to \infty$. For any fixed $r<3$, by means of Taylor expansion up to the second order as that in Step 1 of Case a.1 in part (a) of Theorem \ref{theorem:convergence_rate} (cf. Equation~\eqref{eqn:Taylor_representation}), the following holds
\begin{align}
g(x,\delta_{n}^{(1)}) - g(x,\delta_{n}^{(2)}) = \sum \limits_{\alpha=1}^{2}A_{n,\alpha}\dfrac{\partial^{\alpha}{\phi}}{\partial{\delta^{\alpha}}}(x,-\delta_{n}^{(2)})+ R(x), \nonumber
\end{align}
where $R(x)$ is a combination of Taylor remainders where its detail formulation is postponed to later discussion. Additionally, the formulations of $A_{n,\alpha}$ satisfy $A_{n,1}=0$ and
\begin{align}
A_{n,2} = \dfrac{c}{2}(\delta_{n}^{(2)} - \delta_{n}^{(1)})(\delta_{n}^{(1)}+\delta_{n}^{(2)}) = 0. \nonumber
\end{align}
\paragraph{Step 2 - Hellinger bound and Taylor remainders} Equipped with the above results, we have
\begin{align}
\dfrac{h^2\parenth{g(x,\delta_{n}^{(1)}),g(x,\delta_{n}^{(2)})}}{\abss{\delta_{n}^{(1)}-\delta_{n}^{(2)}}^{2r}} & = \int \dfrac{\parenth{g(x,\delta_{n}^{(1)}) - g(x,\delta_{n}^{(2)})}^2}{2^{r}\abss{\delta_{n}^{(2)}}^{2r}\parenth{\sqrt{g(x,\delta_{n}^{(1)})} + \sqrt{g(x,\delta_{n}^{(2)})}}^2} dx \nonumber \\
& = \int \dfrac{\parenth{R(x)}^2}{2^{r}\abss{\delta_{n}^{(2)}}^{2r}\parenth{\sqrt{g(x,\delta_{n}^{(1)})} + \sqrt{g(x,\delta_{n}^{(2)})}}^2}dx. \nonumber
\end{align}
To validate that the above term goes to 0, we will need to investigate the concrete formulation of $R(x)$. In particular, the formulation of $R(x)$ is
\begin{align}
R(x) = \pi R_{1}(x) + (1-\pi)\sum \limits_{\alpha=1}^{2} \dfrac{c^{\alpha}(\delta_{n}^{(1)} - \delta_{n}^{(2)})^{\alpha}}{\alpha!}R_{2,\alpha}(x)+ (1-\pi)R_{2}(x), \nonumber
\end{align}
where the formulations of Taylor remainders $R_{1}(x)$, $R_{2,\alpha}(x)$, and $R_{2}(x)$ are as follows
\begin{align}
R_{1}(x) = \dfrac{3\parenth{\delta_{n}^{(2)}-\delta_{n}^{(1)}}^{3}}{3!}\int \limits_{0}^{1}(1-t)^2\dfrac{\partial^{3}\phi}{\partial{\delta^3}}\parenth{x,-\delta_{n}^{(2)}+t\parenth{\delta_{n}^{(2)}-\delta_{n}^{(1)}}}dt, \nonumber \\
R_{2}(x) = \dfrac{3c^3\parenth{\delta_{n}^{(1)}-\delta_{n}^{(2)}}^{3}}{3!}\int \limits_{0}^{1}(1-t)^2\dfrac{\partial^{3}\phi}{\partial{\delta^3}}\parenth{x,c\delta_{n}^{(2)}+t\parenth{c\delta_{n}^{(1)}-c\delta_{n}^{(2)}}}dt, \nonumber \\
R_{2,\alpha}(x) = \dfrac{(3-\alpha)(c+1)^{3-\alpha}\parenth{\delta_{n}^{(2)}}^{3-\alpha}}{(3-\alpha)!\alpha !} \int \limits_{0}^{1}(1-t)^{2-\alpha}\dfrac{\partial^{3}\phi}{\partial{\delta^{3}}}\parenth{x,-\delta_{n}^{(2)}+t(c+1)\delta_{n}^{(2)}}dt \nonumber
\end{align}
for any $1 \leq \alpha \leq 2$. 
\paragraph{Step 3 - Taylor remainders control} Now, Holder's inequality leads to
\begin{align}
R_{1}^{2}(x) \leq \dfrac{\parenth{\delta_{n}^{(2)}-\delta_{n}^{(1)}}^{6}}{4}\int \limits_{0}^{1}(1-t)^4\parenth{\dfrac{\partial^{3}\phi}{\partial{\delta^3}}\parenth{x,-\delta_{n}^{(2)}+t\parenth{\delta_{n}^{(2)}-\delta_{n}^{(1)}}}}^{2}dt. \nonumber
\end{align}
Due to the formulation of location Gaussian kernel with variance 1, we can check that
\begin{align}
\sup \limits_{t \in [0,1]} \int \dfrac{\parenth{\dfrac{\partial^{3}\phi}{\partial{\delta^3}}\parenth{x,-\delta_{n}^{(2)}+t\parenth{\delta_{n}^{(2)}-\delta_{n}^{(1)}}}}^{2}}{\phi(x,-\delta_{n}^{(2)})}dx < \infty. \nonumber
\end{align}
Equipped with the above results, the following holds
\begin{align}
\int \dfrac{R_{1}^{2}(x)}{2^{r-1}\abss{\delta_{n}^{(2)}}^{2r}\parenth{\sqrt{g(x,\delta_{n}^{(1)})} + \sqrt{g(x,\delta_{n}^{(2)})}}^2}dx & \leq \int \dfrac{R_{1}^{2}(x)}{2^{r-1}\abss{\delta_{n}^{(2)}}^{2r}\pi \phi(x,-\delta_{n}^{(2)})}dx \nonumber \\
& \lesssim \abss{\delta_{n}^{(2)}}^{6-2r} \to 0 \label{eqn:minimax_location_second}
\end{align} 
as $n \to \infty$ where the first inequality is due to the inequality $\parenth{\sqrt{g(x,\delta_{n}^{(1)})} + \sqrt{g(x,\delta_{n}^{(2)})}}^2 \geq \pi\phi(x,-\delta_{n}^{(2)})$. By means of the similar argument, we also obtain that
\begin{align}
\int \dfrac{R_{2}^{2}(x)}{2^{r-1}\abss{\delta_{n}^{(2)}}^{2r}\parenth{\sqrt{g(x,\delta_{n}^{(1)})} + \sqrt{g(x,\delta_{n}^{(2)})}}^2}dx & \leq \int \dfrac{R_{2}^{2}(x)}{2^{r-1}\abss{\delta_{n}^{(2)}}^{2r}(1-\pi)\phi(x,c\delta_{n}^{(2)})}dx \nonumber \\
& \lesssim \abss{\delta_{n}^{(2)}}^{6-2r} \to 0, \nonumber \\
\int \dfrac{\parenth{\delta_{n}^{(1)}-\delta_{n}^{(2)}}^{\alpha}R_{2,\alpha}^{2}(x)}{2^{r-1}\abss{\delta_{n}^{(2)}}^{2r}\parenth{\sqrt{g(x,\delta_{n}^{(1)})} + \sqrt{g(x,\delta_{n}^{(2)})}}^2}dx & \leq \int \dfrac{\parenth{\delta_{n}^{(1)}-\delta_{n}^{(2)}}^{\alpha} R_{2,\alpha}^{2}(x)}{2^{r-1}\abss{\delta_{n}^{(2)}}^{2r}\pi\phi(x,-\delta_{n}^{(2)})}dx \nonumber \\
& \lesssim \abss{\delta_{n}^{(2)}}^{6-2r} \to 0. \label{eqn:minimax_location_third}
\end{align}
Invoking Cauchy-Schwarz's inequality, the following inequality holds
\begin{align}
R^2(x) \leq 3\parenth{\parenth{\pi R_{1}(x)}^2 + \parenth{(1-\pi)\sum \limits_{\alpha=1}^{2} \dfrac{c^{\alpha}(\delta_{n}^{(1)} - \delta_{n}^{(2)})^{\alpha}}{\alpha!}R_{2,\alpha}(x)}^{2}+ \parenth{(1-\pi)R_{2}(x)}^{2}}. \label{eqn:minimax_location_fourth}
\end{align}
Combining the results from \eqref{eqn:minimax_location_second}, \eqref{eqn:minimax_location_third}, and \eqref{eqn:minimax_location_fourth}, we achieve that 
\begin{align}
\int R^{2}(x)/\parenth{2^{r-1}\abss{\delta_{n}^{(2)}}^{2r}\parenth{\sqrt{g(x,\delta_{n}^{(1)})} + \sqrt{g(x,\delta_{n}^{(2)})}}^2}dx \to 0. \nonumber
\end{align}
As a consequence, we achieve the conclusion with the lower bound of part (a) of the theorem.  

(b) Similar to the proof argument of part (a), to achieve the conclusion of the lower bound of part (b), it is sufficient to demonstrate that
\begin{align}
\inf \limits_{\delta^{(1)},\delta^{(2)} \in \Theta_{2,n}} h\parenth{g(x,\delta^{(1)}),g(x,\delta^{(2)})}/\abss{\abss{\delta^{(1)}} - \abss{\delta^{(2)}}}^{r} = 0 \label{eqn:minimax_location_fifth}
\end{align}
for any $1 \leq r <2$. In particular, we choose two sequences $\left\{\overline{\delta}_{n}^{(1)}\right\}$ and $\left\{\overline{\delta}_{n}^{(2)}\right\}$ such that $\overline{\delta}_{n}^{(1)} = 2\overline{\delta}_{n}^{(2)}$ for all $n \geq 1$ and $\overline{\delta}_{n}^{(1)} \to 0$ as $n \to \infty$. For any $r<2$, invoking Taylor expansion up to the first order as that of Case b.1 in the proof of Theorem \ref{theorem:convergence_rate}, we have
\begin{align}
g(x,\overline{\delta}_{n}^{(1)}) - g(x,\overline{\delta}_{n}^{(2)}) = \overline{R}(x), \nonumber
\end{align}
where the formulation of $\overline{R}(x)$ is
\begin{align}
\overline{R}(x) = \frac{1}{2}\overline{R}_{1}(x) + \frac{1}{2}\parenth{\overline{\delta}_{n}^{(1)}-\overline{\delta}_{n}^{(2)}}\overline{R}_{2,1}(x)+\frac{1}{2}\overline{R}_{2}(x). \nonumber
\end{align}
Here, the detail formulations of Taylor remainders $\overline{R}_{1}(x)$, $\overline{R}_{2,1}(x)$, and $\overline{R}_{2}(x)$ are 
\begin{align}
\overline{R}_{1}(x) = \dfrac{2\parenth{\overline{\delta}_{n}^{(2)}-\overline{\delta}_{n}^{(1)}}^{2}}{2!}\int \limits_{0}^{1}(1-t)\dfrac{\partial^{2}\phi}{\partial{\delta^2}}\parenth{x,-\overline{\delta}_{n}^{(2)}+t\parenth{\overline{\delta}_{n}^{(2)}-\overline{\delta}_{n}^{(1)}}}dt, \nonumber \\
\overline{R}_{2}(x) = \dfrac{2\parenth{\overline{\delta}_{n}^{(1)}-\overline{\delta}_{n}^{(2)}}^{2}}{2!}\int \limits_{0}^{1}(1-t)\dfrac{\partial^{2}\phi}{\partial{\delta^2}}\parenth{x,\overline{\delta}_{n}^{(2)}+t\parenth{\overline{\delta}_{n}^{(1)}-\overline{\delta}_{n}^{(2)}}}dt, \nonumber \\
\overline{R}_{2,1}(x) = 2\overline{\delta}_{n}^{(2)} \int \limits_{0}^{1}\dfrac{\partial^{2}\phi}{\partial{\delta^{2}}}\parenth{x,-\overline{\delta}_{n}^{(2)}+2t\overline{\delta}_{n}^{(2)}}dt. \nonumber
\end{align}
With the choice that $\overline{\delta}_{n}^{(1)} = 2\overline{\delta}_{n}^{(2)} \to 0$ and the same argument as Step 3 in part (a), we can argue that
\begin{align}
\int \overline{R}^2(x)\big/ \parenth{2^{r-1}\abss{\overline{\delta}_{n}^{(2)}}^{2r}\parenth{\sqrt{g(x,\overline{\delta}_{n}^{(1)})}+\sqrt{g(x,\overline{\delta}_{n}^{(2)}}}^2} \to 0 \nonumber
\end{align}
as $n \to \infty$. Therefore, for any $1 \leq r < 2$, we achieve
\begin{align}
h\parenth{g(x,\overline{\delta}_{n}^{(1)}),g(x,\overline{\delta}_{n}^{(2)})}/\biggr|\abss{\overline{\delta}_{n}^{(1)}} - \abss{\overline{\delta}_{n}^{(2)}}\biggr|^{r} \to 0. \nonumber
\end{align}
As a consequence, we achieve the conclusion of part (b) of the theorem.
%%%%%%%%%%%%%%%%%%%%%%%%%%%%%%%%%%%%%%%%%%%
%%%%%%%%%%%%%
\subsection{PROOF OF THEOREM \ref{theorem:convergence_rate_location_scale_Gaussian}}
\label{subsec:proof:theorem:convergence_rate_location_scale_Gaussian}
%%%%%%%%%%%%%
%%%%%%%%%%%%%%%%%%%%%%%%%%%%%%%%%%%%%%%%%%%
For the sake of presentation, we denote $v: = \sigma^{2}$ and $g(x,\delta,v) : = \pi f(x,-\delta,v) + (1 - \pi)f(x,c\delta,v)$ for all $\delta \in \Theta, \sigma \in \Omega$ where $f(x,\delta,v)$ is the density of location-scale Gaussian distribution with location $\delta$ and scale $v$. For the simplicity of the proof argument, we only focus on the proof for the upper bounds of the theorem. The proof for the lower bounds can be argued similarly as that of the lower bounds in Theorem~\ref{theorem:convergence_rate} in Section~\ref{subsub:lower_bound_know_variance}.

(a) By means of the proof argument with the upper bound of Theorem \ref{theorem:convergence_rate}, in order to achieve the upper bound of part (a), it is sufficient to demonstrate that
\begin{eqnarray}
\inf \limits_{\substack{\delta^{(1)},\delta^{(2)} \in \Theta \\ v^{(1)},v^{(2)} \in \Omega}} \dfrac{V\parenth{g(x,\delta^{(1)},v^{(1)}),g(x,\delta^{(2)},v^{(2)})}}{|\delta^{(1)} - \delta^{(2)}|^{3}+|v^{(1)} - v^{(2)}|^{3/2}} > 0,
\end{eqnarray}
where $\Theta = [-1, 1]$ and $\Omega$ is a bounded set containing $\overline{\sigma}$. Assume that the above inequality does not hold. It implies that we can find sequences $\left\{\delta_{n}^{(1)}\right\}$, $\left\{\delta_{n}^{(2)}\right\}$, $\left\{v_{n}^{(1)}\right\}$, and $\left\{v_{n}^{(2)}\right\}$ such that
\begin{eqnarray}
\dfrac{V\parenth{g(x,\delta_{n}^{(1)},v_{n}^{(1)}),g(x,\delta_{n}^{(2)},v_{n}^{(2)})}}{|\delta_{n}^{(1)} - \delta_{n}^{(2)}|^{3}+|v_{n}^{(1)} - v_{n}^{(2)}|^{3/2}} \to 0 \nonumber
\end{eqnarray}
as $n \to \infty$. To simplify the presentation, we only consider the most challenging setting $\delta_{n}^{(1)} \to 0, \delta_{n}^{(2)} \to 0$, $v_{n}^{(1)} \to v_{0}$, $v_{n}^{(2)} \to v_{0}$ for some $v_{0} \in \Omega$. Additionally, we denote 
\begin{align}
D_{n} = |\delta_{n}^{(1)} - \delta_{n}^{(2)}|^{3}+|v_{n}^{(1)} - v_{n}^{(2)}|^{3/2}. \nonumber
\end{align}
Now, we consider the following settings with $\delta_{n}^{(1)}$ and $\delta_{n}^{(2)}$.
\paragraph{Case a.1:} $\delta_{n}^{(1)}/\delta_{n}^{(2)} \not \to 1$ as $n \to \infty$. Similar to the structure of the proof of Theorem~\ref{theorem:convergence_rate}, we also divide the proof argument of this case into two key steps.
\paragraph{Step 1 - Taylor expansion} Under this setting, by means of Taylor expansion up to the third order, we obtain that
\begin{eqnarray}
& & \dfrac{g(x,\delta_{n}^{(1)},v_{n}^{(1)}) - g(x,\delta_{n}^{(2)},v_{n}^{(2)})}{D_{n}} \label{eqn:Taylor_location_scale_Gaussian_first} \\
& & = \dfrac{\pi\biggr(f(x,-\delta_{n}^{(1)},v_{n}^{(1)}) - f(x,-\delta_{n}^{(2)},v_{n}^{(2)})\biggr) + (1-\pi)\biggr(f(x,c\delta_{n}^{(1)},v_{n}^{(1)}) - f(x,c\delta_{n}^{(2)},v_{n}^{(2)})\biggr)}{D_{n}} \nonumber \\
& & = \dfrac{\pi \biggr(\sum \limits_{|\alpha| \leq 3}\dfrac{(\delta_{n}^{(2)} - \delta_{n}^{(1)})^{\alpha_{1}}(v_{n}^{(1)} - v_{n}^{(2)})^{\alpha_{2}}}{\alpha_{1}!\alpha_{2}!}\dfrac{\partial^{|\alpha|}{f}}{\partial{\delta^{\alpha_{1}}}\partial{v^{\alpha_{2}}}}(x,-\delta_{n}^{(2)},v_{n}^{(2)})+R_{1}(x)\biggr)}{D_{n}} \nonumber 
\end{eqnarray}
\begin{eqnarray}
& & + \dfrac{(1-\pi) \biggr(\sum \limits_{|\alpha| \leq 3}\dfrac{c^{\alpha_{1}}(\delta_{n}^{(1)} - \delta_{n}^{(2)})^{\alpha_{1}}(v_{n}^{(1)} - v_{n}^{(2)})^{\alpha_{2}}}{\alpha_{1}!\alpha_{2}!}\dfrac{\partial^{|\alpha|}{f}}{\partial{\delta^{\alpha_{1}}}\partial{v^{\alpha_{2}}}}(x,c\delta_{n}^{(2)},v_{n}^{(2)})+R_{2}(x)\biggr)}{D_{n}} \nonumber \\
& & = \dfrac{\pi \biggr(\sum \limits_{|\alpha| \leq 3}\dfrac{1}{2^{\alpha_{2}}} \dfrac{(\delta_{n}^{(2)} - \delta_{n}^{(1)})^{\alpha_{1}}(v_{n}^{(1)} - v_{n}^{(2)})^{\alpha_{2}}}{\alpha_{1}!\alpha_{2}!}\dfrac{\partial^{\alpha_{1}+2\alpha_{2}}{f}}{\partial{\delta^{\alpha_{1}+2\alpha_{2}}}}(x,-\delta_{n}^{(2)},v_{n}^{(2)})+R_{1}(x)\biggr)}{D_{n}} \nonumber \\
& & \hspace { 1 em} + \dfrac{(1-\pi) \biggr(\sum \limits_{|\alpha| \leq 3}\dfrac{1}{2^{\alpha_{2}}}\dfrac{c^{\alpha_{1}}(\delta_{n}^{(1)} - \delta_{n}^{(2)})^{\alpha_{1}}(v_{n}^{(1)} - v_{n}^{(2)})^{\alpha_{2}}}{\alpha_{1}!\alpha_{2}!}\dfrac{\partial^{\alpha_{1}+2\alpha_{2}}{f}}{\partial{\delta^{\alpha_{1}+2\alpha_{2}}}}(x,c\delta_{n}^{(2)},v_{n}^{(2)})+R_{2}(x)\biggr)}{D_{n}}, \nonumber
\end{eqnarray}
where the last equality is due to the PDE structure of location-scale Gaussian distribution, which is given by
\begin{align}
\dfrac{\partial^{2}{f}}{\partial{\delta^{2}}}(x,\delta,\sigma) = 2 \dfrac{\partial{f}}{\partial{\sigma^{2}}}(x,\delta,\sigma). \nonumber
\end{align}
Additionally, $R_{1}(x)$ and $R_{2}(x)$ are Taylor remainders that satisfy the following inequality
\begin{align}
\max \{\|R_{1}(x)\|_{\infty}, \|R_{2}(x)\|_{\infty}\} = O\parenth{|\delta_{n}^{(1)} - \delta_{n}^{(2)}|^{3+\gamma}+|v_{n}^{(1)} - v_{n}^{(2)}|^{3+\gamma}} \nonumber
\end{align}
for some $\gamma>0$. It implies that $R_{1}(x)/D_{n} \to 0$ and $R_{2}(x)/D_{n} \to 0$ for all $x$ as $n \to \infty$. Now, by means of Taylor expansion up to the third order, we further have
\begin{eqnarray}
\hspace{ 2 em} \dfrac{\partial^{\alpha_{1}+2\alpha_{2}}{f}}{\partial{\delta^{\alpha_{1}+2\alpha_{2}}}}(x,c\delta_{n}^{(2)},v_{n}^{(2)}) = \sum \limits_{\tau = 0}^{3 - |\alpha|} \dfrac{(c+1)^{\tau}(\delta_{n}^{(2)})^{\tau}}{\tau!}\dfrac{\partial^{\alpha_{1}+2\alpha_{2}+\tau}{f}}{\partial{\delta^{\alpha_{1}+2\alpha_{2}+\tau}}}(x,-\delta_{n}^{(2)},v_{n}^{(2)}) + R_{2,\alpha}(x) \label{eqn:Taylor_location_scale_Gaussian_second}
\end{eqnarray}
for each $\alpha = (\alpha_{1},\alpha_{2})$ such that $1 \leq |\alpha| \leq 3$. Here, $R_{2,\alpha}(x)$ is a Taylor remainder that satisfies $\|R_{2,\alpha}(x)\|_{\infty} = O\parenth{|\delta_{n}^{(2)}|^{3-|\alpha|+\gamma}}$ for all $\alpha$. By plugging equations \eqref{eqn:Taylor_location_scale_Gaussian_second} into \eqref{eqn:Taylor_location_scale_Gaussian_first}, the following holds
\begin{eqnarray}
& & \hspace{- 2 em} \dfrac{g(x,\delta_{n}^{(1)},v_{n}^{(1)}) - g(x,\delta_{n}^{(2)},v_{n}^{(2)})}{D_{n}} \nonumber \\
& & \hspace{- 2 em} = \dfrac{\pi \biggr(\sum \limits_{|\alpha| \leq 3}\dfrac{1}{2^{\alpha_{2}}} \dfrac{(\delta_{n}^{(2)} - \delta_{n}^{(1)})^{\alpha_{1}}(v_{n}^{(1)} - v_{n}^{(2)})^{\alpha_{2}}}{\alpha_{1}!\alpha_{2}!}\dfrac{\partial^{\alpha_{1}+2\alpha_{2}}{f}}{\partial{\delta^{\alpha_{1}+2\alpha_{2}}}}(x,-\delta_{n}^{(2)},v_{n}^{(2)})\biggr)}{D_{n}} \nonumber \\
& & \hspace{- 2 em} + \dfrac{(1-\pi) \biggr(\sum \limits_{|\alpha| \leq 3} \sum \limits_{\tau=0}^{3-|\alpha|} \dfrac{1}{2^{\alpha_{2}}}\dfrac{c^{\alpha_{1}}{(c+1)^{\tau}(\delta_{n}^{(2)})^{\tau}(\delta_{n}^{(1)} - \delta_{n}^{(2)})^{\alpha_{1}}(v_{n}^{(1)} - v_{n}^{(2)})^{\alpha_{2}}}}{\tau!\alpha_{1}!\alpha_{2}!}\dfrac{\partial^{\alpha_{1}+2\alpha_{2}}{f}}{\partial{\delta^{\alpha_{1}+2\alpha_{2}}}}(x,-\delta_{n}^{(2)},v_{n}^{(2)})}{D_{n}} \nonumber \\
& & \hspace{- 2 em} + \dfrac{\pi R_{1}(x) + (1-\pi) R_{2}(x) + \sum \limits_{|\alpha| \leq 3} \dfrac{1}{2^{\alpha_{2}}}\dfrac{c^{\alpha_{1}}(\delta_{n}^{(1)} - \delta_{n}^{(2)})^{\alpha_{1}}(v_{n}^{(1)} - v_{n}^{(2)})^{\alpha_{2}}}{\alpha_{1}!\alpha_{2}!}R_{2,\alpha}(x)}{D_{n}} \nonumber \\
& & \hspace{- 2 em} = \dfrac{\sum \limits_{l=1}^{6} A_{n,l}\dfrac{\partial^{l}{f}}{\partial{\delta^{l}}}(x,-\delta_{n}^{(2)},v_{n}^{(2)})+R(x)}{D_{n}}, \nonumber
\end{eqnarray}
where the detail formulations of $A_{n,l}$ and $R(x)$ are as follows
\begin{eqnarray}
A_{n,l} & = & \pi \sum \limits_{\alpha_{1},\alpha_{2}} \dfrac{1}{2^{\alpha_{2}}} \dfrac{(\delta_{n}^{(2)} - \delta_{n}^{(1)})^{\alpha_{1}}(v_{n}^{(1)} - v_{n}^{(2)})^{\alpha_{2}}}{\alpha_{1}!\alpha_{2}!} \nonumber \\
& & \hspace{ 3 em} + (1 - \pi) \sum \limits_{\alpha_{1},\alpha_{2},\tau} \dfrac{1}{2^{\alpha_{2}}}\dfrac{c^{\alpha_{1}}(c+1)^{\tau}(\delta_{n}^{(2)})^{\tau}(\delta_{n}^{(1)}-\delta_{n}^{(2)})^{\alpha_{1}}(v_{n}^{(1)}-v_{n}^{(2)})^{\alpha_{2}}}{\tau!\alpha_{1}!\alpha_{2}!}, \nonumber \\
R(x) & = & \pi R_{1}(x) + (1-\pi) R_{2}(x) + \sum \limits_{|\alpha| \leq 3} \dfrac{1}{2^{\alpha_{2}}}\dfrac{c^{\alpha_{1}}(\delta_{n}^{(1)} - \delta_{n}^{(2)})^{\alpha_{1}}(v_{n}^{(1)} - v_{n}^{(2)})^{\alpha_{2}}}{\alpha_{1}!\alpha_{2}!}R_{2,\alpha}(x) \nonumber
\end{eqnarray} 
for any $1 \leq l \leq 6$ and $x \in \mathbb{R}$. Here, the ranges of $\alpha_{1},\alpha_{2}$ in the first sum of $A_{n,l}$ satisfy $\alpha_{1}+2\alpha_{2} = l$, $1 \leq |\alpha| \leq 3$ while the ranges of $\alpha_{1},\alpha_{2},\tau$ in the second sum of $A_{n,l}$ satisfy $\alpha_{1}+2\alpha_{2}+\tau = l$, $0 \leq \tau \leq 3-|\alpha|$, and $1 \leq |\alpha| \leq 3$. According to the hypothesis $\delta_{n}^{(1)}/\delta_{n}^{(2)} \not \to 1$, we have 
\begin{align}
|\delta_{n}^{(2)}|/|\delta_{n}^{(1)} - \delta_{n}^{(2)}| \not \to \infty. \nonumber
\end{align} 
Therefore, we have 
\begin{eqnarray}
\dfrac{|\delta_{n}^{(1)} - \delta_{n}^{(2)}|^{\alpha_{1}}|v_{n}^{(1)} - v_{n}^{(2)}|^{\alpha_{2}}\|R_{2,\alpha}(x)\|_{\infty}}{D_{n}} = \dfrac{O\parenth{|\delta_{n}^{(1)} - \delta_{n}^{(2)}|^{\alpha_{1}}|v_{n}^{(1)} - v_{n}^{(2)}|^{\alpha_{2}}|\delta_{n}^{(2)}|^{3-|\alpha|+\gamma}}}{D_{n}} \to 0. \nonumber
\end{eqnarray}
As a consequence, we have $\|R(x)\|_{\infty}/D_{n} \to 0$ as $n \to \infty$. 
\paragraph{Step 2 - Non-vanishing coefficients and Fatou's argument} Assume that all the coefficients $A_{n,l}/D_{n} \to 0$ for all $1 \leq l \leq 6$ as $n \to \infty$. We denote the following key term
\begin{eqnarray}
\overline{M}_{n} : = \max \left\{|\delta_{n}^{(1)} - \delta_{n}^{(2)}|,|v_{n}^{(1)} - v_{n}^{(2)}|^{1/2} \right\}. \nonumber
\end{eqnarray} 
As $|\delta_{n}^{(2)}|/|\delta_{n}^{(1)} - \delta_{n}^{(2)}| \not \to \infty$, we also have $|\delta_{n}^{(2)}|/M_{n} \not \to \infty$. Now, we denote $\delta_{n}^{(2)}/M_{n} \to x$, $(\delta_{n}^{(2)} - \delta_{n}^{(1)})/M_{n} \to y$, and $(v_{n}^{(1)} - v_{n}^{(2)})/M_{n}^{2} \to z$ as $n \to \infty$. From the definition of $\overline{M}_{n}$, at least one among $y$ and $z$ is different from 0. By dividing both the numerator and the denominator of $A_{n,l}/D_{n}$ by $\overline{M}_{n}^{l}$ as $1 \leq l \leq 3$, as $n \to \infty$, we have the following system of polynomial equations
\begin{eqnarray}
c y^{2} + z -2c xy = 0, \nonumber \\
\dfrac{\pi(1-2\pi)}{3!(1-\pi)^{2}} y^{3} + \dfrac{1}{2}xz + \dfrac{c^{2}}{2}xy^{2} - \dfrac{\pi}{2(1-\pi)^{2}}x^{2}y = 0. \nonumber
\end{eqnarray}
The above system of polynomial equations leads to $\pi(1-2\pi)y(y^{2}-3xy+3x^{2}) = 0$, which only holds when $y =0$. Therefore, it leads to $z=0$, which is a contradiction. It implies that not all the coefficients $A_{n,l}/D_{n} \to 0$ as $n \to \infty$. Denote $m_{n} = D_{n}/\max \limits_{1 \leq l \leq 6} |A_{n,l}|$. According to the previous result, we have $m_{n} \not \to \infty$. Now, we have that
\begin{eqnarray}
m_{n} \dfrac{g(x,\delta_{n}^{(1)},v_{n}^{(1)}) - g(x,\delta_{n}^{(2)},v_{n}^{(2)})}{D_{n}} \to \sum \limits_{l=1}^{6} \tau_{l}\dfrac{\partial^{l}{f}}{\partial{\delta^{l}}}(x,0,v_{0}) \nonumber
\end{eqnarray}
for some coefficients $\tau_{l}$ such that not all of them are 0. Similar to the proof argument of Theorem \ref{theorem:convergence_rate}, by invoking Fatou's lemma with $V\parenth{g(x,\delta_{n}^{(1)},v_{n}^{(1)}),g(x,\delta_{n}^{(2)},v_{n}^{(2)})}/D_{n} \to 0$, the following equation holds
\begin{align}
\sum \limits_{l=1}^{6} \tau_{l}\dfrac{\partial^{l}{f}}{\partial{\delta^{l}}}(x,0,v_{0}) = 0 \nonumber
\end{align}
for almost surely $x$. However, due to the linear independence of $\left\{\dfrac{\partial^{l}{f}}{\partial{\delta^{l}}}(x,0,v_{0})\right\}$, we have $\tau_{l} = 0$ for all $1 \leq l \leq 6$, which is a contradiction. Therefore, Case a.1 does not hold.
\paragraph{Case a.2:} $\delta_{n}^{(1)}/\delta_{n}^{(2)} \to 1$ as $n \to \infty$. It implies that $|\delta_{n}^{(2)}|/|\delta_{n}^{(1)} - \delta_{n}^{(2)}| \to \infty$. Similar to Case a.2 in the proof of Theorem \ref{theorem:convergence_rate}, the main challenge with that setting is that $R(x)/D_{n}$ does not converge to 0; therefore, we cannot hinge upon the previous argument in Case a.1 to argue the contradiction with this case. To be able to deal with that problem, we will demonstrate two key properties under that setting: $\max \limits_{1 \leq l \leq 6} \left\{|A_{n,l}|\right\}/D_{n} \not \to 0$ and $\|R(x)\|_{\infty}/\max \limits_{1 \leq l \leq 6} |A_{n,l}| \to 0$. Indeed, we have the following possibilities regarding $\delta_{n}^{(1)}, \delta_{n}^{(2)}, v_{n}^{(1)}$, and $v_{n}^{(2)}$.
\paragraph{Case a.2.1:} $|v_{n}^{(1)}- v_{n}^{(2)}|/\biggr\{|\delta_{n}^{(1)}-\delta_{n}^{(2)}||\delta_{n}^{(1)}+\delta_{n}^{(2)}|\biggr\} \to \infty$. Assume by the contrary that the following term $\max \limits_{1 \leq l \leq 6} \left\{|A_{n,l}|\right\}/D_{n} \to 0$. From the formulation of $A_{n,2}$, we have
\begin{eqnarray}
|A_{n,2}| = \dfrac{1}{2}\biggr|(v_{n}^{(1)} - v_{n}^{(2)}) - c(\delta_{n}^{(2)}- \delta_{n}^{(1)})(\delta_{n}^{(2)}+\delta_{n}^{(1)})\biggr| \gtrsim |v_{n}^{(1)} - v_{n}^{(2)}|, \nonumber
\end{eqnarray}
as $n$ is sufficiently large due to the assumption of Case a.2.1. Since $A_{n,2}/D_{n} \to 0$, it implies that $(v_{n}^{(1)} - v_{n}^{(2)})/D_{n} \to 0$. Therefore, it leads to $(\delta_{n}^{(1)} - \delta_{n}^{(2)})(\delta_{n}^{(2)} + \delta_{n}^{(1)})/D_{n} \to 0$. As $|\delta_{n}^{(2)}|/|\delta_{n}^{(1)} - \delta_{n}^{(2)}| \to \infty$, the previous limit implies that $|\delta_{n}^{(1)} - \delta_{n}^{(2)}|^{2} / D_{n} \to 0$. These results mean that
\begin{eqnarray}
1 = \dfrac{|v_{n}^{(1)} - v_{n}^{(2)}|^{3/2}+ |\delta_{n}^{(1)} - \delta_{n}^{(2)}|^{3}}{D_{n}} \to 0, \nonumber
\end{eqnarray}
which is a contradiction. Therefore, we have $\max \limits_{1 \leq l \leq 6} \left\{|A_{n,l}|\right\}/D_{n} \not \to 0$. Now, for any $1 \leq |\alpha| \leq 3$, as $n$ is sufficiently large, we have
\begin{eqnarray}
\dfrac{|\delta_{n}^{(1)} - \delta_{n}^{(2)}|^{\alpha_{1}}|v_{n}^{(1)} - v_{n}^{(2)}|^{\alpha_{2}}\|R_{2,\alpha}(x)\|_{\infty}}{\max \limits_{1 \leq l \leq 6} \left\{|A_{n,l}|\right\}} \leq \dfrac{O(|\delta_{n}^{(1)} - \delta_{n}^{(2)}|^{\alpha_{1}}|v_{n}^{(1)} - v_{n}^{(2)}|^{\alpha_{2}}|\delta_{n}^{(2)}|^{3-|\alpha|+\gamma})}{|v_{n}^{(1)} - v_{n}^{(2)}|} \to 0. \nonumber
\end{eqnarray}
Hence, we achieve that $\|R(x)\|_{\infty}/\max \limits_{1 \leq l \leq 6} \left\{|A_{n,l}|\right\} \to 0$ for all $x \in \mathbb{R}$. 
\paragraph{Case a.2.2:} $|v_{n}^{(1)}- v_{n}^{(2)}|/\biggr\{|\delta_{n}^{(1)}-\delta_{n}^{(2)}||\delta_{n}^{(1)}+\delta_{n}^{(2)}|\biggr\} \to \overline{c} \neq c$. Under that assumption, we have
\begin{eqnarray}
|A_{n,2}| = \dfrac{1}{2}\biggr|(v_{n}^{(1)} - v_{n}^{(2)}) - c(\delta_{n}^{(2)}- \delta_{n}^{(1)})(\delta_{n}^{(2)}+\delta_{n}^{(1)})\biggr| \gtrsim |\delta_{n}^{(1)} - \delta_{n}^{(2)}||\delta_{n}^{(1)}+\delta_{n}^{(2)}| \nonumber
\end{eqnarray}
when $n$ is sufficiently large. If we have $\max \limits_{1 \leq l \leq 6} \left\{|A_{n,l}|\right\}/D_{n} \to 0$, then $\abss{A_{n,2}}/D_{n}$ leads to both $(v_{n}^{(1)}-v_{n}^{(2)})/D_{n} \to 0$ and $\parenth{\delta_{n}^{(1)} - \delta_{n}^{(2)}}\parenth{\delta_{n}^{(1)}+\delta_{n}^{(2)}}/D_{n} \to 0$, which does not hold according to the argument of Case a.2.1. Therefore, $\max \limits_{1 \leq l \leq 6} \left\{|A_{n,l}|\right\}/D_{n} \not \to 0$. On the other hand, for any $1 \leq |\alpha| \leq 3$, as $n$ is sufficiently large, we have
\begin{eqnarray}
\dfrac{|\delta_{n}^{(1)} - \delta_{n}^{(2)}|^{\alpha_{1}}|v_{n}^{(1)} - v_{n}^{(2)}|^{\alpha_{2}}\|R_{2,\alpha}(x)\|_{\infty}}{\max \limits_{1 \leq l \leq 6} \left\{|A_{n,l}|\right\}} & \leq & \dfrac{O\parenth{|\delta_{n}^{(1)} - \delta_{n}^{(2)}|^{\alpha_{1}}|v_{n}^{(1)} - v_{n}^{(2)}|^{\alpha_{2}}|\delta_{n}^{(2)}|^{3-|\alpha|+\gamma}}}{|\delta_{n}^{(1)}-\delta_{n}^{(2)}||\delta_{n}^{(1)}+\delta_{n}^{(2)}|} \nonumber \\
& = & \dfrac{O\parenth{|\delta_{n}^{(1)} - \delta_{n}^{(2)}|^{|\alpha|}|\delta_{n}^{(2)}|^{3-\alpha_{1}+\gamma}}}{|\delta_{n}^{(1)}-\delta_{n}^{(2)}||\delta_{n}^{(1)}+\delta_{n}^{(2)}|} \to 0. \nonumber
\end{eqnarray}
Hence, we achieve that $\|R(x)\|_{\infty}/\max \limits_{1 \leq l \leq 6} \left\{|A_{n,l}|\right\} \to 0$ for all $x \in \mathbb{R}$.
\paragraph{Case a.2.3:} $|v_{n}^{(1)}- v_{n}^{(2)}|/\biggr\{|\delta_{n}^{(1)}-\delta_{n}^{(2)}||\delta_{n}^{(1)}+\delta_{n}^{(2)}|\biggr\} \to c$. Without loss of generality, we assume that $(v_{n}^{(1)} - v_{n}^{(2)})/(\delta_{n}^{(1)} - \delta_{n}^{(2)})(\delta_{n}^{(1)}+\delta_{n}^{(2)}) \to c$ as the argument when this ratio goes to $-c$ is similar. Under this assumption, we have
\begin{eqnarray}
\dfrac{|A_{n,3}|}{|\delta_{n}^{(1)}-\delta_{n}^{(2)}||\delta_{n}^{(1)}+\delta_{n}^{(2)}||\delta_{n}^{(2)}|} \to \biggr|\dfrac{c}{2} - \dfrac{(1-\pi)c(c+1)^{2}}{4}\biggr| > 0. \nonumber
\end{eqnarray}
Therefore, as $n$ is sufficiently large, we have $|A_{n,3}| \gtrsim |\delta_{n}^{(1)} - \delta_{n}^{(2)}||\delta_{n}^{(1)}+\delta_{n}^{(2)}||\delta_{n}^{(2)}|$. If we have $\max \limits_{1 \leq l \leq 6} \left\{|A_{n,l}|\right\}/D_{n} \to 0$, then $\abss{A_{n,3}}/D_{n} \to 0$ leads to $|\delta_{n}^{(1)} - \delta_{n}^{(2)}||\delta_{n}^{(1)}+\delta_{n}^{(2)}||\delta_{n}^{(2)}|/D_{n} \to 0$. Therefore, the following holds
\begin{eqnarray}
|v_{n}^{(1)} - v_{n}^{(2)}|^{3/2}/\biggr\{|\delta_{n}^{(1)}-\delta_{n}^{(2)}||\delta_{n}^{(1)}+\delta_{n}^{(2)}||\delta_{n}^{(2)}|\biggr\} \to \infty, \nonumber
\end{eqnarray} 
which means $|v_{n}^{(1)} - v_{n}^{(2)}|/|\delta_{n}^{(2)}|^{2} \to \infty$ --- a contradiction to the assumption of Case a.2.3. Hence, $\max \limits_{1 \leq l \leq 6} \left\{|A_{n,l}|\right\}/D_{n} \not \to 0$. On the other hand, for any $1 \leq |\alpha| \leq 3$, as $n$ is sufficiently large, we have
\begin{eqnarray}
\dfrac{|\delta_{n}^{(1)} - \delta_{n}^{(2)}|^{\alpha_{1}}|v_{n}^{(1)} - v_{n}^{(2)}|^{\alpha_{2}}\|R_{2,\alpha}(x)\|_{\infty}}{\max \limits_{1 \leq l \leq 6} \left\{|A_{n,l}|\right\}} & \leq & \dfrac{O\parenth{|\delta_{n}^{(1)} - \delta_{n}^{(2)}|^{\alpha_{1}}|v_{n}^{(1)} - v_{n}^{(2)}|^{\alpha_{2}}|\delta_{n}^{(2)}|^{3-|\alpha|+\gamma}}}{|\delta_{n}^{(1)}-\delta_{n}^{(2)}||\delta_{n}^{(1)}+\delta_{n}^{(2)}||\delta_{n}^{(2)}|} \nonumber \\
& = & \dfrac{O\parenth{|\delta_{n}^{(1)} - \delta_{n}^{(2)}|^{|\alpha|}|\delta_{n}^{(2)}|^{3-\alpha_{1}+\gamma}}}{|\delta_{n}^{(1)}-\delta_{n}^{(2)}||\delta_{n}^{(1)}+\delta_{n}^{(2)}||\delta_{n}^{(2)}|} \to 0. \nonumber
\end{eqnarray}
Thus, we obtain that $\|R(x)\|_{\infty}/\max \limits_{1 \leq l \leq 6} \left\{|A_{n,l}|\right\} \to 0$ for all $x \in \mathbb{R}$.

Governed by the results from Case a.2.1, Case a.2.2, and Case a.2.3, we finally achieve that $\max \limits_{1 \leq l \leq 6} \left\{|A_{n,l}|\right\}/D_{n} \not \to 0$ and $\|R(x)\|_{\infty}/\max \limits_{1 \leq l \leq 6} |A_{n,l}| \to 0$. Denote $m_{n}' = D_{n}/\max \limits_{1 \leq l \leq 6} \left\{|A_{n,l}|\right\}$. Then, we will have $m_{n}' \not \to \infty$. Thus, the following limit holds
\begin{eqnarray}
m_{n}' \dfrac{g(x,\delta_{n}^{(1)},v_{n}^{(1)}) - g(x,\delta_{n}^{(2)},v_{n}^{(2)})}{D_{n}} \to \sum \limits_{l=1}^{6} \tau_{l}'\dfrac{\partial^{l}{f}}{\partial{\delta^{l}}}(x,0,v_{0}), \nonumber
\end{eqnarray}
for some coefficients $\tau_{l}'$ such that not all of them are 0. By means of Fatou's lemma with the ratio $V\parenth{g(x,\delta_{n}^{(1)},v_{n}^{(1)}),g(x,\delta_{n}^{(2)},v_{n}^{(2)})}/D_{n} \to 0$, we obtain that
\begin{align}
\sum \limits_{l=1}^{6} \tau_{l}'\dfrac{\partial^{l}{f}}{\partial{\delta^{l}}}(x,0,v_{0}) = 0. \nonumber
\end{align}
However, due to the linear independence of $\left\{\dfrac{\partial^{l}{f}}{\partial{\delta^{l}}}(x,0,v_{0})\right\}$, we will have $\tau_{l}' = 0$ for all $1 \leq l \leq 6$, which is a contradiction. Therefore, Case a.2 does not hold. As a consequence, we achieve the conclusion with the upper bound of part (a) of the theorem. 

(b) Similar to the proof argument of part (a), it is sufficient to demonstrate that
\begin{eqnarray}
\inf \limits_{\substack{\delta^{(1)},\delta^{(2)} \in \Theta \\ v^{(1)},v^{(2)} \in \Omega}} \dfrac{V\parenth{g(x,\delta^{(1)},v^{(1)}),g(x,\delta^{(2)},v^{(2)})}}{\biggr||\delta^{(1)}| - |\delta^{(2)}|\biggr|^{4} +|v^{(1)} - v^{(2)}|^{2}} > 0, \nonumber
\end{eqnarray}
where $\Theta = [-1, 1]$ and $\Omega$ is a bounded set containing $\overline{\sigma}$. Assume that the above inequality does not hold. It implies that we can find sequences $\left\{\delta_{n}^{(1)}\right\}$, $\left\{\delta_{n}^{(2)}\right\}$, $\left\{v_{n}^{(1)}\right\}$, and $\left\{v_{n}^{(2)}\right\}$ such that
\begin{eqnarray}
\dfrac{V\parenth{g(x,\delta_{n}^{(1)},v_{n}^{(1)}),g(x,\delta_{n}^{(2)},v_{n}^{(2)})}}{\biggr||\delta_{n}^{(1)}| - |\delta_{n}^{(2)}|\biggr|^{4} +|v_{n}^{(1)} - v_{n}^{(2)}|^{2}} \to 0 \nonumber
\end{eqnarray}
as $n \to \infty$. Similar the proof argument of part (a), we only consider the most challenging setting $\delta_{n}^{(1)} \to 0, \delta_{n}^{(2)} \to 0$, $v_{n}^{(1)} \to v_{0}$, $v_{n}^{(2)} \to v_{0}$ for some $v_{0} \in \Omega$. For the convenience of presentation, we denote 
\begin{eqnarray}
\overline{D}_{n} = \biggr||\delta_{n}^{(1)}| - |\delta_{n}^{(2)}|\biggr|^{4} +|v_{n}^{(1)} - v_{n}^{(2)}|^{2}.  \nonumber
\end{eqnarray}
Now, we have three settings with $\delta_{n}^{(1)}$ and $\delta_{n}^{(2)}$ in the proof of part (b). 
\paragraph{Case b.1:} $\delta_{n}^{(1)}/\delta_{n}^{(2)} \not \to 1$ as $n \to \infty$ and $\delta_{n}^{(1)}\delta_{n}^{(2)} \geq 0$ for all $n$. Under this case, we have
\begin{eqnarray}
\overline{D}_{n}  = |\delta_{n}^{(1)} - \delta_{n}^{(2)}|^{4} + |v_{n}^{(1)} - v_{n}^{(2)}|^{2}. \nonumber
\end{eqnarray}
To facilitate the proof argument of this case, we also divide it into two key steps.
\paragraph{Step 1 - Taylor expansion} Using the similar argument as that of part (a), by means of Taylor expansion up to the fourth order, we get the following representation
\begin{eqnarray}
\dfrac{g(x,\delta_{n}^{(1)},v_{n}^{(1)}) - g(x,\delta_{n}^{(2)},v_{n}^{(2)})}{\overline{D}_{n}} = \dfrac{\sum \limits_{l=1}^{8} B_{n,l}\dfrac{\partial^{l}{f}}{\partial{\delta^{l}}}(x,-\delta_{n}^{(2)},v_{n}^{(2)}) + \overline{R}(x)}{\overline{D}_{n}}, \nonumber
\end{eqnarray}
where the formulations of $B_{n,l}$ and $\overline{R}(x)$ are as follows
\begin{eqnarray}
B_{n,l} & = & \dfrac{1}{2} \sum \limits_{\alpha_{1},\alpha_{2}} \dfrac{1}{2^{\alpha_{2}}} \dfrac{(\delta_{n}^{(2)} - \delta_{n}^{(1)})^{\alpha_{1}}(v_{n}^{(1)} - v_{n}^{(2)})^{\alpha_{2}}}{\alpha_{1}!\alpha_{2}!} \nonumber \\
& & \hspace{5 em} + \dfrac{1}{2} \sum \limits_{\alpha_{1},\alpha_{2},\tau} \dfrac{1}{2^{\alpha_{2}}}\dfrac{2^{\tau}(\delta_{n}^{(2)})^{\tau}(\delta_{n}^{(1)}-\delta_{n}^{(2)})^{\alpha_{1}}(v_{n}^{(1)}-v_{n}^{(2)})^{\alpha_{2}}}{\tau!\alpha_{1}!\alpha_{2}!}, \nonumber \\
\overline{R}(x) & = & \dfrac{1}{2} \overline{R}_{1}(x) + \dfrac{1}{2} \overline{R}_{2}(x) + \sum \limits_{|\alpha| \leq 4} \dfrac{1}{2^{\alpha_{2}}}\dfrac{(\delta_{n}^{(1)} - \delta_{n}^{(2)})^{\alpha_{1}}(v_{n}^{(1)} - v_{n}^{(2)})^{\alpha_{2}}}{\alpha_{1}!\alpha_{2}!}\overline{R}_{2,\alpha}(x). \nonumber
\end{eqnarray}
Here, the ranges of $\alpha_{1},\alpha_{2}$ in the first sum of $B_{n,l}$ satisfy $\alpha_{1}+2\alpha_{2} = l$, $1 \leq |\alpha| \leq 4$ while the ranges of $\alpha_{1},\alpha_{2},\tau$ in the second sum of $B_{n,l}$ satisfy $\alpha_{1}+2\alpha_{2}+\tau = l$, $0 \leq \tau \leq 4-|\alpha|$, and $1 \leq |\alpha| \leq 4$. Additionally, $\overline{R}_{1}(x)$ is a Taylor remainder from expanding $f(x,-\delta_{n}^{(1)},v_{n}^{(1)})$ around $f(x,-\delta_{n}^{(2)},v_{n}^{(2)})$ up to the fourth order, $\overline{R}_{2}(x)$ is Taylor remainder from expanding $f(x,c\delta_{n}^{(1)},v_{n}^{(1)})$ around $f(x,c\delta_{n}^{(2)},v_{n}^{(2)})$ up to the fourth order, and $\overline{R}_{2,\alpha}(x)$ is Taylor remainder from expanding $\dfrac{\partial^{\alpha_{1}+2\alpha_{2}}{f}}{\partial{\delta^{\alpha_{1}+2\alpha_{2}}}}(x,c\delta_{n}^{(2)},v_{n}^{(2)})$ around $\dfrac{\partial^{\alpha_{1}+2\alpha_{2}}{f}}{\partial{\delta^{\alpha_{1}+2\alpha_{2}}}}(x,-\delta_{n}^{(2)},v_{n}^{(2)})$ up to the order $4-|\alpha|$. Similar to the argument of Case a.1, the assumption of Case b.1 is sufficient to guarantee that $\overline{R}(x)/\overline{D}_{n} \to 0$. 
\paragraph{Step 2 - Non-vanishing coefficients and Fatou's argument} Assume that all the coefficients $B_{n,l}/\overline{D}_{n} \to 0$ for all $1 \leq l \leq 8$ as $n \to \infty$. Remind from part (a) that we denote
\begin{eqnarray}
\overline{M}_{n} : = \max \left\{|\delta_{n}^{(1)} - \delta_{n}^{(2)}|,|v_{n}^{(1)} - v_{n}^{(2)}|^{1/2} \right\}. \nonumber
\end{eqnarray} 
Additionally, we also denote $\delta_{n}^{(2)}/\overline{M}_{n} \to x$, $(\delta_{n}^{(2)} - \delta_{n}^{(1)})/\overline{M}_{n} \to y$, and $(v_{n}^{(1)} - v_{n}^{(2)})/\overline{M}_{n}^{2} \to z$ as $n \to \infty$ where at least one from $y$ and $z$ is different from 0. Due to the assumption that $\delta_{n}^{(1)}\delta_{n}^{(2)} \geq 0$, we have $x(x-y) \geq 0$. Now, by dividing both the numerator and the denominator of $B_{n,l}/D_{n}$ by $\overline{M}_{n}^{l}$ as $1 \leq l \leq 4$, as $n \to \infty$, we have the following system of polynomial equations
\begin{eqnarray}
y^{2} + z -2xy = 0, \nonumber \\
\dfrac{y^{4}}{4!}+\dfrac{y^{2}z}{4} + \dfrac{z^{2}}{8} - \dfrac{xyz}{2} + \dfrac{x^{2}z}{2} - \dfrac{xy^{3}}{6} + \dfrac{x^{2}y^{2}}{2} - \dfrac{2x^{3}y}{3} = 0. \nonumber
\end{eqnarray}
When $x=0$, the above system of polynomial equations leads to $y = z = 0$, which is a contradiction with the assumption that at least one of $y,z$ is different from 0. When $x \neq 0$, the above system of polynomial equations leads to $y^{3} - 4xy^{2} + 6x^{2}y - 4x^{3} = 0$, which leads to $y=2x$ ---  a contradiction to the condition $x(x-y) \geq 0$ and $x \neq 0$. Therefore, not all of the coefficients $B_{n,l}/\overline{D}_{n} \to 0$ as $n \to \infty$. From here, using the same proof argument as that of Case a.1 in part (a), we achieve the conclusion that Case b.1 cannot hold. 
\paragraph{Case b.2:} $\delta_{n}^{(1)}/\delta_{n}^{(2)} \not \to 1$ as $n \to \infty$ and $\delta_{n}^{(1)}\delta_{n}^{(2)} < 0$ for all $n$. Under this case, we have
\begin{eqnarray}
\overline{D}_{n} = |\delta_{n}^{(1)} + \delta_{n}^{(2)}|^{4} + |v_{n}^{(1)} - v_{n}^{(2)}|^{2}. \nonumber
\end{eqnarray}
By means of Taylor expansion up to the fourth order, we obtain the following representation
\begin{eqnarray}
& & \hspace{- 4 em} \dfrac{g(x,\delta_{n}^{(1)},v_{n}^{(1)}) - g(x,\delta_{n}^{(2)},v_{n}^{(2)})}{\overline{D}_{n}} \nonumber \\
& = & \dfrac{\dfrac{1}{2}(f(x,-\delta_{n}^{(1)},v_{n}^{(1)}) - f(x,\delta_{n}^{(2)},v_{n}^{(2)})) + \dfrac{1}{2}(f(x,\delta_{n}^{(1)},v_{n}^{(1)}) - f(x,-\delta_{n}^{(2)},v_{n}^{(2)}))}{\overline{D}_{n}} \nonumber \\
& = & \dfrac{\sum \limits_{\alpha=1}^{8} C_{n,l} \dfrac{\partial^{l}{f}}{\partial{\delta^{l}}}(x,\delta_{n}^{(2)},v_{n}^{(2)})+\widetilde{R}(x)}{\overline{D}_{n}}, \nonumber
\end{eqnarray}
where the formulations of $C_{n,l}$ and $\overline{R}_{1}(x)$ are as follows
\begin{eqnarray}
C_{n,l} & = & \dfrac{1}{2} \sum \limits_{\alpha_{1},\alpha_{2}} \dfrac{1}{2^{\alpha_{2}}} \dfrac{(-\delta_{n}^{(2)} - \delta_{n}^{(1)})^{\alpha_{1}}(v_{n}^{(1)} - v_{n}^{(2)})^{\alpha_{2}}}{\alpha_{1}!\alpha_{2}!} \nonumber \\
& & + \dfrac{1}{2} \sum \limits_{\alpha_{1},\alpha_{2},\tau} \dfrac{1}{2^{\alpha_{2}}}\dfrac{2^{\tau}(-\delta_{n}^{(2)})^{\tau}(\delta_{n}^{(1)}+\delta_{n}^{(2)})^{\alpha_{1}}(v_{n}^{(1)}-v_{n}^{(2)})^{\alpha_{2}}}{\tau!\alpha_{1}!\alpha_{2}!}, \nonumber \\
\widetilde{R}(x) & = & \dfrac{1}{2} \widetilde{R}_{1}(x) + \dfrac{1}{2} \widetilde{R}_{2}(x) + \sum \limits_{|\alpha| \leq 4} \dfrac{1}{2^{\alpha_{2}}}\dfrac{c^{\alpha_{1}}(\delta_{n}^{(1)} + \delta_{n}^{(2)})^{\alpha_{1}}(v_{n}^{(1)} - v_{n}^{(2)})^{\alpha_{2}}}{\alpha_{1}!\alpha_{2}!}\widetilde{R}_{2,\alpha}(x). \nonumber
\end{eqnarray}
Here, the ranges of $\alpha_{1},\alpha_{2}$ in the first sum of $C_{n,l}$ satisfy $\alpha_{1}+2\alpha_{2} = l$, $1 \leq |\alpha| \leq 4$ while the ranges of $\alpha_{1},\alpha_{2},\tau$ in the second sum of $C_{n,l}$ satisfy $\alpha_{1}+2\alpha_{2}+\tau = l$, $0 \leq \tau \leq 4-|\alpha|$, and $1 \leq |\alpha| \leq 4$. Additionally, $\widetilde{R}_{1}(x)$ is a Taylor remainder from expanding $f(x,-\delta_{n}^{(1)},v_{n}^{(1)})$ around $f(x,\delta_{n}^{(2)},v_{n}^{(2)})$ up to the fourth order, $\widetilde{R}_{2}(x)$ is a Taylor remainder from expanding $f(x,\delta_{n}^{(1)},v_{n}^{(1)})$ around $f(x,-\delta_{n}^{(2)},v_{n}^{(2)})$ up to the fourth order, and $\widetilde{R}_{2,\alpha}(x)$ is a Taylor remainder from expanding $\dfrac{\partial^{\alpha_{1}+2\alpha_{2}}{f}}{\partial{\delta^{\alpha_{1}+2\alpha_{2}}}}(x,-\delta_{n}^{(2)},v_{n}^{(2)})$ around $\dfrac{\partial^{\alpha_{1}+2\alpha_{2}}{f}}{\partial{\delta^{\alpha_{1}+2\alpha_{2}}}}(x,\delta_{n}^{(2)},v_{n}^{(2)})$ up to the order $4-|\alpha|$. Due to the assumption of Case b.2, we can check that $\|\widetilde{R}(x)\|_{\infty}/\overline{D}_{n} \to 0$ as $n \to \infty$. 

Assume that all the coefficients $C_{n,l}/\overline{D}_{n} \to 0$ for all $1 \leq l \leq 8$ as $n \to \infty$. We denote
\begin{eqnarray}
\widetilde{M}_{n} : = \max \left\{|\delta_{n}^{(1)} + \delta_{n}^{(2)}|,|v_{n}^{(1)} - v_{n}^{(2)}|^{1/2} \right\}. \nonumber
\end{eqnarray} 
From the definition of $\widetilde{M}_{n}$, we can denote $\delta_{n}^{(2)}/\widetilde{M}_{n} \to x_{1}$, $(\delta_{n}^{(2)} + \delta_{n}^{(1)})/\widetilde{M}_{n} \to y_{1}$, and $(v_{n}^{(1)} - v_{n}^{(2)})/\widetilde{M}_{n}^{2} \to z_{1}$ as $n \to \infty$ where at least one from $y_{1}$ and $z_{1}$ is different from 0. Due to the assumption that $\delta_{n}^{(1)}\delta_{n}^{(2)} < 0$, we have $x_{1}(y_{1}-x_{1}) \leq 0$. Now, by dividing both the numerator and the denominator of $C_{n,l}/D_{n}$ by $\widetilde{M}_{n}^{4}$ as $1 \leq l \leq 4$, as $n \to \infty$, we have the following system of polynomial equations
\begin{eqnarray}
y_{1}^{2} + z_{1} -2x_{1}y_{1} = 0, \nonumber \\
\dfrac{y_{1}^{4}}{4!}+\dfrac{y_{1}^{2}z_{1}}{4} + \dfrac{z_{1}^{2}}{8} - \dfrac{x_{1}y_{1}z_{1}}{2} + \dfrac{x_{1}^{2}z_{1}}{2} - \dfrac{x_{1}y_{1}^{3}}{6} + \dfrac{x_{1}^{2}y_{1}^{2}}{2} - \dfrac{2x_{1}^{3}y_{1}}{3} = 0. \nonumber
\end{eqnarray}
If $x_{1}=0$, the above system leads to $y_{1}=z_{1}=0$, which is a contradiction with the assumption of $y_{1},z_{1}$. As $x_{1} \neq 0$, the above system of polynomial equations leads to $y_{1}=2x_{1}$ ---  a contradiction to the condition $x_{1}(y_{1}-x_{1}) \leq 0$ and $x_{1} \neq 0$. Therefore, not all of the coefficients $C_{n,l}/\overline{D}_{n} \to 0$ as $n \to \infty$. From here, using the same proof argument as that of Case a.1 in part (a), we achieve the conclusion that Case b.2 cannot hold. 
\paragraph{Case b.3:} $\delta_{n}^{(1)}/\delta_{n}^{(2)} \to 1$ as $n \to \infty$. Under this assumption, we have $\delta_{n}^{(1)}\delta_{n}^{(2)} >0$ as $n$ is sufficiently large. Without loss of generality, we assume that $\delta_{n}^{(1)}\delta_{n}^{(2)} > 0$ for all $n$. Therefore, we have
\begin{eqnarray}
\overline{D}_{n} = |\delta_{n}^{(1)} - \delta_{n}^{(2)}|^{4} + |v_{n}^{(1)} - v_{n}^{(2)}|^{2}. \nonumber
\end{eqnarray}
Remind from case b.1 that we have the following representation
\begin{eqnarray}
\dfrac{g(x,\delta_{n}^{(1)},v_{n}^{(1)}) - g(x,\delta_{n}^{(2)},v_{n}^{(2)})}{\overline{D}_{n}} = \dfrac{\sum \limits_{l=1}^{8} B_{n,l}\dfrac{\partial^{l}{f}}{\partial{\delta^{l}}}(x,-\delta_{n}^{(2)},v_{n}^{(2)}) + \overline{R}(x)}{\overline{D}_{n}}. \nonumber
\end{eqnarray}
The main challenge in Case b.3 is that $\|\overline{R}(x)\|_{\infty}/\overline{D}_{n} \not \to 0$ as $n \to \infty$. To avoid this issue, we will utilize the technique in Case a.2 of the proof of Theorem \ref{theorem:convergence_rate_location_scale_Gaussian}. In particular, we will demonstrate two key properties: $\|\overline{R}(x)\|_{\infty}/\max \limits_{1 \leq l \leq 8} |B_{n,l}| \to 0$ and $\max \limits_{1 \leq l \leq 8} |B_{n,l}|/\overline{D}_{n} \not \to 0$ as $n \to \infty$. 

Under the settings of Case a.2.1 and Case a.2.2 in the proof of part (a), with the same argument as that in these cases, we have $|B_{n,2}|/\overline{D}_{n} \not \to 0$ and $\|\overline{R}(x)\|_{\infty}/|B_{n,2}| \to 0$. Therefore, we have $\overline{R}(x)/\max \limits_{1 \leq l \leq 8} |B_{n,l}| \to 0$ and $\max \limits_{1 \leq l \leq 8} |B_{n,l}|/\overline{D}_{n} \not \to 0$ under the settings of Case a.2.1 and Case a.2.2. It implies that we only need to focus on the setting that 
\begin{align}
|v_{n}^{(1)} - v_{n}^{(2)}|/\biggr\{|\delta_{n}^{(1)} - \delta_{n}^{(2)}||\delta_{n}^{(1)} + \delta_{n}^{(2)}|\biggr\} \to 1. \nonumber
\end{align}
Without loss of generality, we assume that $(v_{n}^{(1)} - v_{n}^{(2)})/\biggr\{(\delta_{n}^{(1)}-\delta_{n}^{(2)})(\delta_{n}^{(1)}+\delta_{n}^{(2)})\biggr\} \to 1$ as the argument for the setting that this ratio goes to -1 is similar. Under this setting, we can easily check that
\begin{eqnarray}
|B_{n,4}|/\biggr\{|\delta_{n}^{(1)} - \delta_{n}^{(2)}||\delta_{n}^{(2)}|^{3}\biggr\} \to 4. \nonumber
\end{eqnarray}
Therefore, as $n$ is sufficiently large, we have 
\begin{align}
|B_{n,4}| \gtrsim |\delta_{n}^{(1)} - \delta_{n}^{(2)}||\delta_{n}^{(2)}|^{3}. \nonumber
\end{align} 
If we have $\max \limits_{1 \leq l \leq 8} |B_{n,l}|/\overline{D}_{n} \to 0$, then $\abss{B_{n,4}}/\overline{D}_{n} \to 0$ leads to $|\delta_{n}^{(1)} - \delta_{n}^{(2)}||\delta_{n}^{(2)}|^{3}/\overline{D}_{n} \to 0$. Therefore, the following holds
\begin{eqnarray}
|v_{n}^{(1)} - v_{n}^{(2)}|^{2} / \biggr\{|\delta_{n}^{(1)} - \delta_{n}^{(2)}||\delta_{n}^{(2)}|^{3}\biggr\} \to \infty, \nonumber
\end{eqnarray}
which means $|v_{n}^{(1)} - v_{n}^{(2)}|/|\delta_{n}^{(2)}|^{2} \to \infty$, which is a contradiction to the assumption that $(v_{n}^{(1)} - v_{n}^{(2)})/\biggr\{(\delta_{n}^{(1)}-\delta_{n}^{(2)})(\delta_{n}^{(1)}+\delta_{n}^{(2)})\biggr\} \to 1$. Thus, we have $\max \limits_{1 \leq l \leq 8} |B_{n,l}|/\overline{D}_{n} \not \to 0$. On the other hand, as $n$ is sufficiently large, we have
\begin{eqnarray}
\dfrac{|\delta_{n}^{(1)} - \delta_{n}^{(2)}|^{\alpha_{1}}|v_{n}^{(1)} - v_{n}^{(2)}|^{\alpha_{2}}\|\overline{R}_{2,\alpha}(x)\|_{\infty}}{\max \limits_{1 \leq l \leq 8} \left\{|B_{n,l}|\right\}} & \leq & \dfrac{O\parenth{|\delta_{n}^{(1)} - \delta_{n}^{(2)}|^{\alpha_{1}}|v_{n}^{(1)} - v_{n}^{(2)}|^{\alpha_{2}}|\delta_{n}^{(2)}|^{4-|\alpha|+\gamma}}}{|\delta_{n}^{(1)}-\delta_{n}^{(2)}||\delta_{n}^{(2)}|^{3}} \nonumber \\
& = & \dfrac{O\parenth{|\delta_{n}^{(1)} - \delta_{n}^{(2)}|^{|\alpha|}|\delta_{n}^{(2)}|^{4-\alpha_{1}+\gamma}}}{|\delta_{n}^{(1)}-\delta_{n}^{(2)}||\delta_{n}^{(2)}|^{3}} \to 0. \nonumber
\end{eqnarray}
It implies that $\|\overline{R}(x)\|_{\infty}/\max \limits_{1 \leq l \leq 8} \left\{|B_{n,l}|\right\} \to 0$. From here, using the same argument as that of Case a.2.3, we obtain the contradiction, which leads to the conclusion that Case b.3 cannot hold. As a consequence, we achieve the conclusion of part (b) of the theorem. 
\subsection{Proof of extra results}
\label{subsec:extra_results}
In this appendix, we provide proof for an additional result with the non-polynomial convergence rate of MLE $\widehat{\delta}_{n}^{\mle}$ under the known variances setting~\eqref{eq: block}.
\begin{proposition}
Under the symmetric regime of the true model~\eqref{eq: block}, we have
\begin{align*}
	\sup \limits_{\delta_{n} \in \Theta} \mathbb{E}_{\delta_{n}} \abss{\widehat{\delta}_{n}^{\mle} - \delta_{n}} \gtrsim n^{-1/r},
\end{align*}
where $\Theta = [-1, 1]$. Here, $\mathbb{E}_{\delta_{n}}$ denotes the expectation taken with respect to product measure with mixture density of $Y_{1},\ldots,Y_{n}$ under the model~\eqref{eq: block}.
\end{proposition}
\begin{proof}
We divide our argument for the proof of this result into two key parts.
\paragraph{Part 1 - Upper bound of Hellinger distance between mixing densities in terms of their corresponding parameters} To obtain the conclusion for this inequality, we first prove the following key result
\begin{align}
\inf \limits_{\delta^{(1)},\delta^{(2)} \in \Theta} h\parenth{g(x,\delta^{(1)}),g(x,\delta^{(2)})}/\abss{\delta^{(1)} - \delta^{(2)}}^{r} = 0 \label{eqn:inequality_third}
\end{align}
for any $r \geq 1$. In fact, we construct two sequences $\left\{\delta_{n}^{(1)}\right\}$ and $\left\{\delta_{n}^{(2)}\right\}$ such that $\delta_{n}^{(1)} = -\delta_{n}^{(2)}$ for all $n \geq 1$. Then, it is clear that $h\parenth{g(x,\delta_{n}^{(1)}),g(x,\delta_{n}^{(2)})} = 0$ for all $n \geq 1$. Therefore, it is straightforward that $h\parenth{g(x,\delta_{n}^{(1)}),g(x,\delta_{n}^{(2)})} \leq \abss{\delta_{n}^{(1)} - \delta_{n}^{(2)}}^{r}$ for any $r \geq 1$. As a consequence, we achieve the conclusion of \eqref{eqn:inequality_third}. 
\paragraph{Part 2 - Le Cam's argument for minimax lower bound} Now, we follow the traditional Le Cam's argument for minimax lower bound to achieve the conclusion with non-polynomial convergence rate of $\widehat{\delta}_{n}^{\mle}$ to $\delta_{n}$~\cite{Yu-97}. In particular, due to the result from \eqref{eqn:inequality_third}, for any $\epsilon_{n}>0$ sufficiently small and any fixed $r \geq 1$, we can find $\delta_{n}^{(1)}$ and $\delta_{n}^{(2)}$ such that $\abss{\delta_{n}^{(1)}-\delta_{n}^{(2)}} = 2\epsilon_{n}$ and $h\parenth{g(x,\delta_{n}^{(1)}),g(x,\delta_{n}^{(2)})} \leq C\epsilon_{n}^{r}$ where $C$ is a fixed positive constant. Invoking Lemma 1 from~\cite{Yu-97}, the following inequality holds
\begin{align}
\sup \limits_{\delta_{n} \in \Theta} \mathbb{E}_{\delta_{n}} \abss{\widehat{\delta}_{n}^{\mle} - \delta_{n}} \geq \sup \limits_{\delta_{n} \in \left\{\delta_{n}^{(1)},\delta_{n}^{(2)}\right\}} \mathbb{E}_{\delta_{n}} |\widehat{\delta}_{n} - \delta_{n}| \geq \epsilon_{n}\brackets{1-V\parenth{g^{n}\parenth{x,\delta_{n}^{(1)}},g^{n}\parenth{x,\delta_{n}^{(2)}}}}, \label{eqn:inequality_fourth}
\end{align}
where $g^{n}\parenth{x,\delta_{n}^{(1)}}$ denotes the density of $n$ i.i.d. samples $Y_{1},\ldots,Y_{n}$. By means of classical inequality between total variation distance and Hellinger distance $V \leq h$, we obtain that
\begin{align}
V\parenth{g^{n}(x,\delta_{n}^{(1)}),g^{n}(x,\delta_{n}^{(2)})} \leq h\parenth{g^{n}(x,\delta_{n}^{(1)}),g^{n}(x,\delta_{n}^{(2)})} \leq \sqrt{1 - \parenth{1-C^2\epsilon_{n}^{2r}}^{n}}. \nonumber
\end{align}
By choosing $C^2\epsilon_{n}^{2r} = 1/n$, it is clear that
\begin{align}
\epsilon_{n}\brackets{1-V\parenth{g^{n}\parenth{x,\delta_{n}^{(1)}},g^{n}\parenth{x,\delta_{n}^{(2)}}}} \gtrsim \epsilon_{n} \gtrsim n^{-1/2r}. \label{eqn:inequality_fifth} 
\end{align}
Combining the results from \eqref{eqn:inequality_fourth} and \eqref{eqn:inequality_fifth}, we achieve the conclusion that 
\begin{align}
\sup \limits_{\delta_{n} \in \Theta} \mathbb{E}_{\delta_{n}} \abss{\widehat{\delta}_{n}^{\mle} - \delta_{n}} \gtrsim n^{-1/r} \nonumber
\end{align}
for any $r \geq 2$.  
\end{proof}

\end{document}